\pdfoutput=1
\documentclass[ALICE,manyauthors]{cernphprep}

\usepackage[comma,square,numbers,sort&compress]{natbib}
\usepackage{hyperref}
\usepackage{booktabs}
\usepackage{lineno}
\usepackage{color}
\usepackage[T1]{fontenc}
\newcommand{\ppip}{\ensuremath{\pi^{+}}\xspace}
\newcommand{\ppim}{\ensuremath{\pi^{-}}\xspace}
\newcommand{\ppi}{\ensuremath{\pi}\xspace}
\newcommand{\ppipm}{\ensuremath{\pi^{\pm}}\xspace}
\newcommand{\ppis}{\ensuremath{\ppip + \ppim}\xspace}

\newcommand{\pkap}{\ensuremath{{\rm K}^{+}}\xspace}
\newcommand{\pkam}{\ensuremath{{\rm K}^{-}}\xspace}
\newcommand{\pka}{\ensuremath{{\rm K}}\xspace}
\newcommand{\pkapm}{\ensuremath{{\rm K}^{\pm}}\xspace}
\newcommand{\pkas}{\ensuremath{\pkap + \pkam}\xspace}

\newcommand{\pkastar}{\ensuremath{{\rm K}^{*}}\xspace}
\newcommand{\pkastarm}{\ensuremath{\overline{\pkastar}}\xspace}
\newcommand{\pkazero}{\ensuremath{{\rm K}^{0}_{\rm S}}\xspace}

\newcommand{\pprp}{\ensuremath{{\rm p}}\xspace}
\newcommand{\pprm}{\ensuremath{\overline{\rm{p}}}\xspace}
\newcommand{\ppr}{\ensuremath{{\rm p}}\xspace}
\newcommand{\pprpm}{\ensuremath{(\pprm) \pprp }\xspace}
\newcommand{\pprs}{\ensuremath{\pprp + \pprm}\xspace}

\newcommand{\dcaXY}{\ensuremath{{\rm DCA}_{\it{xy}}}\xspace}
\newcommand{\dcaZ}{\ensuremath{{\rm DCA}_{\it{z}}}\xspace}
\newcommand{\pbpb}{\ensuremath{{\rm Pb}-{\rm Pb}}\xspace}
\newcommand{\ppb}{\ensuremath{{\rm p}-{\rm Pb}}\xspace}
\newcommand{\pp}{\ensuremath{\rm pp}\xspace}

\newcommand{\s}{\ensuremath{\sqrt{s}}\xspace}
\newcommand{\sF}{\ensuremath{\s~=~5.02}\xspace}
\newcommand{\snn}{\ensuremath{\sqrt{s_{\rm{NN}}}}\xspace}
\newcommand{\snnT}{\ensuremath{\snn~=~2.76}\xspace}
\newcommand{\snnF}{\ensuremath{\snn~=~5.02}\xspace}
\newcommand{\pt}{\ensuremath{p_{\rm{T}}}\xspace}
\newcommand{\meanpt}{\ensuremath{\langle \pt \rangle}\xspace}
\newcommand{\mt}{\ensuremath{m_{\rm{T}}}\xspace}
\newcommand{\qt}{\ensuremath{q_{\rm{T}}}\xspace}

\newcommand{\AAaa}{\ensuremath{{\rm{AA}}}\xspace}
\newcommand{\raa}{\ensuremath{R_{\AAaa}}\xspace}
\newcommand{\taa}{\ensuremath{T_{\AAaa}}\xspace}
\newcommand{\ncoll}{\ensuremath{N_{\rm{coll}}}\xspace}
\newcommand{\dedx}{\ensuremath{{\rm d}E/{\rm d}x}\xspace}

\newcommand{\dndy}{\ensuremath{{\rm d}N/{\rm d}y}\xspace}
\newcommand{\Nch}{\ensuremath{{\it N}_{\rm{ch}}}\xspace}
\newcommand{\dNchdeta}{\ensuremath{\rm{d}\Nch/\rm{d}\eta}\xspace}
\newcommand{\avdNchdeta}{\ensuremath{\langle\dNchdeta\rangle}\xspace}

\newcommand{\gevc}{\ensuremath{{\rm GeV}/c}\xspace}
\newcommand{\mevc}{\ensuremath{{\rm MeV}/c}\xspace}

\newcommand{\Tch}{\ensuremath{T_{\rm{ch}}}\xspace}
\newcommand{\Tc}{\ensuremath{T_{\rm{c}}}\xspace}
\newcommand{\Tkin}{\ensuremath{T_{\rm{kin}}}\xspace}
\newcommand{\Bt}{\ensuremath{\beta_{\rm{T}}}\xspace}
\newcommand{\avBt}{\ensuremath{\langle\Bt\rangle}\xspace}
\newcommand{\EcrossB}{\ensuremath{E\times B}\xspace}

\newcommand{\Nsigma}{\ensuremath{N_{\sigma}}\xspace}
\newcommand{\nsigma}{\ensuremath{\Nsigma}\xspace}
\newcommand{\Refs}{Refs.\xspace}
\newcommand{\Ref}{Ref.\xspace}
\newcommand{\Figs}{Figs.\xspace}
\newcommand{\Fig}{Fig.\xspace}
\newcommand{\Ustat}{Stat. Uncert.\xspace}
\newcommand{\Usyst}{Syst. Uncert.\xspace}
\newcommand{\BR}{B.R.\xspace}

\newcommand{\mc}[3]{\multicolumn{#1}{#2}{#3}}

\begin{document}%

%%%%%%%%%%%%%%%  Title page %%%%%%%%%%%%%%%%%%%%%%%%
\begin{titlepage}
  % CERN-EP-2019-208
  \PHyear{2019}
  \PHnumber{208}      % required, will be obtained from PH
  \PHdate{01 October}  % required, will be obtained from PH
  %

  %%% Put your own title + short title here:
  \title{Production of charged pions, kaons and (anti-)protons in \pbpb and inelastic pp collisions at \textbf{\snn}~=~5.02 TeV}
  \ShortTitle{\ppi$/$K$/$p production in \pbpb and MB pp collisions at \snnF\,TeV}   % appears on right page headers

  %%% Do not change the next lines
  \Collaboration{ALICE Collaboration\thanks{See Appendix~\ref{app:collab} for the list of collaboration members}}
  \ShortAuthor{ALICE Collaboration} % appears on left page headers, do not change

  \begin{abstract}

    Mid-rapidity production of \ppipm, \pkapm and \pprpm measured by the ALICE experiment at the LHC, in \pbpb and inelastic pp collisions at \snnF\,TeV, is presented.
    The invariant yields are measured over a wide transverse momentum (\pt) range from hundreds of \mevc up to 20 \gevc.
    The results in \pbpb collisions are presented as a function of the collision centrality, in the range 0$-$90\%.
    The comparison of the \pt-integrated particle ratios, i.e.\ proton-to-pion (p/\ppi) and kaon-to-pion (K/\ppi) ratios, with similar measurements in \pbpb collisions at \snnT\,TeV show no significant energy dependence.
    Blast-wave fits of the \pt spectra indicate that in the most central collisions radial flow is slightly larger at 5.02~TeV with respect to 2.76~TeV.
    Particle ratios (p/\ppi, K/\ppi) as a function of \pt show pronounced maxima at \pt~$\approx$~3~\gevc in central \pbpb collisions.
    At high \pt, particle ratios at 5.02~TeV are similar to those measured in pp collisions at the same energy and in \pbpb collisions at  \snnT\,TeV.
    Using the pp reference spectra measured at the same collision energy of 5.02~TeV, the nuclear modification factors for the different particle species are derived.
    Within uncertainties, the nuclear modification factor is particle species independent for high \pt and compatible with measurements at \snnT\,TeV.
    The results are compared to state-of-the-art model calculations, which are found to describe the observed trends satisfactorily.
  \end{abstract}

\end{titlepage}
\setcounter{page}{2}

\section{Introduction}
 \label{intro}
 Previous observations at the Relativistic Heavy-Ion Collider (RHIC) %~\cite{Trainor:2013bma}
 %~\cite{Arsene:2004fa, Adcox:2004mh, Back:2004je,Adams:2005dq}
 and at the CERN Large Hadron Collider (LHC) %~\cite{Bala:2016hlf}
 %~\cite{Aamodt:2010pa,Aamodt:2010jd, Abelev:2012rv,Aad:2010bu,Chatrchyan:2011sx}
 demonstrated that in high-energy heavy-ion (A--A) collisions, a strongly interacting Quark--Gluon Plasma (sQGP)  \cite{Arsene:2004fa, Adcox:2004mh, Back:2004je, Adams:2005dq, Schukraft:2011na} is formed.
 It behaves as a strongly-coupled near-perfect liquid with a small viscosity-to-entropy ratio $\eta$/s~\cite{Heinz:2013th}.
 The experimental results have led to the development and adoption of a standard theoretical framework for describing the bulk properties of the QGP in these collisions~\cite{Braun-Munzinger:2015hba}.
 In this paradigm, the beam energy dependence is mainly encoded in the initial energy density (temperature) of the QGP.
 After formation, the QGP expands hydrodynamically as a near perfect liquid before it undergoes a chemical freeze-out.
 The chemical freeze-out temperature is nearly beam-energy independent for center-of-mass energy per nucleon pair larger than 10 GeV~\cite{Braun-Munzinger:2015hba,Adamczyk:2017iwn}.
 The hadronic system continues to interact (elastically) until kinetic freeze-out.
 We report in this paper a comprehensive study of bulk particle production at the highest beam energy for A$-$A collisions available at the LHC.
 We probe the highest QGP temperature, to further study this paradigm and address its open questions.

 Transverse momentum distributions of identified particles in \pbpb collisions provide information on the transverse expansion of the QGP and the freeze-out properties of the ensuing hadronic phase.
 By analyzing the \pt-integrated yields in \pbpb collisions it has been shown that hadron yields in high-energy nuclear interactions can be described assuming their production at thermal and chemical equilibrium~\cite{Andronic:2017pug, Becattini:2014hla,Becattini:2004rq,Acharya:2017bso}, with a single chemical freeze-out temperature, \Tch~$\approx$~156~MeV, close to the one predicted by lattice QCD calculations for the QGP-hadronic phase transition, \Tc~=~(154~$\pm$~9)~MeV \cite{Bazavov:2014pvz}.
 Indeed, the \pbpb data from LHC Run~1 \cite{Abelev:2013vea} showed an excellent agreement with the statistical hadronization model with the exception of the proton and antiproton, (\pkastarm)\pkastar and multi-strange particle yields~\cite{Andronic:2017pug,Acharya:2017bso}.
 The deviation could be in part due to interactions in the hadronic phase, which result in baryon-antibaryon annihilation that is most significant for (anti-)protons~\cite{Stock:2013iua, Karpenko:2012yf, Becattini:2012xb, Steinheimer:2012rd}.
 Proposed explanations for the observed discrepancy with respect to the thermal model predictions can be found in \Refs \cite{Steinheimer:2012rd, Becattini:2016xct, Chatterjee:2013yga, Alba:2016hwx, PhysRevC.88.034907}.
 Moreover, at \snnT\,TeV the proton-to-pion ($(\pprs)/(\ppis) \equiv \ppr/\ppi$) ratio exhibits a slight decrease with centrality and a slightly lower value than measured at RHIC.
 New measurements at \snnF\,TeV, which exploit the currently highest medium density, could provide an improved understanding of the particle production mechanisms~\cite{PhysRevC.88.034907}.

 The spectral shapes at low \pt (\pt~$<$~2~\gevc) in central \pbpb collisions at \snnT\,TeV showed a stronger radial flow than that measured at RHIC energies, in agreement with the expectation based on hydrodynamic models~\cite{Bozek:2012qs,Abelev:2013vea}.
 The results for identified particle production at low \pt and higher \snn are useful to further test hydrodynamic predictions.
 %Moreover, the measurement of pion yields down to very low transverse momentum could be used to search for Bose$-$Einstein condensation effects.
 %They have been searched for using heavy-ion data at lower energies \cite{Begun:2015ifa}.
 %According to theoretical predictions, about 5\% of all pions may form the Bose$-$Einstein condensate, producing a significant enhancement of the pion spectra for \pt $<$ 0.1 \gevc.
 %This is however at or below the lowest \pt limit reached in the analysis presented in this paper.

 % At intermediate \pt (2$-$10\,\gevc) the spectra are sensitive to different hadronization mechanisms like quark recombination~\cite{Fries:2003kq, Pop:2004dq,Brodsky:2008qp}.
 %In the most central \pbpb collisions at \snnT\,TeV, the ${\rm p}/\pi$ ratio reaches values larger than 0.8 for \pt $\approx$~3~\gevc, which surpass those for inelastic pp collisions at the same energy~\cite{Abelev:2013xaa,Adam:2015kca}.
 %The enhancement of baryon-to-meson ratios is attributed to the collective hydrodynamic expansion of the system~\cite{Shen:2014vra, Zhao:2017yhj,McDonald:2016vlt}.
 %In the coalescence models~\cite{Greco:2003xt,Fries:2003vb,Minissale:2015zwa},  this is a consequence of hadronization via recombination of quarks from the medium.
 %Therefore, the enhancement at intermediate \pt is an effect of the coalescence of lower~\pt quarks that leads to a larger production of baryons than mesons.
 %The recombination mechanism would anyway require a certain amount of radial flow to describe the baryon-to-meson ratios.

 At intermediate \pt (2$-$10\,\gevc), the particle ratios experimentally show the largest variation and in particular for the baryon-to-meson enhancement several new hadronization mechanisms have been proposed~\cite{Fries:2003kq, Pop:2004dq,Brodsky:2008qp}. In the most central \pbpb collisions at \snnT\,TeV, the ${\rm p}/\pi$ ratio reaches values larger than 0.8 for  \pt $\approx$~3~\gevc, which surpass those for inelastic pp collisions at the same energy~\cite{Abelev:2013xaa,Adam:2015kca}. An intermediate \pt enhancement of heavier hadrons over lighter hadrons is expected from the collective hydrodynamic expansion of the system alone~\cite{Shen:2014vra, Zhao:2017yhj,McDonald:2016vlt}. In coalescence models~\cite{Greco:2003xt,Fries:2003vb,Minissale:2015zwa}, which requires radial flow as well, baryon-to-meson ratios are further enhanced at intermediate \pt by the coalescence of lower \pt quarks that leads to a production of baryons (3 quarks) with larger \pt than for mesons (2 quarks).
 The baryon-to-meson ratio decreases at high \pt and reaches the values observed in pp collisions as a consequence of the increasing importance of parton fragmentation.
 The observation of a qualitatively similar enhancement of the kaon-to-pion ($(\pkas)/(\ppis) \equiv \pka/\ppi$) ratio in central \pbpb collisions with respect to inelastic pp collisions~\cite{Abelev:2014laa,Adam:2015kca} supports an interpretation based on the collective radial expansion of the system that affects heavier particles more.

 For high \pt (\pt~$>10$\,\gevc), measurements of the production of identified particles in \pbpb collisions relative to inelastic pp collisions contribute to the study of hard probes propagating through the medium.
 This offers the possibility to determine the properties of the QGP like the transport coefficient $\hat{q}$ \cite{Burke:2013yra} and the space-time profile of the bulk medium in terms of local temperature and fluid velocity~\cite{Wang:2014xda}.
 The modification of particle production is quantified with the nuclear modification factor, \raa, defined as:

 \begin{equation}
   \raa = \frac{{\rm d}^{2}N^{\AAaa}/({\rm d}y{\rm d}\pt)}{\langle \taa \rangle {\rm d}^{2}\sigma^{\pp}/({\rm d}y{\rm d}\pt)},
 \end{equation}

 where ${\rm d}^{2}N^{\AAaa}/({\rm d}y{\rm d}\pt)$ is the particle yield in nucleus-nucleus collisions and $\sigma^{\pp}$
 is the production cross section in pp collisions.
 The average nuclear overlap function is represented by $\langle \taa\rangle$ and is obtained from a Glauber model calculation~\cite{Abelev:2013qoq}.
 It is related to the average number of binary nucleon-nucleon collisions $\langle \ncoll \rangle$, and the total inelastic nucleon$-$nucleon cross section, $\sigma^{\rm NN}_{\rm{INEL}}$  = ($67.6 \pm 0.6$)\,mb at \snnF\,TeV \cite{Loizides:2017ack}, as $\langle \taa\rangle$~=~$\langle \ncoll\rangle$/$\sigma^{\rm NN}_{\rm{INEL}}$. \\
Several measurements of \raa at high \pt for different \snn~\cite{Adler:2003au,Adams:2003kv,Aamodt:2010jd,CMS:2012aa,Aad:2015wga,Abelev:2007ra,Agakishiev:2011dc} support the formation of a dense partonic medium in heavy-ion collisions where hard partons lose energy via a combination of elastic and inelastic collisions with the constituents of the sQGP~\cite{Qin:2015srf}.
 Results from \pbpb collisions at \snnT~TeV showed that within uncertainties, the suppression is the same for pions, kaons and (anti-)protons~\cite{Adam:2015kca}.
 Moreover, the inclusive charged-particle nuclear modification factor measured in \pbpb collisions at 5.02\,TeV shows that the suppression continues to diminish for \pt above 100\,\gevc~\cite{Khachatryan:2016odn} while the suppression of jets saturates at a value of 0.5~\cite{Khachatryan:2016jfl}.
 Particle production at high transverse momentum has also been studied as a function of the Bjorken energy density~\cite{Adare:2015cua} and path length~\cite{Christiansen:2013hya, Ortiz:2017cul, Acharya:2018eaq}.
 The results show interesting scaling properties which can be further tested using LHC data at higher energies.

 In this paper, the measurement of \pt spectra of \ppipm, \pkapm and \pprpm in inelastic pp and \pbpb collisions at \snnF~TeV over a wide \pt range, from 100 \mevc for pions, 200~\mevc for kaons, and 300~\mevc for (anti-)protons, up to 20 \gevc for all species, are presented.
 Particles are identified by combining several particle identification (PID) techniques based on specific ionization energy loss (\dedx) and time-of-flight measurements, Cherenkov radiation detection and the identification of the weak decays of charged kaons via their kink-topology.
 The article is organized as follows: Sec.~\ref{sec2} outlines the analysis details including the track and event selections as
 well as the particle identification strategies.
 The obtained results are discussed in Sec.~\ref{sec3}.
 Section~\ref{sec4} presents the comparison of data with model predictions.
 Finally, Sec.~\ref{sec5} contains a summary of the main results.

\section{Data analysis} \label{sec2}

 In this paper the measurements obtained with the central barrel of the ALICE detector, which has full azimuthal coverage around mid-rapidity, $|\eta| < $ 0.8~\cite{Aamodt:2008zz}, are presented.
 A detailed description of the ALICE detector can be found in \Ref~\cite{Abelev:2014ffa}.
 
The pp results were obtained from the analysis of  $\approx$~1.2$\times$10$^8$ minimum bias pp collisions, collected in 2015. The \pbpb analysis with ITS and TOF uses $\approx$~5$\times$10$^6$ minimum bias \pbpb collisions, collected in 2015. The \pbpb analysis where PID is provided by the TPC, the High Momentum Particle Identification (HMPID) detector and the kink decay topology requires more statistics and uses the full data sample collected in 2015 corresponding to $\approx$~6.5$\times$10$^7$ \pbpb collisions.
% The results were obtained from the analysis of  $\approx$~1.2$\times$10$^8$ and $\approx$~5$\times$10$^6$ minimum bias pp and \pbpb collisions,  respectively, collected in 2015.
% The \pbpb analysis where PID is provided by the TPC, the High Momentum Particle Identification (HMPID) detector and the kink decay topology requires larger statistics and uses the full data sample collected in 2015 corresponding to $\approx$~6.5$\times$10$^7$ \pbpb collisions.

 Both in pp and \pbpb collisions, the interaction trigger is provided by a pair of forward scintillator hodoscopes, the V0 detectors, which cover the pseudorapidity ranges 2.8~$< \eta <$~5.1 (V0A) and -~3.7~$< \eta <$~-~1.7 (V0C)~\cite{Abbas:2013taa}.
 The minimum bias trigger is defined as a coincidence between the V0A  and the V0C trigger signals.
 The V0 detector signals, which are proportional to the charged-particle multiplicities, are used to divide the \pbpb event sample into centrality classes, defined in terms of percentiles of the hadronic cross section~\cite{Abelev:2013qoq}.
 A Glauber Monte Carlo model is fitted to the V0 amplitude distribution to compute the fraction of the hadronic cross section corresponding to any given range of V0 amplitudes.
 The 90$-$100\% centrality class has substantial contributions from QED processes ($\approx$ 20\%) \cite{Abelev:2013qoq} and its low track multiplicity presents difficulties in the extraction of the trigger inefficiency; it is therefore not included in the results presented here.
 Also, an offline event selection is used to remove beam background events.
 It employs the information from two Zero Degree Calorimeters (ZDCs) positioned at 112.5\,m on either side of the nominal interaction point.
 Beam background events are removed by using the V0 timing information and the correlation between the sum and the difference of times measured in each of the ZDCs~\cite{Abelev:2014ffa}. \\

 The central barrel detectors are located inside a solenoidal magnet providing a magnetic field of 0.5~T and are used for tracking and particle identification.
 The innermost barrel detector is the Inner Tracking System (ITS)~\cite{Aamodt:2010aa}, which consists of six layers of silicon devices grouped in three detector systems (from the innermost outwards): the Silicon Pixel Detector (SPD), the Silicon Drift Detector (SDD) and the Silicon Strip Detector (SSD).
 The Time Projection Chamber (TPC), the main central-barrel tracking device, follows outwards.
 The results are presented for primary particles, defined as particles with a mean proper lifetime $\tau~>~1$~cm/$c$ which are either produced directly in the
 interaction or from decays of particles with $\tau~<~1$~cm/$c$, restricted to decay chains leading to the interaction~\cite{ALICE-PUBLIC-2017-005}.
 To limit the contamination due to secondary particles and tracks with wrongly associated hits and to ensure high tracking efficiency,  tracks are required to cross at least 70 TPC readout rows with a $\chi^2$ normalized to the number of TPC space-points (``clusters''), $\chi^2$/NDF, lower than 2.
 Tracks are also required to have at least two hits reconstructed in the ITS out of which at least one is in the SPD layers and to have a Distance of Closest Approach (DCA) to the interaction vertex in the direction parallel to the beam axis ($z$), $|\dcaZ|~<$~2~cm.
 A \pt-dependent selection on the DCA in the transverse plane (\dcaXY) of the selected tracks to the primary vertex is also applied \cite{Adam:2015qaa}.
 Furthermore, the tracks associated with the decay products of weakly decaying kaons (``kinks'') are rejected.
 The latter selection is not applied in the study of kaon production from kink decay topology.
 The primary vertex position is determined from tracks, including short track segments reconstructed in the SPD~\cite{ALICE:2012xs}.
 The position of the primary vertex along the beam axis is required to be within 10 cm from the nominal interaction point.
 The position along $z$ of the SPD and track vertices are required to be compatible within 0.5 cm.
 This ensures a uniform acceptance and reconstruction efficiency in the pseudorapidity region $|\eta| <$ 0.8 and rejects pileup events in pp collisions.
 Different PID detectors are used for the identification of the different particle species.
 Ordering by \pt, from lowest to highest, the results are obtained using the \dedx measured in the ITS and the TPC~\cite{Alme:2010ke}, the time of flight measured in the Time-Of-Flight (TOF) detector~\cite{Akindinov:2013tea}, the Cherenkov angle measured in the High-Momentum Particle IDentification detector (HMPID)~\cite{Martinengo:2011zza} and the TPC \dedx in the relativistic rise region of the Bethe$-$Bloch curve.
 The performance of these devices is reported in \Ref~\cite{Abelev:2014ffa}.

 \subsection{Particle identification strategy}

   For the analysis presented here, pions, kaons and (anti-)protons have been identified following the same analysis techniques as in the previous ALICE measurements.
   The ITS, TPC (low \pt) and TOF analyses are described in \cite{Abelev:2012wca, Abelev:2013vea, Abelev:2013haa}, while the HMPID and TPC (high \pt) analyses are documented in \cite{Abelev:2014laa, Adam:2015kca, Adam:2016dau}.
   The kink analysis is described in \cite{Adam:2015qaa}.
   In this paper, only the most relevant aspects of each specific analysis are described.

   In most analyses, the yield is extracted from the number-of-sigma (\nsigma) distribution.
   This quantity is defined as:

   \begin{equation}
     \nsigma^{i} = \frac{(signal - \langle signal \rangle_{i})}{\sigma_{i}}
   \end{equation}

   where $i$ refers to a given particle species ($i = \pi$, K, p), $signal$ is the detector PID signal (e.g. \dedx), and $\langle signal \rangle_{i}$ and $\sigma_{i}$ are the expected average PID signals in a specific detector and its standard deviation, respectively.

\begin{figure}[htb]
     \centerline{
       \includegraphics[width=1.0\columnwidth]{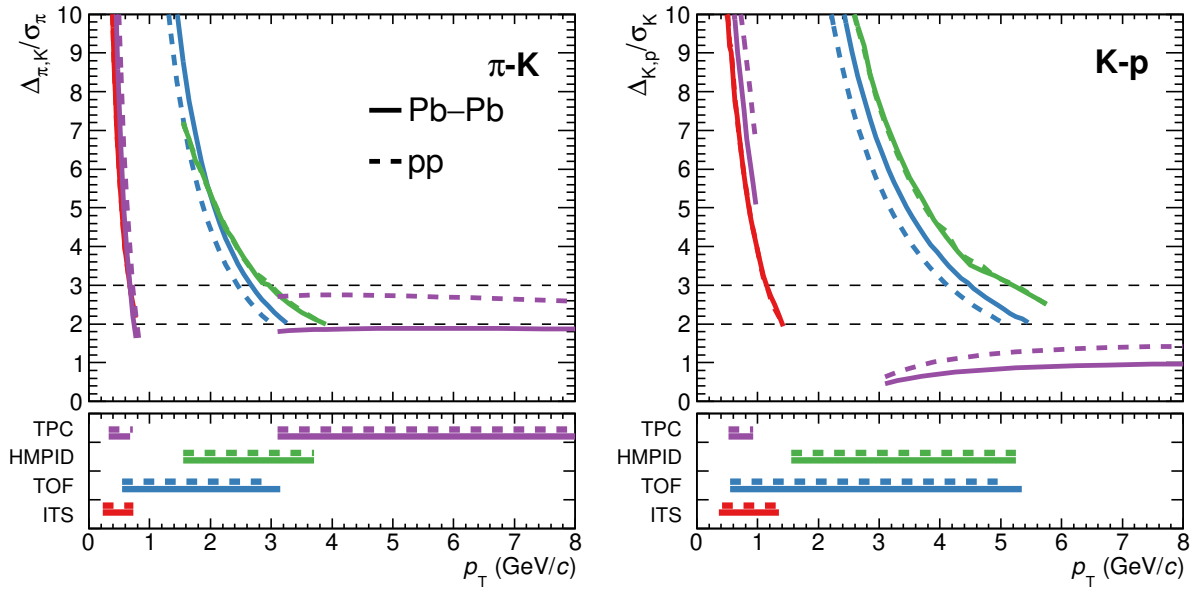}}
     \caption{Separation power of hadron identification in the ITS (red), TPC (magenta), TOF (blue), and HMPID (green) as a function of \pt at mid-rapidity for inelastic pp and 0$-$90\% \pbpb collisions at \snnF TeV.
       The left (right) panel shows the separation of pions and kaons (kaons and protons), expressed as the distance between the expected average PID signal divided by the resolution for the pion (kaon) (see eq. \ref{separ}), averaged over $|\eta| <$~0.5.
       The lower panels show the range in which the ITS, TPC, TOF, and HMPID provide a separation power $\approx$ 2$\sigma$ or larger.}
     \label{separation}
   \end{figure}

   Figure \ref{separation} shows the pion$-$kaon and kaon$-$proton separation power as a function of \pt for ITS, TPC, TOF and HMPID.
   The separation power is defined as follows:

   \begin{equation} \label{separ}
     Sep_{(\pi,\pka)} = \frac{\Delta_{\pi,\pka}}{\sigma_{\pi}} = \frac{|{\langle signal \rangle_{\pi} - \langle signal \rangle_{\pka}|}}{\sigma_{\pi}};  \hspace{0.5cm} Sep_{(\pka,\ppr)} = \frac{\Delta_{\pka,\ppr}}{\sigma_{\pka}} = \frac{|{\langle signal \rangle_{\pka} - \langle signal \rangle_{\ppr}|}}{\sigma_{\pka}}
   \end{equation}

   %= \Delta_{\pi,\rm{K}}/\sigma_{\pi}

   Note that the response for the individual detectors is momentum ($p$) dependent.
   However, since results are reported in transverse momentum bins, the separation power as a function of \pt has been evaluated, averaging the momentum-dependent response over the pseudorapidity range $|\eta| <$ 0.5.
   In Tab.~\ref{tablePIDRange} the transverse momentum ranges covered with each PID technique in the analysis are reported for pions, kaons and (anti-)protons.

   % Requires the booktabs if the memoir class is not being used
   \begin{table}[htbp]
     \centering
     \caption{Transverse momentum ranges (in \gevc) and the corresponding PID methods for pions, kaons and (anti-)protons.
       Values in parenthesis refer to pp analysis.
     }
     %\topcaption{Table captions are better up top} % requires the topcapt package
     \begin{tabular}{lccccc} % Column formatting, @{} suppresses leading/trailing space
       \toprule
       %\multicolumn{3}{c}{Particle species} \\
       %\cmidrule(l){2-3} % Partial rule. (r) trims the line a little bit on the right; (l) & (lr) also possible
       Technique      & \ppipm [\gevc] & \pkapm [\gevc]     & p and  \pprm [\gevc] \\
       \midrule
       ITS            & 0.1$-$0.7      & 0.2$-$0.5(0.6)     & 0.3$-$0.6 (0.65)     \\
       TPC (low \pt)  & 0.25$-$0.7     & 0.25$-$0.45        & 0.4$-$0.8            \\
       TPC (high \pt) & 3.0$-$20.0     & 4.0$-$20.0         & 4.0$-$20.0           \\
       %   TOF                                & 0.6$-$2.5 (3.5)    &  1.00 (0.65)$-$2.5 (3.5)  &  0.8$-$4.0 (4.5)  \\
       TOF            & 0.6$-$ 3.5     & 1.00 (0.65)$-$ 3.5 & 0.8$-$4.5            \\
       HMPID          & 1.5$-$4.0      & 1.50$-$4.0         & 1.5$-$6.0            \\
       Kinks          & $-$            & 0.5$-$6.0 (4.0)    & $-$                  \\
       \bottomrule
     \end{tabular}
     \label{tablePIDRange}
   \end{table}

   \paragraph{ITS analysis.}
     The four outer layers of the ITS provide specific energy-loss measurements.
     The dynamic range of the analog readout of the detector is large enough \cite{Abelevetal:2014dna} to provide \dedx measurements for highly ionizing particles.
     Therefore, the ITS can be used as a standalone low-\pt PID detector in the non-relativistic region where the \dedx is proportional to 1/$\beta^2$.
     For each track, the energy loss fluctuation effects are reduced by using a truncated mean: the average of the lowest two \dedx values in case four values are measured, or a weighted sum of the lowest (weight 1) and the second lowest (weight 1/2), in case only three values are available.
     %Particle identification in the ITS follows the \nsigma strategy.
     The plane ($p$; \dedx) is divided into identification regions where each point is assigned a unique particle identity.
     The identity of a track is assigned based on which \dedx curve the track is closest to, removing in this way the sensitivity to the \dedx resolution.
     To reject electrons, a selection on $|\nsigma^{\pi}| < 2$, is applied.
     % The selection criterium $|\nsigma^{\ppr}| <$ 1 is chosen to reject deuterons.
     Using this strategy, it is possible to identify \ppi and K with an efficiency of about 96-97\% above \pt = 0.3~\gevc, and \pprpm with an efficiency of 91-95\% in the entire \pt range of interest.
     In the lowest \pt bin, the PID efficiency reaches $\approx$~60\%, $\approx$~80\% and $\approx$~91\% for pions, kaons and (anti-)protons, respectively.
     By means of this technique it is possible to identify \ppipm, \pkapm and \pprpm in \pbpb (pp) collisions in the \pt ranges 0.1$-$0.7~\gevc,
     0.2--0.5~(0.6)~\gevc and 0.3--0.6~(0.65)~\gevc, respectively.

     % TOF
   \paragraph{TOF analysis.}
     The analysis with the TOF detector uses the sub-sample of tracks for which a time measurement with TOF is available.
     The time of flight $t_{\rm{TOF}}$ is the difference between the measured particle arrival time $\tau_{\rm{TOF}}$ and the event time $t_{0}$, namely $t_{\rm{TOF}}$~=~$\tau_{\rm{TOF}}$ - $t_0$.
     In the ALICE experiment, the $t_{0}$ value can be obtained with different techniques~\cite{Adam:2016ilk}.
     The best precision on the $t_0$ evaluation is obtained by using the TOF detector itself.
     In this case, the $t_0$ is obtained on an event-by-event basis by using a combinatorial algorithm that compares the measured $\tau_{\rm{TOF}}$ with the expected one under different mass hypotheses.
     The procedure to evaluate $t_0$  with the TOF detector is fully efficient if enough reconstructed tracks are available, which is the case of the 0-80\% \pbpb collisions.
     The resolution on the $t_0$ evaluated with the TOF detector is better than 20 ps if more than 50 tracks are used for its determination.
     This improvement with respect to Run~1 performance~\cite{Adam:2016ilk} is due to improved calibration procedures carried out during Run~2.
     Overall the TOF signal resolution is about 60~ps in central \pbpb collisions.
     In pp and 80-90\% \pbpb collisions the measurement of the event time relies on the T0 detector ($\sigma_{t_{\rm ev}^{\rm T0}}~\approx$~50~ps) \cite{Adam:2016ilk} or, in case it is not available, on the bunch crossing time, which has the worst resolution ($\approx$~200 ps).
     The PID procedure is based on a statistical unfolding of the time-of-flight \nsigma distribution.
     For each \pt bin, the expected shapes for \ppipm, \pkapm and \pprpm are fitted to the $t_{\rm TOF}$ distributions, allowing the three particles to be distinguished when the separation is as low as $\approx 2\sigma$.
     An additional template is needed to account for the tracks that are wrongly associated with a hit in the TOF.
     The templates are built from data as described in \Ref~\cite{Abelev:2013vea}.
     For this purpose the length of measured tracks is used to compute a realistic distribution of the expected time of arrival for each mass hypothesis and the signal shape is reproduced by sampling the parametrized TOF response function (described by a Gaussian with an exponential tail) obtained from data.
     Since the rapidity of a track depends on the particle mass, the fit is repeated for each mass hypothesis.
     TOF analysis makes identification of \ppipm, \pkapm and \pprpm in \pbpb (pp) collisions possible in the \pt ranges 0.60$-$3.50~\gevc, 1.00 (0.65)$-$3.50~\gevc and 0.80-4.50~\gevc, respectively.

     % TPC
   \paragraph{TPC analysis.}
     The TPC provides information for particle identification over a wide momentum range via the specific energy loss~\cite{Abelev:2014ffa}.
     Up to 159 space-points per trajectory can be measured.
     A truncated mean, utilizing 60\% of the available clusters, is employed in the  \dedx determination \cite{Alme:2010ke}.
     The \dedx resolution for the Minimum Ionizing Particle (MIP)  is $\approx$~5.5\% in peripheral and $\approx$~6.5\% in central \pbpb collisions.
     Particle identification on a track-by-track basis is possible in the region of momentum where particles are well separated by more than $3\sigma$.
     This allows the identification of pions, kaons and (anti-)protons within the transverse momentum ranges 0.25-0.70~\gevc, 0.25-0.45~\gevc and 0.45-0.90~\gevc, respectively.

     The TPC \dedx signal in the relativistic rise region  (3~$< \beta\gamma <<$~1000), where the average energy loss increases as $\rm{ln}(\beta\gamma)$, allows identification of charged pions, kaons, and (anti-)protons from  \pt~$\approx$~2$-$3~\gevc up to \pt =~20~\gevc.
     The first step of the TPC high-\pt analysis is the calibration of the PID signal; a detailed description of the the \dedx calibration procedure can be found in \Ref \cite{Abelev:2014laa,Adam:2015kca}.
     Particle identification requires precise knowledge of the $\langle$\dedx$\rangle$ response and resolution $\sigma$.
     This is achieved using the PID signals of pure samples of secondary pions and protons originating from \pkazero and $\Lambda$ decays as well as a sample of tracks selected with TOF.
     In addition, measured \pkazero spectra are used to further constrain the TPC charged kaon response \cite{Adam:2015kca}.
     For different momentum intervals, a sum of four Gaussian functions associated with the pion, kaon, proton and electron signals is fitted to the \dedx distribution.

     % HMPID
   \paragraph{HMPID analysis}
     The HMPID performs identification of charged hadrons based on the measurement of the emission angle of  Cherenkov radiation.
     Starting from the association of a track to the MIP cluster centroid one has to reconstruct the photon emission angle.
     Background, due to other tracks, secondaries and electronic noise, is discriminated exploiting the Hough Transform Method (HTM)~\cite{DiBari:2003wy}.
     Particle identification with the HMPID is based on statistical unfolding.
     In pp collisions, a negligible background allows for the extraction of the particle yields from a three-Gaussian fit to the Cherenkov angle distributions in a narrow transverse momentum range.
     In the case of \pbpb collisions, 
     the Cherenkov angle distribution for a narrow
transverse momentum bin is described by the sum of three Gaussian distributions for \ppipm, \pkapm and \pprpm
for the signal and a sixth-order polynomial function for the background~\cite{Adam:2015kca}.
     %    the three Gaussian distributions for \ppipm, \pkapm and \pprpm in a narrow transverse momentum bin are summed to a sixth-order polynomial function of the Cherenkov angle that strongly increases     with this and that is used to fit the background distribution~\cite{Adam:2015kca}.
     This background is due to misidentification in the high occupancy events: the larger the angle, the larger the probability to find background clusters arising from other tracks or photons in the same event.
     This background is uniformly distributed on the chamber plane.
     The resolution in \pbpb events is the same as in pp collisions ($\approx$ 4 mrad at $\beta \approx$ 1).
     In this analysis, the HMPID provides results in pp and \pbpb collisions in the transverse momentum ranges 1.5$-$4.0 \gevc for \ppipm~and \pkapm, and in 1.5$-$6.0 \gevc for (\pprm)p.

\begin{figure}[htb]
     \centering
             \subfigure[]{\label{fig:a}\includegraphics[scale=0.35]{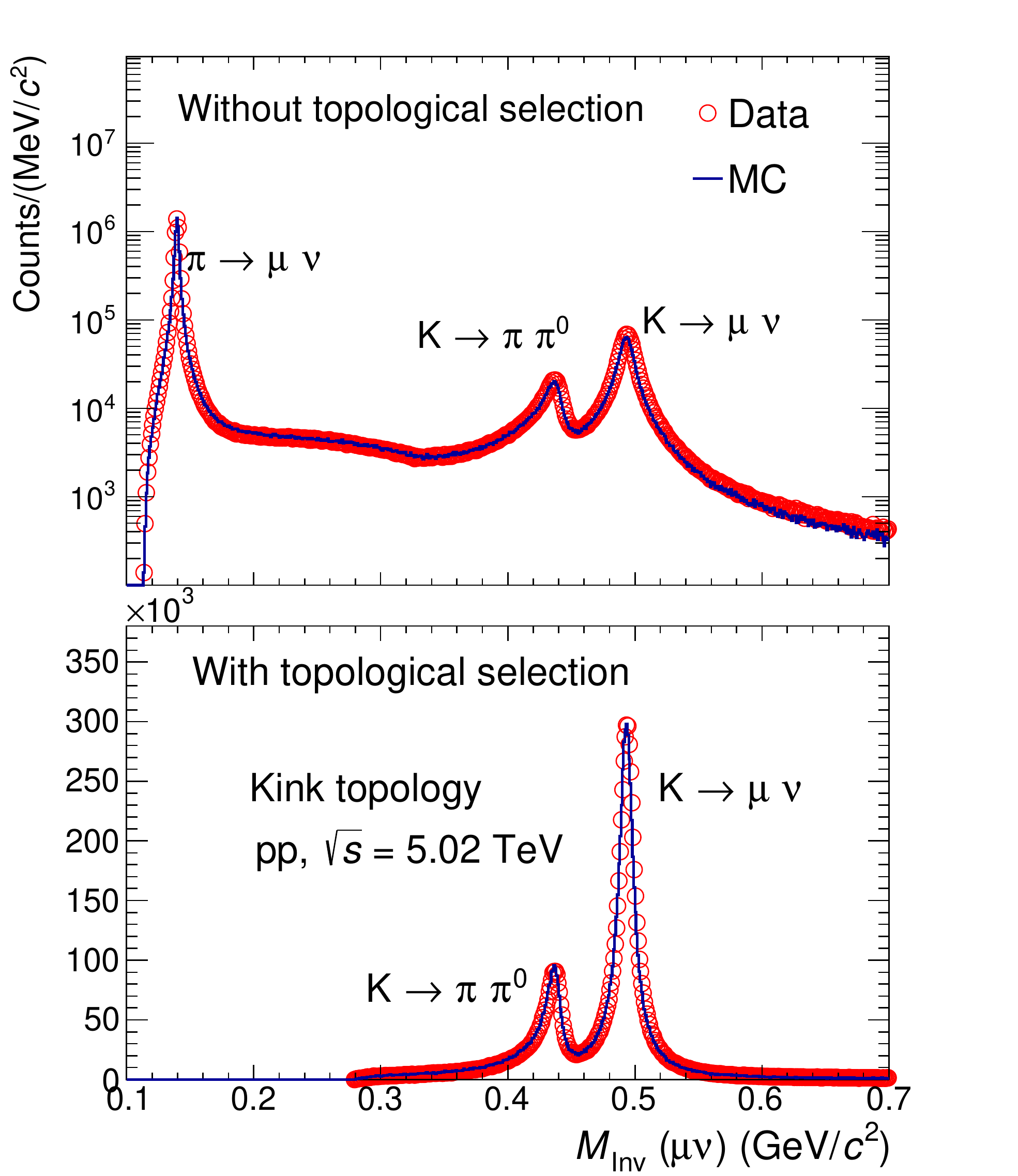}}
             \subfigure[]{\label{fig:b}\includegraphics[scale=0.35]{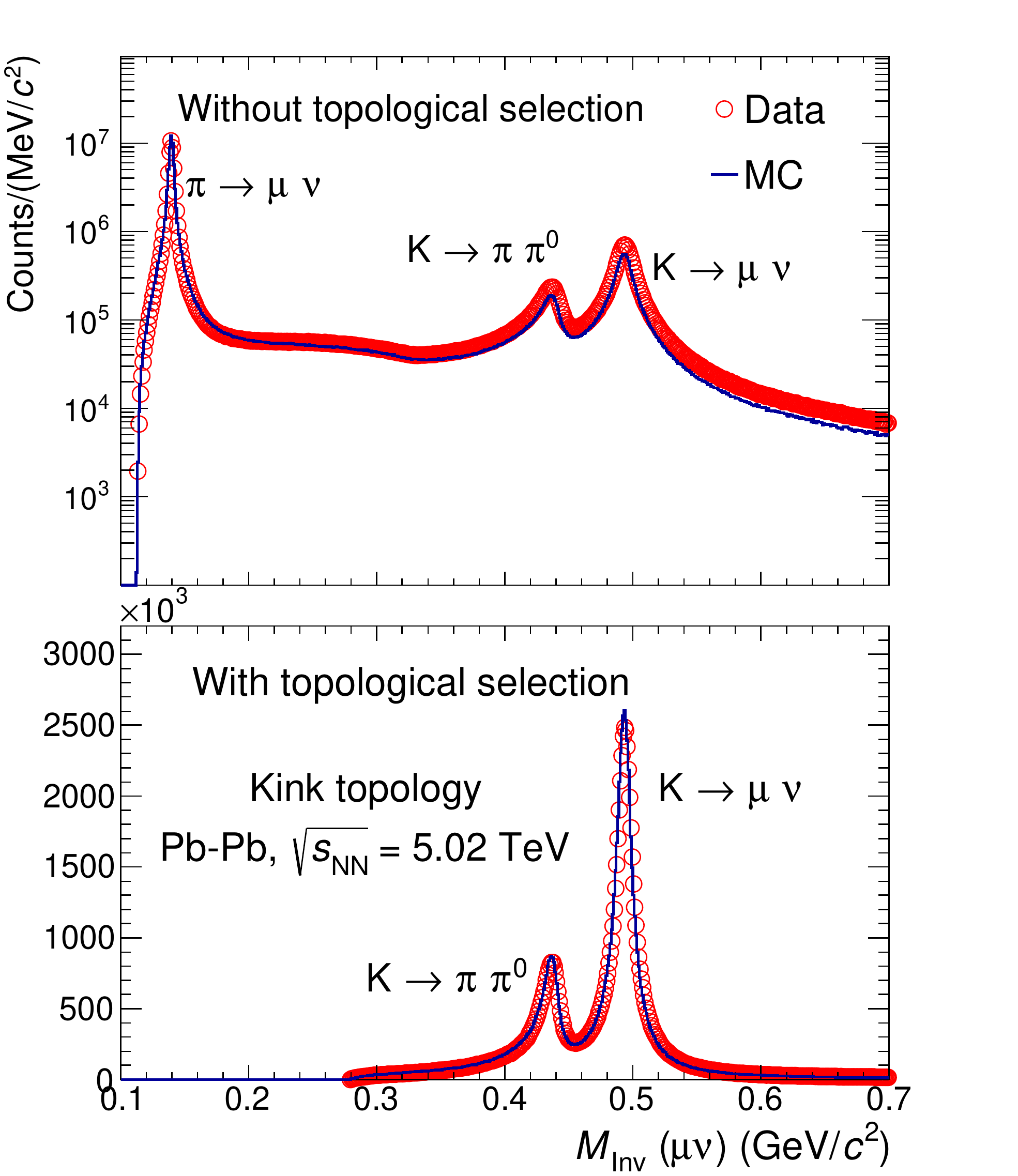}}
       \caption{Invariant mass distribution of identified charged kaons from their decay products in pp (a) and \pbpb collisions (b) at \snnF~TeV.
         The red circles and blue lines represent the experimental data and Monte Carlo simulation, respectively, before (upper) and after (lower) the topological  selection.
         The peak centered at $M_{\mu\nu}~=~0.49$~GeV/$\it{c}^2$ is for the decay channel K $\rightarrow \mu + \nu_{\mu}$ (\BR $= 63.55\%$), whereas the peak centered at $M_{\mu\nu}$~=~0.43~GeV/$\it{c}^2$ is for the decay channel K $\rightarrow \pi + \pi^0$ (\BR = 20.66\%), whose invariant mass is calculated with the wrong mass hypothesis.}
       \label{PbPbKinksperf}
     \end{figure}

     % Kinks
   \paragraph{Kink analysis.}
     In addition to the particle identification techniques mentioned above, charged kaons can also be identified in the TPC using the kink topology of their two-body decay mode (e.g. K$\rightarrow \mu + \nu_{\mu}$) \cite{Adam:2015qaa}.
     With the available statistics, this technique extends PID of charged kaons up to 4 \gevc in pp collisions and up to 6 GeV/c in \pbpb collisions.
     The kink analysis reported here is applied for the first time to \pbpb data.
     For the reconstruction of kaon kink decays, the algorithm is implemented within the fiducial volume of the TPC detector ($130 < R < 200$ cm), to ensure that an adequate number of clusters is found to reconstruct
     the tracks of both the mother and the daughter with the necessary precision to be able to identify the particles.
     The mother tracks of the kinks are selected using similar criteria as for other primary tracks, except that the minimum number of TPC clusters required are 30
instead of 70, because they are shorter compared to the primary ones.
     Assuming the neutrino to be massless, the invariant mass of the decayed particle ($M_{\mu\nu}$) is estimated from the charged decay product track and the momentum of the neutrino as reported in \Ref \cite{Adam:2015qaa}.
     The main background is from charged pion decays, $\pi \rightarrow \mu + \nu_{\mu}$  (\BR = $99.99\%$), which also gives rise to a kink topology.
     A proper \qt selection, where  \qt is the transverse momentum of the daughter track with respect to the mother's direction at the kink, can separate most of the pion kink background from the kaon kinks.
     Since the upper limit of \qt values for the decay channels  $\pi \rightarrow \mu + \nu_{\mu}$ and K $\rightarrow \mu + \nu_{\mu}$ are 30 MeV/$\it{c}$ and 236 MeV/$\it{c}$ respectively, a selection of $\qt >$~120~MeV/$\it{c}$ rejects more than 80\% (85\% in pp collisions) of the pion background.
     For further removal of the contamination from pion decays, an additional selection on kink opening angle, as reported in \Ref \cite{Adam:2015qaa}, has been implemented.
     Finally, the TPC \dedx of the mother tracks is required to have $|\nsigma^{K}| <$ 3, which improves the purity of the sample.
     After these selections, the purity ranges from 99\% at low \pt to 92\% (96\% in pp collisions) at high $\pt$ according to Monte Carlo studies. The remaining very low background is coming from random
associations of charged tracks reconstructed as fake kinks.
     After applying all these topological selection criteria, the invariant mass of kaons ($M_{\mu\nu}$) obtained from the reconstruction of their decay products integrated over the above mentioned mother momentum ranges for pp and \pbpb collisions are shown in \Fig \ref{PbPbKinksperf}.

 \subsection{Correction of raw spectra}

   In order to obtain the \pt distributions of primary \ppipm, \pkapm and (\pprm)p, the raw spectra are corrected for PID efficiency, misidentification probability, acceptance and tracking efficiencies, following the procedures described in \Ref \cite{Abelev:2013vea} for the ITS, TPC (low \pt) and TOF, in \Ref \cite{Adam:2015kca} for the HMPID and TPC (high \pt) and in \Ref \cite{Adam:2015qaa}  for the kink analysis.
   The acceptance, reconstruction, and tracking efficiencies are obtained from Monte Carlo simulated events generated with
   PYTHIA 8.1 (Monash 2013 tune)~\cite{Skands:2014pea} for pp collisions and with HIJING~\cite{Wang:1991hta} for \pbpb collisions.
   The particles are propagated through the detector using the GEANT 3 transport code~\cite{Brun:1119728}, where the detector geometry and response, as well as the data taking conditions, are reproduced in detail.
   Since GEANT 3 does not describe well the interaction of low-momentum \pprm and \pkam with the material, a correction to the efficiencies is estimated using GEANT 4 and FLUKA respectively, which are known to describe such processes better ~\cite{Agostinelli:2002hh,Abelev:2013vea,Aamodt:2010dx, Battistoni:2007zzb}.
   The PID efficiency and the misidentification probability are evaluated by performing the analysis on the Monte Carlo simulation, which requires that the simulated data are first tuned
   to reproduce the real PID response for each PID technique.
   The contamination due to weak decays of light flavor hadrons (mainly \pkazero affecting \ppipm spectra, $\Lambda$ and $\Sigma^+$ affecting (\pprm)p spectra) and interactions with the material has to be computed and subtracted from the raw spectra.
   Since strangeness production is underestimated in the event generators and the interactions of low \pt particles with the material are not properly modeled in the transport codes, the secondary-particle contribution is evaluated with a data-driven approach.
   For each PID technique and species, the contribution of feed-down in a given \pt interval is extracted by fitting the measured distributions of \dcaXY of the tracks identified as the given hadron species.
   The \dcaXY distributions are modeled with three contributions: primary particles, secondary particles from weak decays of strange hadrons and secondary particles produced in the interactions with the detector material.
   Their shapes are extracted for each \pt interval and particle species from the Monte Carlo simulation described above.
   The contribution of secondaries is different for each PID analysis due to the different track and PID selections and is more important at low \pt.
   The measured \pbpb spectra are then normalized to the number of events in each centrality class.

   The spectra measured in pp collisions are also normalized to the number of inelastic collisions obtained from the number of analyzed minimum bias events corrected with an inelastic normalization factor of 0.757 ($\pm$~2.51\%), defined as the ratio between the V0 visible cross section and the inelastic pp cross section at \sF~TeV~\cite{Loizides:2017ack}.

   % Requires the booktabs if the memoir class is not being used
   \begin{table}[htbp]
     \centering
     \caption{Main sources and values of the relative systematic uncertainties (expressed in \%) of the \pt-differential yields of \ppipm, \pkapm and (\pprm)p obtained in the analysis of \pbpb collisions.
       When two values are reported, these correspond to the lowest and highest \pt bin of the corresponding analysis, respectively.
       If only one value is reported, the systematic uncertainty is not \pt-dependent.
       If not specified, the uncertainty is not centrality-dependent.
       The first three systematic uncertainties are common to all PID techniques.
     The maximum (among centrality classes) total systematic uncertainties and the centrality-independent ones are also shown.}
     %\topcaption{Table captions are better up top} % requires the topcapt package
     \resizebox{\columnwidth}{!}{%
       \begin{tabular}{lccccc} % Column formatting, @{} suppresses leading/trailing space
         \toprule
         %\multicolumn{3}{c}{Particle species} \\
         %\cmidrule(l){2-3} % Partial rule. (r) trims the line a little bit on the right; (l) & (lr) also possible
         Effect                                              & $\ppipm (\%)$ & K$^{\pm}(\%)$ & p and  \pprm (\%) & K/\ppi (\%) & p/\ppi (\%) \\
         \midrule
         %     Event selection                                       & 0.14   & 0.14   & 0.14    & -   & - \\
         Event selection                                     & 0.1           & 0.1           & 0.1               & -           & -           \\
         ITS$-$TPC matching efficiency                       & 0.2$-$1.2     & 0.2$-$1.2     & 0.2$-$1.2         & -           & -           \\
         Material budget                                     & 1.6$-$0.2     & 1.3$-$0.4     & 2.9$-$0.1         & 2.0$-$0.4   & 3.2$-$0.3   \\
         Hadronic interaction cross section                  & 2.5$-$2.4     & 2.7$-$1.8     & 4.6               & 3.3$-$3.0   & 5.0$-$5.2   \\
         \midrule
         ITS PID                                             & 1.9$-$5.7     & 0.8$-$3.1     & 3.4$-$2.7         & 1.8$-$3.8   & 4.1$-$4.4   \\
         Track selection                                     & 2.0$-$2.1     & 2.6$-$2.3     & 4.9$-$4.4         & 1.6$-$1.1   & 4.1$-$3.5   \\
         \EcrossB                                            & 3.0           & 3.0           & 3.0               & 4.2         & 4.2         \\
         Feed-down correction                                & 1.1           & $-$           & 0.4               & 1.1         & 1.2         \\
         Matching efficiency (0-5\%)                         & 2.8           & 2.8           & 2.8               & $-$         & $-$         \\
         Matching efficiency (40-50\%)                       & 1.9           & 1.9           & 1.9               & $-$         & $-$         \\
         Matching efficiency (80-90\%)                       & 0.5           & 0.5           & 0.5               & $-$         & $-$         \\
         Hadronic interaction cross section (ITS tracks)     & 2.0           & 2.7$-$1.5     & 4.6$-$2.0         & 3.3$-$2.5   & 5.0$-$2.8   \\
         \midrule
         Low-\pt TPC PID  (0-5\%)                            & 2.7$-$8.3     & 3.0$-$10.0    & 3.2$-$13.6        & 6.0$-$16.0  & 8.0$-$18.0  \\
         Low-\pt TPC PID  (40-50\%)                          & 2.2$-$6.0     & 2.5$-$6.0     & 2.1$-$9.3         & 2.0$-$11.0  & 3.0$-$13.0  \\
         Low-\pt TPC PID  (80-90\%)                          & 4.5$-$6.8     & 3.0$-$6.8     & 3.3$-$8.3         & 4.0$-$11.0  & 8.0$-$11.0  \\
         Track selection                                     & 1.0$-$5.0     & 1.0$-$5.0     & 1.0$-$5.0         & $-$         & $-$         \\
         Feed-down correction                                & 1.0           & $-$           & 2.5               & 1.2$-$0.4   & 10.0$-$5.0  \\
         \midrule
         TOF PID                                             & 3.0$-$12.0    & 3.0$-$18.0    & 2.0$-$20.0        & 2.0$-$15.0  & 2.0$-$20.0  \\
         Track selection                                     & 1.5           & 1.5           & 1.8               & 2.0         & 1.5         \\
         Matching efficiency                                 & 4.0           & 4.0           & 4.0               & $-$         & $-$         \\
         Feed-down correction                                & 0.5$-$0.2     & $-$           & 1.0$-$0.5         & 0.5$-$0.2   & 0.5$-$1.5   \\
         \midrule
         HMPID PID                                           & 3.0$-$11.0    & 2.0$-$11.0    & 2.0$-$11.0        & 3.0$-$11.5  & 2.0$-$11.5  \\
         Track selection                                     & 4.5           & 4.5           & 4.5               & 3.6         & 3.6         \\
         PID efficiency correction                           & 5.0           & 5.0           & 5.0               & 5.0         & 5.0         \\
         Distance selection correction (matching efficiency) & 2.0           & 2.0           & 4.0 - 2.0         & 1.0         & 1.0         \\
         Feed-down correction                                & 0.1           & $-$           & 0.3               & 0.2         & 0.3         \\
         Background (0-5\%)                                  & 18.0$-$6.0    & 10.0$-$2.0    & 10.0$-$1.5        & 10.0$-$2.0  & 10.0$-$2.0  \\
         Background (30-40\%)                                & 10.0$-$1.0    & 5.0$-$1.0     & 5.0$-$1.0         & 6.0$-$1.5   & 6.0$-$1.5   \\
         Background (60-70\%)                                & 4.0$-$1.0     & 2.0$-$1.0     & 2.0$-$1.0         & 3.0$-$1.0   & 3.0$-$1.0   \\
         \midrule
         High-\pt TPC Bethe$-$Bloch param. (0-5\%)           & 4.2$-$2.0     & 22.3$-$8.5    & 13.1$-$8.0        & 21.9$-$8.0  & 11.4$-$10.0 \\
         High-\pt TPC Bethe$-$Bloch param. (40-50\%)         & 4.3$-$2.0     & 17.0$-$8.5    & 16.3$-$8.0        & 17.1$-$8.0  & 15.6$-$10.0 \\
         High-\pt TPC Bethe$-$Bloch param. (80-90\%)         & 2.9$-$2.0     & 11.4$-$8.5    & 21.1$-$7.9        & 11.9$-$8.0  & 20.3$-$10.0 \\
         Track selection (0-5\%)                             & 1.5$-$1.1     & 1.5$-$1.1     & 1.5$-$1.1         & $-$         & $-$         \\
         Track selection (40-50\%)                           & 1.0$-$0.7     & 1.0$-$0.7     & 1.0$-$0.7         & $-$         & $-$         \\
         Track selection (80-90\%)                           & 0.7$-$1.7     & 0.7$-$1.7     & 0.7$-$1.7         & $-$         & $-$         \\
         \pt resolution                                      & 0.0$-$0.3     & 0.0$-$0.3     & 0.0$-$0.3         & $-$         & $-$         \\
         Feed-down correction                                & 0.4$-$0.4     & $-$           & 3.0$-$2.6         & $-$         & 3.0$-$2.6   \\
         %Feed-down correction (40-50\%)                                 & 0.3$-$0.3 & $-$  & 3.1$-$3.1  & $-$ & 3.1$-$3.0  \\
         %Feed-down correction (80-90\%)                                 & 0.3$-$0.3 & $-$  & 1.9$-$1.7  & $-$ & 1.9$-$1.7   \\
         \midrule
         Kink PID + reconstruction efficiency (0-5\%)        & $-$           & 1.0$-$10.4    & $-$               & $-$         & $-$         \\
         Kink PID + reconstruction efficiency (30-40\%)      & $-$           & 0.5$-$4.5     & $-$               & $-$         & $-$         \\
         Kink PID + reconstruction efficiency (80-90\%)      & $-$           & 0.7$-$5.5     & $-$               & $-$         & $-$         \\
         Track selection                                     & $-$           & 3.0           & $-$               & $-$         & $-$         \\
         Contamination (0-5\%)                               & $-$           & 0.6$-$5.0     & $-$               & $-$         & $-$         \\
         Contamination (30-40\%)                             & $-$           & 0.6$-$5.0     & $-$               & $-$         & $-$         \\
         Contamination (80-90\%)                             & $-$           & 0.6$-$4.0     & $-$               & $-$         & $-$         \\
         \midrule
         Total                                               & 7.3$-$3.9     & 5.9$-$9.8     & 9.7$-$9.2         & 7.7$-$8.0   & 9.9$-$11.0  \\
         Total (\Nch-independent)                            & 7.0$-$2.7     & 5.5$-$9.4     & 9.2$-$8.7         & 7.2$-$8.0   & 9.4$-$9.2   \\      \bottomrule
       \end{tabular}
     }
     \label{table1}
   \end{table}

   % Requires the booktabs if the memoir class is not being used
   \begin{table}[htbp]
     \centering
     \caption{Main sources and values of the relative systematic uncertainties (expressed in \%) of the \pt-differential yields of \ppipm, \pkapm and (\pprm)p obtained in the analysis of pp collisions.
       When two values are reported, these correspond to the lowest and highest \pt bin of the corresponding analysis, respectively.
       If only one value is reported, the systematic uncertainty is not \pt-dependent.
       The first three systematic uncertainties are common to all PID techniques.
     In the last row, the total systematic uncertainty is reported.}
     %\topcaption{Table captions are better up top} % requires the topcapt package
     \resizebox{\columnwidth}{!}{%
       \begin{tabular}{lccccc} % Column formatting, @{} suppresses leading/trailing space
         \toprule
         %\multicolumn{3}{c}{Particle species} \\
         %\cmidrule(l){2-3} % Partial rule. (r) trims the line a little bit on the right; (l) & (lr) also possible
         Effect                                              & $\ppipm (\%)$ & K$^{\pm}(\%)$ & p and  \pprm (\%) & K/\ppi (\%) & p/\ppi (\%) \\
         \midrule
         Event selection                                     & 0.5           & 0.5           & 0.5               & -           & -           \\
         ITS$-$TPC matching efficiency                       & 0.0$-$1.1     & 0.0$-$1.1     & 0.0$-$1.1         & -           & -           \\
         Material budget                                     & 1.6$-$0.2     & 2.0$-$0.4     & 2.9$-$0.1         & 2.4$-$0.4   & 3.2$-$0.3   \\
         Hadronic interaction cross section                  & 2.5$-$2.4     & 2.7$-$1.8     & 4.6               & 3.3$-$3.0   & 5.0$-$5.2   \\
         \midrule
         ITS PID                                             & 1.5$-$6.4     & 0.4$-$5.7     & 1.2$-$1.5         & 0.9$-$7.4   & 1.5$-$1.9   \\
         Track selection                                     & 2.6$-$2.1     & 2.5$-$3.8     & 3.0$-$2.0         & 1.8$-$0.5   & 2.5$-$1.7   \\
         Feed-down correction                                & $-$           & $-$           & 1.6               & $-$         & 1.6         \\
         \EcrossB                                            & 3.0           & 3.0           & 3.0               & 4.2         & 4.2         \\
         Hadronic interaction cross section (ITS tracks)     & 2.0           & 2.7$-$1.8     & 4.6$-$2.0         & 3.3$-$2.7   & 5.0$-$2.8   \\
         \midrule
         Low-\pt TPC PID                                     & 5.7$-$8.3     & 4.6$-$7.9     & 9.2$-$13.2        & 5.0$-$9.0   & 10.0$-$15.0 \\
         Track selection                                     & 1.0$-$4.0     & 1.0$-$4.0     & 1.0$-$4.0         & $-$         & $-$         \\
         Feed-down correction                                & 1.0           & $-$           & 2.0               & 1.1$-$0.6   & 5.0$-$2.0   \\
         % Tracking efficiency                              & 1 - 5   &  1 - 5 & 1 - 5   & ? - ?& ? - ?\\
         \midrule
         TOF PID                                             & 1.0$-$8.0     & 1.2$-$15.0    & 1.0$-$15.0        & 2.0$-$20.0  & 2.0$-$20.0  \\
         Track selection                                     & 1.5           & 1.5           & 2.0               & 2.0         & 3.0         \\
         Matching efficiency                                 & 1.0           & 1.0           & 1.0               & $-$         & $-$         \\
         Feed-down correction                                & 0.5$-$0.1     & $-$           & 1$-$0.5           & 0.5$-$0.1   & 0.2$-$0.5   \\
         \midrule
         HMPID PID                                           & 3.0$-$11.0    & 2.0$-$11.0    & 2.0$-$11.0        & 3.0$-$11.5  & 2.0$-$11.5  \\
         Track selection                                     & 4.5           & 4.5           & 4.5               & 3.6         & 3.6         \\
         Distance selection correction (matching efficiency) & 2.0           & 2.0           & 4.0$-$2.0         & 1.0         & 1.0         \\
         Feed-down correction                                & 0.1           & $-$           & 0.3               & 0.2         & 0.3         \\
         \midrule
         High-\pt TPC Bethe$-$Bloch parameterization         & 2.4$-$2.0     & 14.5$-$8.0    & 22.0$-$12.0       & 15.1$-$8.0  & 22.5$-$15.0 \\
         Track selection                                     & 0.9$-$1.7     & 0.9$-$1.7     & 0.9$-$1.7         & $-$         & $-$         \\
         %   Tracking efficiency                                              & 1 - 5    &  1 - 5 & 1 - 5   &  ? - ?    &  ? - ?  \\
         \pt resolution                                      & 0.0$-$0.3     & 0.0$-$0.3     & 0.0$-$0.3         & $-$         & $-$         \\
         Feed-down correction                                & 0.0$-$0.3     & $-$           & 1.9$-$1.7         & $-$         & 1.9$-$1.7   \\
         \midrule
         Kink PID + reconstruction efficiency                & $-$           & 4.3           & $-$               & $-$         & $-$         \\
         Track selection                                     & $-$           & 3.0           & $-$               & $-$         & $-$         \\
         Contamination                                       & $-$           & 0.2$-$3.2     & $-$               & $-$         & $-$         \\
         \midrule
         Total                                               & 6.4$-$3.4     & 4.6$-$9.2     & 6.9$-$12.5        & 4.9$-$8.0   & 6.7$-$15.1  \\
         \bottomrule
       \end{tabular}
     }
     \label{table2}
   \end{table}

 \subsection{Systematic uncertainties}

   The evaluation of systematic uncertainties follows the procedures described in \Ref\ \cite{Abelev:2013vea} for the ITS, TPC (low \pt) and TOF analyses, in \Ref \cite{Adam:2015kca} for the HMPID and TPC (high \pt) analyses~and in \Ref\ \cite{Adam:2015qaa}  for the kink analysis.
   The main sources of systematic uncertainties, for each analysis, are summarized in Tabs.~\ref{table1} and~\ref{table2}, for the~\pbpb~and pp analyses, respectively.
   Sources of systematic effects such as the different PID techniques, the feed-down correction, the imperfect description of the material budget in the Monte Carlo simulation, the knowledge of the hadronic interaction cross section in the detector material, the TPC-TOF and ITS-TPC matching efficiency and the track selection have been taken into account.
   The systematic uncertainties related to track selection were evaluated by varying the criteria used to select single tracks (number of reconstructed crossed rows in the TPC, number of available clusters in the ITS, \dcaXY and \dcaZ, $\chi^{2}/\rm{NDF}$ of the reconstructed track).
   The ratio of the corrected spectra with modified selection criteria to the default case is computed to estimate the systematic uncertainty for a given source.
   A similar approach is used for the evaluation of the systematic uncertainties related to the PID procedure.
   The uncertainties due to the imperfect description of the material budget in the Monte Carlo simulation is estimated varying the material budget in the simulation by $\pm$7\%.
   To account for the effect related to the imperfect knowledge of the hadronic interaction cross section in the detector material, different transport codes (GEANT3, GEANT4, and FLUKA) are compared.
   Finally, the uncertainties due to the feed-down correction procedure are estimated for all analyses by varying the range of the \dcaXY fit, by using different track selections, by applying different cuts on the (longitudinal) \dcaZ, and by varying the particle composition of the Monte Carlo templates used in the fit. \\

   For the ITS analysis, the standard $N_{\rm \sigma}$ method is compared with the yields obtained with a Bayesian PID technique \cite{Adam:2016acv}.
   Moreover, the Lorentz force causes shifts of the cluster position in the ITS, pushing the charge in opposite directions when switching the polarity of the magnetic field of the experiment (\EcrossB effect) \cite{Abelev:2013vea}.
   This effect is not fully reproduced in the Monte Carlo simulation and has been estimated by analyzing data samples collected with different magnetic field polarities.
   To estimate possible systematic effects deriving from signal extraction in the low \pt TPC analysis, the yield was computed by varying the selection based on the number of TPC crossed rows from 70 to 90
   and the yield was computed from the sum of the bin content of the \Nsigma distribution in the range $[-3, 3]$, instead of fitting.
   %   and the yield was obtained using the nominal technique, that consists in fitting the \Nsigma distributions.
   %   Alternatively, the yield was also computed from the sum of the bin content of the \nsigma distribution in the range $[-3, 3]$.
   The systematic uncertainty was obtained from the comparison to the nominal yield.
   Regarding the TPC analysis at high \pt, the imprecise knowledge of both the Bethe$-$Bloch and resolution parametrizations constitutes the most significant source of systematic uncertainties associated with the signal extraction.
   To quantify the size of the uncertainty, the relative variations of \dedx and resolution with respect to the original parameterizations were used.
   The TOF analysis estimates the PID systematic uncertainties by comparing the standard spectra with the ones extracted from a statistical deconvolution, which is based on templates generated from a TOF time response function with varied parameters.
   For the HMPID analysis, the selection on the distance between the extrapolated track point at the HMPID chamber planes and the corresponding MIP cluster centroid, $d_{\rm{MIP-trk}}$, is varied by $\pm1$~cm to check its systematic effect on the matching of tracks with HMPID signals.
   Moreover, the systematic bias due to the background fitting, which represents the largest source, is estimated by changing the fitting function: from a sixth-order polynomial to a power law of the tangent of the Cherenkov angle.
   This function is derived from geometrical considerations \cite{Beole:1998yq}.
   For the kink analysis, the systematic uncertainties are estimated by comparing the standard spectra with the ones obtained by varying the selection on decay product transverse momentum, the minimum number of TPC clusters, kink radius and TPC \Nsigma values of the mother tracks. \\

   By using the same methods as for the spectra, the systematic uncertainties for the \pt-dependent particle ratios were computed to take into account the correlated sources of uncertainty (mainly due to PID and tracking efficiency). Finally, for both \pt-dependent spectra and ratios the particle-multiplicity-dependent systematic uncertainties, those that are uncorrelated across different centrality bins, were determined.

The improved reconstruction and track selection in the analysis of pp and \pbpb data at \snnF~TeV  lead to reduced systematic uncertainties as compared to previously published results at \snnT~TeV.

   %   By using the same methods for the spectra, the systematic uncertainties for the \pt-dependent particle ratios were computed to take into account the correlated sources of uncertainty (mainly due to PID and to the tracking efficiency). \\
   %   For both \pt-dependent spectra and ratios, the study of systematic uncertainties was repeated for the different centrality bins to separate the sources of uncertainty which are dependent on particle multiplicity and uncorrelated across different centrality bins.

   %\newpage

\section{Results and discussion} \label{sec3}

\begin{figure}[htb]
   \centerline{
     \includegraphics[width=1.1\columnwidth]{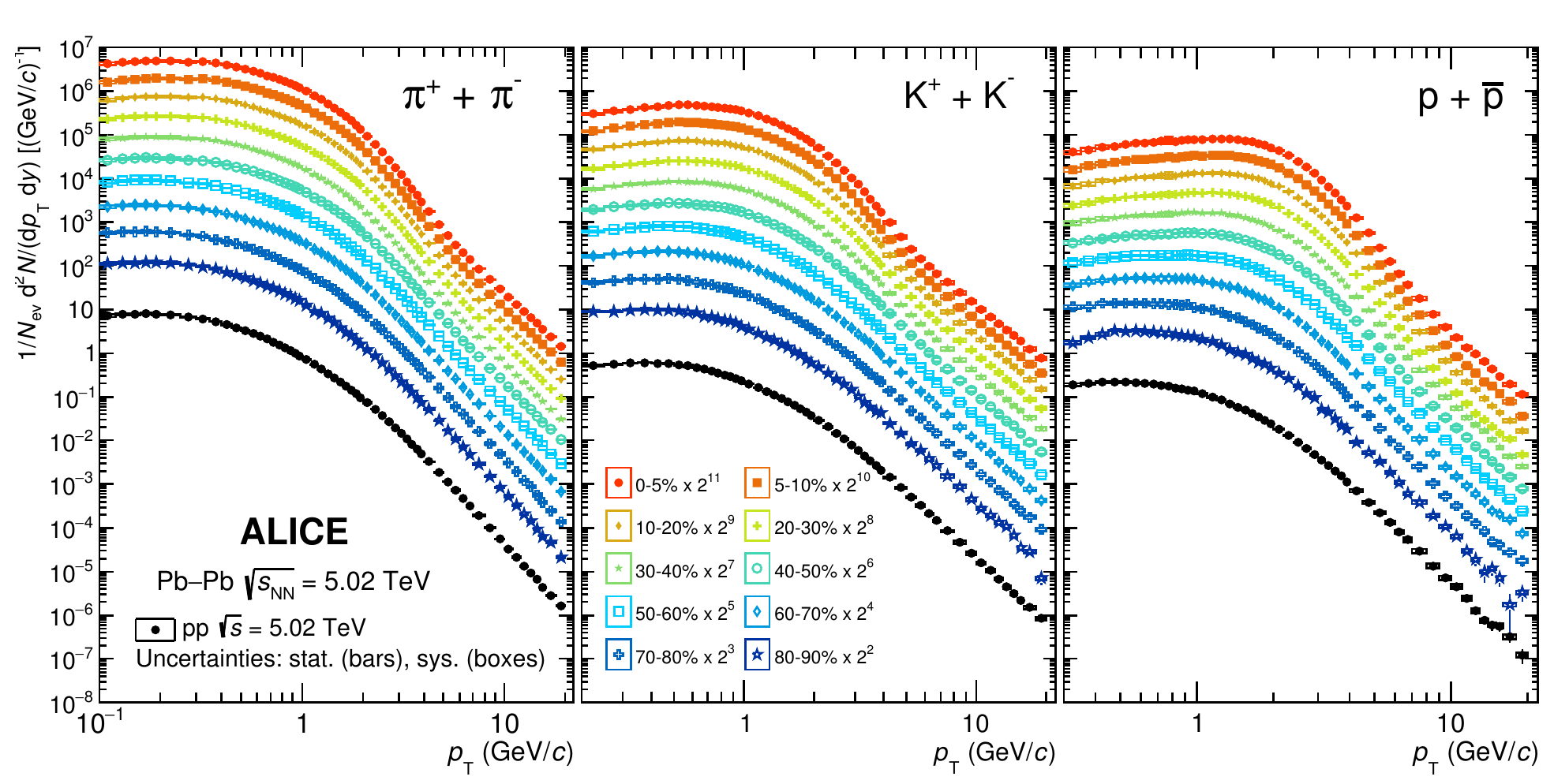}}
   \caption{Transverse momentum spectra of pions (left), kaons (middle) and (anti-)protons (right) measured in \pbpb collisions at \snnF\,TeV for different centrality classes.
     Scale factors are applied for better visibility.
     The results are compared with the spectra measured in inelastic pp collisions at \sF\,TeV.
   Statistical and systematic uncertainties are displayed as error bars and boxes around the data points, respectively.}
   \label{PbPbSpectra}
 \end{figure}

 The measured \pt spectra of \ppipm, \pkapm, and \pprpm from the independent analyses have to be combined in the overlapping ranges using a weighted average with the systematic and statistical uncertainties as weights.
 All the systematic uncertainties are considered to be uncorrelated across the different PID techniques apart from those related to the ITS-TPC matching efficiency and the event selection.
 The correlated systematic uncertainties have been added in quadrature after the spectra have been combined.
 For a given hadron species, the spectra of particles and antiparticles are found to be compatible, and therefore all spectra reported in this section are shown for summed charges. \\
 Figure~\ref{PbPbSpectra} shows the combined \pt spectra of \ppipm, \pkapm and \pprpm measured in 0$-$90\% \pbpb and inelastic pp collisions at \snnF~TeV.
 Results for \pbpb collisions are presented for different centrality classes.
 Scaling is applied in the plots to improve spectra visibility.
 In the low \pt region, the maximum of the spectra is pushed towards higher momenta while going from peripheral to central \pbpb events.
 This effect is mass dependent and can be interpreted as a signature of radial flow~\cite{Abelev:2013vea}.
 For high \pt, the spectra follow a power-law shape, as expected from perturbative QCD (pQCD) calculations~\cite{Kretzer:2000yf}.

 The \pt-integrated yields, \dndy, and the average transverse momentum, \meanpt,
 are determined for the different centrality classes using an extrapolation to \pt = 0.
 The extrapolation procedure is performed after fitting the measured spectra with Boltzmann-Gibbs~Blast-Wave~\cite{Schnedermann:1993ws} (for~\pbpb) or the L\'evy-Tsallis~\cite{Tsallis:1987eu, Abelev:2006cs} (for pp) functions.
 In the most central \pbpb collisions (0$-$5\%), the extrapolated fractions of the total yields are 5.84\%, 5.20\% and 3.72\%, for pions, kaons and (anti-)protons, respectively.
 The fractions increase as centrality decreases, reaching 8.63\%, 9.36\% and 10.73\% in the most peripheral collisions (80$-$90\%).
 In pp collisions the fractions are 8.59\%, 9.98\% and 12.61\% for pions, kaons and (anti-)protons, respectively.
 The systematic uncertainties are then propagated to the \pt-integrated yields and mean transverse momentum.
 For the uncertainty on \dndy, the fit is performed with all data points shifted up by their full systematic uncertainties.
 To estimate the uncertainty on \meanpt, points in the 0$-$3~\gevc range are shifted up and down within their systematic uncertainty to obtain the softest and hardest spectra.
 The maximum difference (in absolute value) between the integrated quantities obtained with the standard and modified spectra are included as part of the systematic uncertainty.
 %In addition to this contribution, different functions\footnote{L\'evy-Tsallis (\pbpb only); Boltzmann-Gibbs~Blast-Wave (pp only); \mt-exponential: $A x \times exp(-\sqrt{x^2 + m^2}/T)$, where $A$ is a normalization constant, $T$ the temperature and $m$ the mass; Fermi-Dirac $A x \times 1 / (exp(\sqrt{x^2 + m^2}/T)+1)$; Boltzmann $A x \times \sqrt{x^2+m^2} \times exp(-\sqrt{x^2 + m^2}/T)$} were used to perform the extrapolation.
 Additionally, different functions\footnote{L\'evy-Tsallis (\pbpb only); Boltzmann-Gibbs~Blast-Wave (pp only); \mt-exponential: $A x \times exp(-\sqrt{x^2 + m^2}/T)$, where $A$ is a normalization constant, $T$ the temperature and $m$ the mass; Fermi-Dirac $A x \times 1 / (exp(\sqrt{x^2 + m^2}/T)+1)$; Bose-Einstein $A x \times 1 / (exp(\sqrt{x^2 + m^2}/T)-1)$; Boltzmann $A x \times \sqrt{x^2+m^2} \times exp(-\sqrt{x^2 + m^2}/T)$} were used to perform the extrapolation and the largest differences were added to the previous contributions. 
%adding the maximum variation of the integrated quantities to the previous contribution.
 %the maximum variation from the Blast-Wave (L\'evy-Tsallis for pp) of the integrated quantities are added to the previous contribution.
% The statistical uncertainties on the \dndy and \meanpt values are directly provided from the fit procedure as uncertainties on the fit parameters.
 The statistical uncertainties on the \dndy and \meanpt values are evaluated propagating the uncertainties on the fit parameters obtained directly from the fit procedure.
 The procedure described above is repeated using the systematic uncertainties uncorrelated across different centrality bins to extract the centrality uncorrelated part of the systematic uncertainties on the \pt-integrated particle yields and the average transverse momenta.

 In Table 4, %, \ref{table4} and \ref{table5}
 the \dndy and \meanpt are shown for \pbpb and pp collisions, respectively.
 For \pbpb collisions the values are given for different centrality ranges.

 %\newpage

 \subsection{Particle production at low transverse momentum} \label{sec3.1}

   The Boltzmann-Gibbs blast-wave function is a three-parameter simplified hydrodynamic model in which particle production is given by~\cite{Schnedermann:1993ws}:

   \begin{equation}
     E\frac{{\rm d}^3N}{{\rm d}p^3} \propto \int_{0}^{R} \mt I_0 \left(\frac{\pt \sinh(\rho)}{\Tkin}\right) K_1 \left(\frac{\mt\cosh(\rho)}{\Tkin}\right) r\,{\rm d}r
     \label{eq:bwfit}
   \end{equation}

The velocity profile $\rho$ is given by:

   \begin{equation}
     \rho = \tanh^{-1} \beta_{\rm T}=\tanh^{-1} \bigg[ \bigg( \frac{r}{R} \bigg)^{n}\beta_{\rm s} \bigg]
   \end{equation}

   where \Bt is the radial expansion velocity, \mt the transverse mass ($\mt = \sqrt{m^2+{\pt}^2}$) and \Tkin the temperature at the kinetic freeze-out, $I_{0}$ and $K_{1}$ are the modified Bessel functions, $r$ is the radial distance in the transverse plane, $R$ is the radius of the fireball, $\beta_{\rm s}$ is the transverse expansion velocity at the surface, and $n$ is the exponent of the velocity profile.

\begin{figure}[htb]
     \centerline{
       \includegraphics[width=0.6\columnwidth]{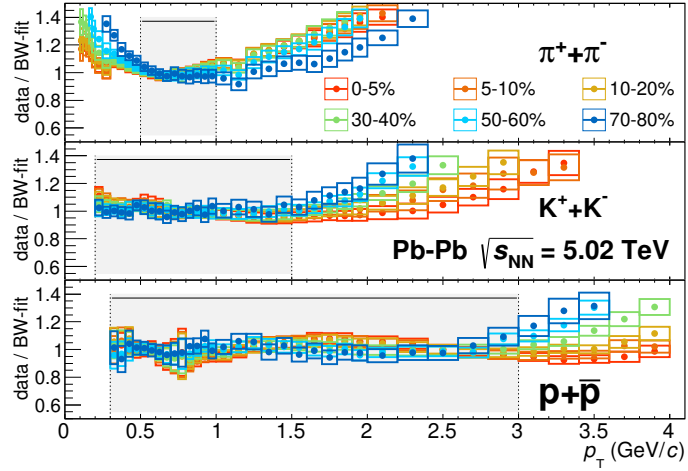}}
     \caption{Ratios of centrality-dependent \pt spectra to model (blast-wave parameterization) predictions in \pbpb collisions at \snnF\,TeV for pions (top), kaons (middle) and protons (bottom).
     The fit ranges are indicated as grey shaded areas.}
     \label{FitOverData}
   \end{figure}

   To quantify the centrality dependence of spectral shapes at low \pt, the Boltzmann-Gibbs blast-wave function has been simultaneously fitted to the charged pion, kaon and (anti-)proton \pt spectra, using a common set of parameters but different normalization factors and masses.
   Although the absolute values of the parameters have a strong dependence on the \pt range used for the fit \cite{Abelev:2013vea}, the evolution of the parameters with \snn~can still be compared across different collision energies by using the same fitting ranges.
   The present analysis uses the same \pt intervals employed for fitting as in a previous publication~\cite{Abelev:2013vea}, namely, $0.5-1$\,\gevc, $0.2-1.5$\,\gevc and $0.3-3$\,\gevc for charged pions, kaons and (anti-)protons, respectively.
   Figure~\ref{FitOverData} shows the ratios of the spectra to results of the fits for different centrality classes and particle species.
   If the shape of the \pt distributions over the full measured \pt range was purely driven by the collective radial expansion of the system, the functions determined by fitting the data in a limited \pt range would be expected to describe the spectral shapes in the full \pt range.
   Within uncertainties, this is only observed for the proton \pt spectra (up to 4 \gevc) in 0$-$20\% \pbpb collisions.
   A different situation is observed for pions where, due to their small mass and the large centrality-dependent feed-down contribution from resonance decays, the agreement with the model is worse than that observed for kaons and (anti-)protons.
   %By some authors this is also partially attributed to the Bose--Einstein condensation effect of pions \cite{Begun:2015ifa} that is more relevant at low \pt (see Sec.\ \ref{intro}).
   The \pt interval where the model describes the data within uncertainties gets wider going from peripheral to central \pbpb collisions.

   \begin{table}
     {
       %\newcommand{\mc}[3]{\multicolumn{#1}{#2}{#3}}
       %\begin{center}
       \centering
       \caption{\dndy and \meanpt measured in \pbpb and pp collisions at \snnF~TeV. \pbpb results are shown for the different centrality classes.
       Statistical and systematic uncertainties are also reported.}
       \resizebox{\columnwidth}{!}{%
         \begin{tabular}{lllllll}
           \midrule
           \mc{7}{c}{\ppis}\\
           \midrule
           \midrule
           Centrality Class & \dndy  & \Ustat & \Usyst & \meanpt & \Ustat & \Usyst \\
           \midrule
           0-5\%            & 1699.80 & 0.88   & 116.91 & 0.5682  & 0.0002 & 0.0320 \\
           5-10\%           & 1377.49 & 0.71   & 66.90  & 0.5711  & 0.0002 & 0.0181 \\
           10-20\%          & 1039.47 & 0.46   & 47.36  & 0.5704  & 0.0002 & 0.0174 \\
           20-30\%          & 712.92  & 0.34   & 36.06  & 0.5615  & 0.0002 & 0.0192 \\
           30-40\%          & 467.76  & 0.26   & 23.97  & 0.5525  & 0.0002 & 0.0198 \\
           40-50\%          & 292.91  & 0.19   & 15.80  & 0.5389  & 0.0003 & 0.0206 \\
           50-60\%          & 171.14  & 0.18   & 9.77   & 0.5214  & 0.0004 & 0.0215 \\
           60-70\%          & 88.82   & 0.10   & 5.21   & 0.5082  & 0.0004 & 0.0205 \\
           70-80\%          & 41.69   & 0.07   & 2.49   & 0.4924  & 0.0006 & 0.0203 \\
           80-90\%          & 16.31   & 0.04   & 0.91   & 0.4775  & 0.0008 & 0.0178 \\

           \midrule
           \mc{7}{c}{\pkas}\\
           \midrule
           \midrule
           %Centrality Class & $\langle dN/dy \rangle$ & \Ustat & \Usyst & \meanpt & \Ustat & \Usyst \\
           Centrality Class & \dndy   & \Ustat & \Usyst & \meanpt & \Ustat & \Usyst \\
           \midrule
           0-5\%            & 273.41  & 0.35   & 11.62  & 0.9177  & 0.0009 & 0.0140 \\
           5-10\%           & 222.48  & 0.54   & 9.37   & 0.9214  & 0.0018 & 0.0130 \\
           10-20\%          & 168.16  & 0.24   & 6.89   & 0.9193  & 0.0010 & 0.0126 \\
           20-30\%          & 114.70  & 0.15   & 4.67   & 0.9052  & 0.0008 & 0.0114 \\
           30-40\%          & 75.00   & 0.09   & 2.96   & 0.8919  & 0.0008 & 0.0106 \\
           40-50\%          & 46.36   & 0.06   & 1.88   & 0.8685  & 0.0009 & 0.0113 \\
           50-60\%          & 26.38   & 0.05   & 1.09   & 0.8369  & 0.0011 & 0.0132 \\
           60-70\%          & 13.38   & 0.03   & 0.64   & 0.8165  & 0.0015 & 0.0138 \\
           70-80\%          & 6.01    & 0.02   & 0.30   & 0.7881  & 0.0019 & 0.0160 \\
           80-90\%          & 2.27    & 0.01   & 0.12   & 0.7541  & 0.0032 & 0.0179 \\
           \midrule
           \mc{7}{c}{\pprs}\\
           \midrule
           \midrule
           Centrality Class & \dndy   & \Ustat & \Usyst & \meanpt & \Ustat & \Usyst \\
           \midrule
           0-5\%            & 74.56   & 0.06   & 3.75   & 1.4482  & 0.0007 & 0.0244 \\
           5-10\%           & 61.51   & 0.07   & 2.93   & 1.4334  & 0.0009 & 0.0224 \\
           10-20\%          & 47.40   & 0.04   & 2.20   & 1.4143  & 0.0007 & 0.0216 \\
           20-30\%          & 33.17   & 0.04   & 1.50   & 1.3768  & 0.0008 & 0.0199 \\
           30-40\%          & 22.51   & 0.03   & 1.01   & 1.3209  & 0.0010 & 0.0177 \\
           40-50\%          & 14.46   & 0.02   & 0.66   & 1.2570  & 0.0012 & 0.0179 \\
           50-60\%          & 8.71    & 0.02   & 0.40   & 1.1822  & 0.0016 & 0.0151 \\
           60-70\%          & 4.74    & 0.01   & 0.27   & 1.1004  & 0.0022 & 0.0184 \\
           70-80\%          & 2.30    & 0.01   & 0.14   & 1.0181  & 0.0030 & 0.0221 \\
           80-90\%          & 0.92    & 0.01   & 0.06   & 0.9464  & 0.0053 & 0.0277 \\
           \midrule
           \mc{7}{c}{pp collisions}\\
           \midrule
           \midrule
         Particle specie & \dndy & \Ustat & \Usyst & \meanpt & \Ustat & \Usyst \\
         \midrule
         \midrule
           \ppis            & 4.1342  & 0.0005 & 0.3032 & 0.4582  & 0.0001 & 0.0284 \\
           \midrule
           \pkas            & 0.5343  & 0.0014 & 0.0273 & 0.7412  & 0.0008 & 0.0296 \\
           \midrule
           \pprs            & 0.2331  & 0.0002 & 0.0205 & 0.8820  & 0.0006 & 0.0498 \\
         \end{tabular}
       }
       %\end{center}
    } 
     \label{table4}
   \end{table}

%   \begin{table}
%     {
       %\newcommand{\mc}[3]{\multicolumn{#1}{#2}{#3}}
       %\begin{center}
 %      \centering
%       \caption{\dndy and \meanpt  for the different centrality classes as measured in pp collisions at \sF~TeV.
 %      Statistical and systematic uncertainties are also reported.}
 %      \begin{tabular}{lllllll}
 %        \midrule
 %        Particle specie & \dndy & \Ustat & \Usyst & \meanpt & \Ustat & \Usyst \\
 %        \midrule
  %       \midrule
  %       \ppis           & 4.13  & 0.0005 & 0.27   & 0.458   & 0.0001 & 0.025  \\
 %        \midrule
 %        \pkas           & 0.534 & 0.0013 & 0.029  & 0.742   & 0.0008 & 0.029  \\
  %       \midrule
  %       \pprs           & 0.233 & 0.0002 & 0.017  & 0.882   & 0.0005 & 0.046  \\
  %     \end{tabular}
       %\end{center}
  %   }
  %   \label{table5}
  % \end{table}

   %\input{dndytable.tex}

   \begin{figure}[htb]
     \centerline{
       \includegraphics[width=0.6\columnwidth]{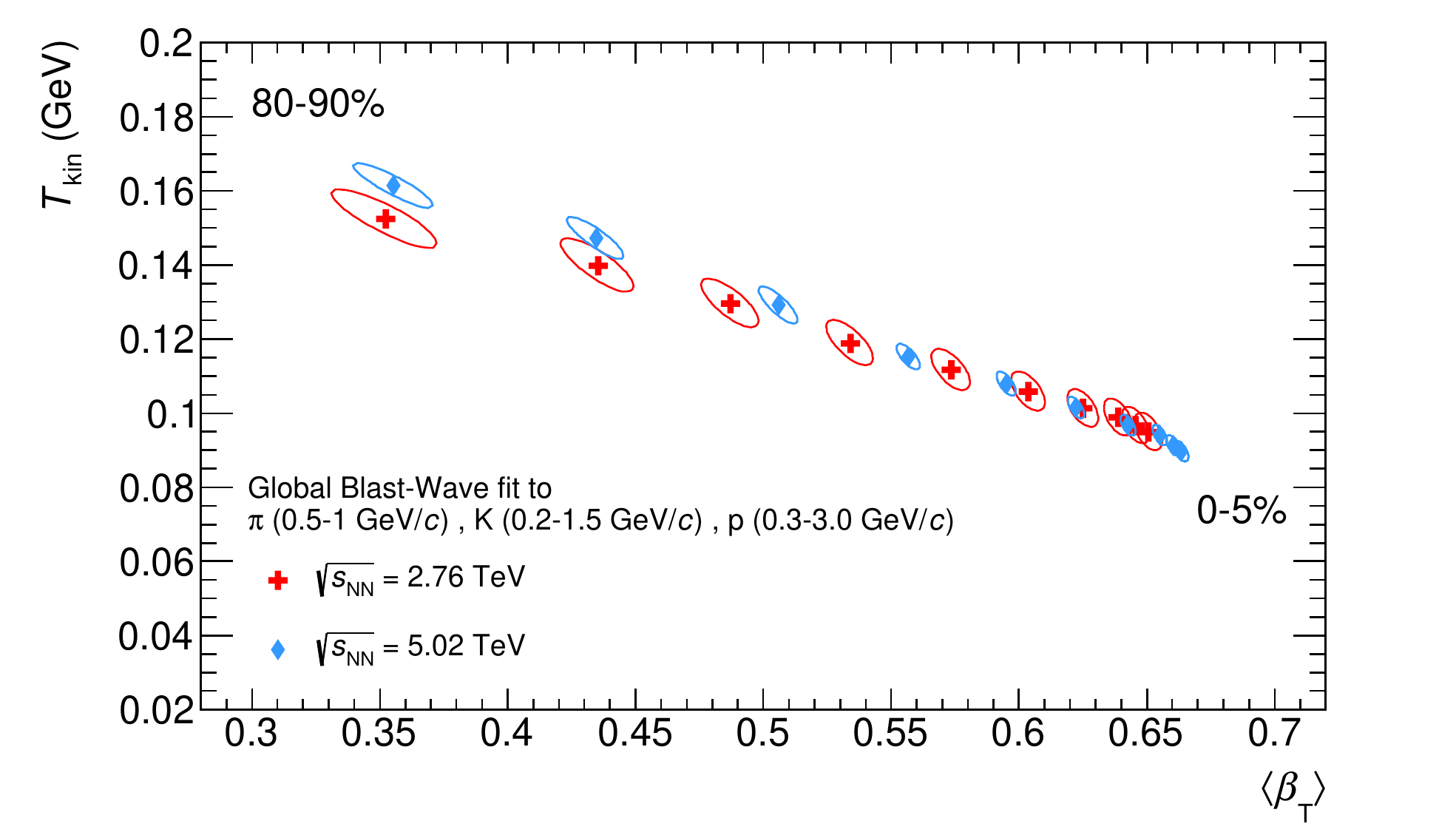}}
     \caption{Average expansion velocity (\avBt) and kinetic freeze-out temperature (\Tkin) progression from the simultaneous Boltzmann-Gibbs blast-wave fit to \ppipm, \pkapm and p(\pprm) spectra measured in \pbpb collisions at \snnF and 2.76 TeV \cite{Abelev:2013vea}.
       The correlated uncertainties from the global fit are shown as ellipses.
       The elliptic contours correspond to 1$\sigma$ uncertainties, with statistical and systematic uncertainties being added in quadrature.
     }
     \label{blastwavefit}
   \end{figure}

   In Table \ref{table3} the blast-wave fit parameters \avBt, \Tkin and $n$ in \pbpb collisions at \snnF~TeV, for different centrality classes, are listed.
   Figure~\ref{blastwavefit} shows the correlation between \avBt and \Tkin, both obtained from the simultaneous fit for \pbpb collisions at \snnT\,TeV and 5.02 TeV.
   For \pbpb collisions at \snnF\,TeV, \avBt increases with centrality, reaching $\avBt=0.663 \pm 0.003$ in 0-5$\%$ central collisions, while \Tkin decreases from $\Tkin~=~(0.161~\pm~0.006$)~GeV to $\Tkin~=~(0.090~\pm~0.003$)~GeV, similarly to what was observed at lower energies.
   This can be interpreted as a possible indication of a more rapid expansion with increasing centrality~\cite{Adams:2005dq, Abelev:2013vea}.
   In peripheral collisions this is consistent with the expectation of a shorter lived
   fireball with stronger radial pressure gradients~\cite{Heinz:2004qz}. % essentially the energy conservation negatively correlates the radial flow velocity \avBt and \Tkin.
   The value of the exponent of the velocity profile of the expansion, $n$, is about 0.74 in central collisions and it increases up to 2.52 in peripheral collisions (see Table~\ref{table3}).
   The values of $n$ in peripheral collisions increase with respect to those in central collisions to reproduce the power-law tail of the \pt spectra.
   Finally, in the most central \pbpb (0$-$5\%) collisions the difference of the average transverse velocity between the two collision energies is $\approx$~2.4 standard deviations. 
   %The value at 5.02~TeV is $\approx$ 2\% larger at 5.02~TeV than that measured at 2.76~TeV,
The value at 5.02 TeV is $\approx$ 2\% larger than that measured at 2.76 TeV, whereas the kinetic freeze-out temperature results are slightly smaller at larger collision energy but the difference is not significative. Just for the most peripheral collisions the kinetic freeze-out temperature is slightly higher at 5.02~TeV than that at 2.76~TeV. 
This is in contrast with our interpretation for central collisions where a larger volume has the kinetic freeze out later allowing the kinetic temperature to decrease further. It is worth questioning
whether the blast wave formalism is applicable also for these smaller system and it will be interesting to see if models, which can also describe small systems, can explain this changing
pattern. Moreover, we note that event and geometry biases may also play a role in the peripheral \pbpb collisions \cite{Morsch:2017brb}.
   
% This is in contrast with our interpretation where larger volume should freeze out later so at lower temperature, but we should take into account that for peripheral \pbpb collisions event and geometry biases may play a role~\cite{Morsch:2017brb}.

\begin{figure}[htb]
     \centerline{
       \includegraphics[width=0.8\columnwidth]{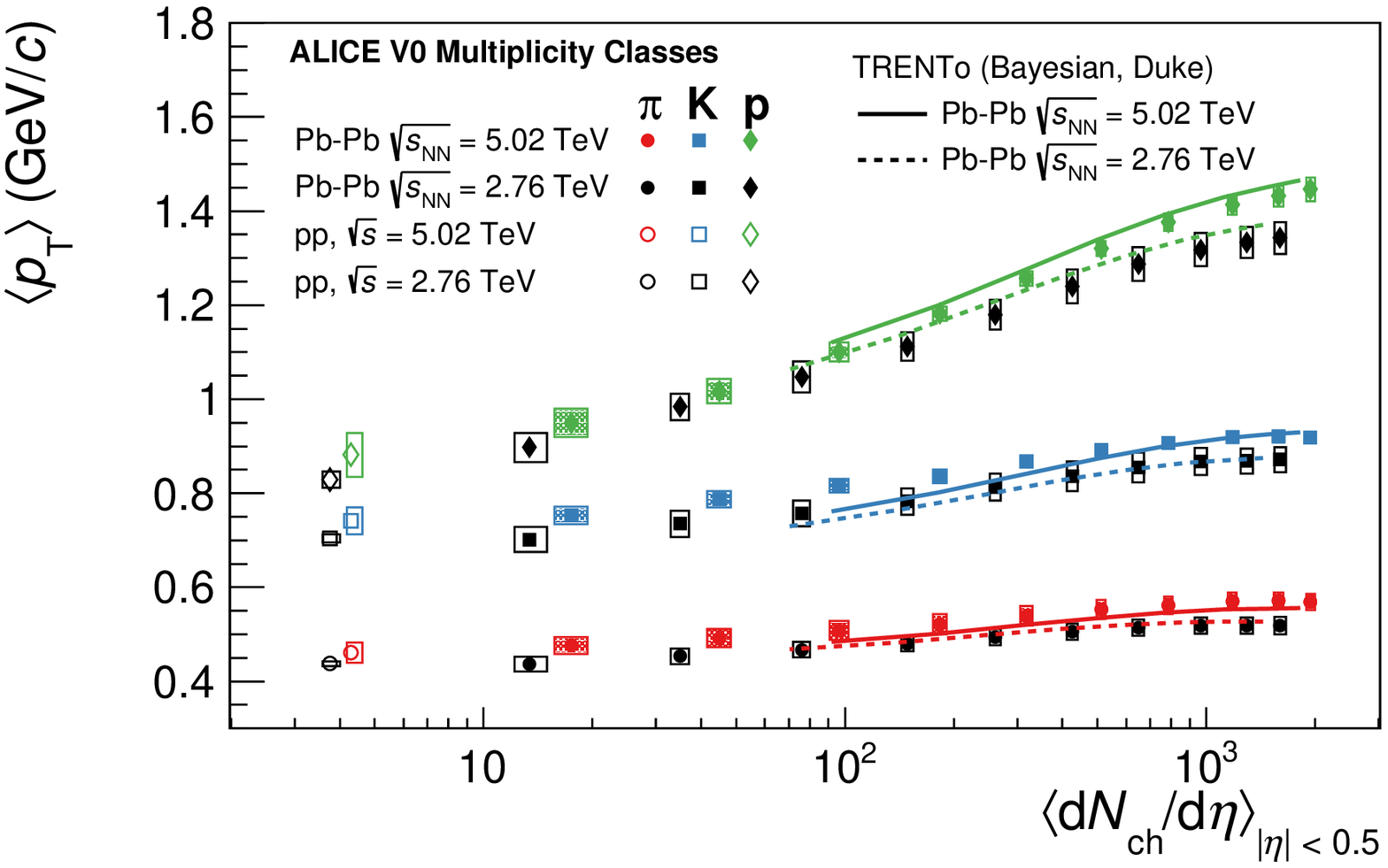}}
     \caption{Mean transverse momentum as a function of \avdNchdeta for \ppipm, \pkapm and \pprpm in \pbpb collisions at \snnF (full color markers) and 2.76 TeV \cite{Abelev:2013vea} (full black markers) and in inelastic pp collisions at \sF\ and 2.76~TeV (empty color markers) \cite{Adam:2015qaa}.
       The empty boxes show the total systematic uncertainty; the shaded boxes indicate the contribution uncorrelated across centrality bins (not estimated in \pbpb collisions at \snnT~TeV).
       Continuous lines represent the Bayesian analysis predictions.%{\bf Systematic uncertainty for 5.02 TeV seems too small, why?}
     }
     \label{ptvsmultiplicity_bayesian}
   \end{figure}

   Figure~\ref{ptvsmultiplicity_bayesian} shows the \meanpt for charged pions, kaons and (anti-)protons as a function of the charged particle multiplicity density $\langle\rm{d}\it{N}_{\rm{ch}}$/d$\eta\rangle$ at mid-rapidity in \pbpb  collisions at \snnF and 2.76 TeV.
   Going from inelastic pp collisions to peripheral and central \pbpb collisions, the \meanpt increases with $\langle\rm{d}\it{N}_{\rm{ch}}$/d$\eta\rangle$.
   The rise of the average \pt gets steeper with increasing hadron mass, this effect is consistent with the presence of radial flow.
   Within uncertainties and for comparable charged particle multiplicity densities, the results for both energies are consistent for 20$-$90\% \pbpb collisions.
   For 0$-$20\% \pbpb collisions, \meanpt is slightly higher at 5.02\,TeV than at 2.76\,TeV.
   The increase originates from the low \pt part of the spectra.
   Again, this is an effect consistent with a stronger radial flow in \pbpb collisions at the highest collision energy.

   % Requires the booktabs if the memoir class is not being used
   \begin{table}[htbp]
     \centering
     \caption{Results of the combined Boltzmann-Gibbs blast-wave fits to the particle spectra measured in \pbpb collisions at \snnF~TeV, in the \pt ranges 0.5$-$1 \gevc, 0.2$-$1.5 \gevc, and 0.3$-$3.0 \gevc for \ppipm, \pkapm and \pprpm, respectively.
       Values in parenthesis refer to the ratios to the values in \pbpb collisions at \snnT~TeV \cite{Abelev:2013vea}.
       The charged particle multiplicity values are taken from \Refs \cite{Adam:2016ddh, ALICE-PUBLIC-2015-008}.}
     %\topcaption{Table captions are better up top} % requires the topcapt package
     \begin{tabular}{lccccc} % Column formatting, @{} suppresses leading/trailing space
       \toprule
       %\multicolumn{3}{c}{Particle species} \\
       %\cmidrule(l){2-3} % Partial rule. (r) trims the line a little bit on the right; (l) & (lr) also possible
       Centrality & \avdNchdeta    & \avBt                    & \Tkin (GeV)            & $n$                      \\
       \midrule
       0$-$5\%    & 1943 $\pm$ 56  & (1.018)0.663 $\pm$ 0.003 & (0.947)0.090 $\pm$ 0.003 & (1.032)0.735 $\pm$ 0.013 \\
       5$-$10\%   & 1587 $\pm$ 47  & (1.022)0.660 $\pm$ 0.003 & (0.938)0.091 $\pm$ 0.003 & (1.005)0.736 $\pm$ 0.013 \\
       10$-$20\%  & 1180 $\pm$ 31  & (1.025)0.655 $\pm$ 0.003 & (0.949)0.094 $\pm$ 0.003 & (1.001)0.739 $\pm$0.013  \\
       20$-$30\%  & 786 $\pm$ 20   & (1.029)0.643 $\pm$ 0.003 & (0.960)0.097 $\pm$ 0.003 & (0.990)0.771 $\pm$0.014  \\
       30$-$40\%  & 512 $\pm$ 15   & (1.030)0.622 $\pm$ 0.003 & (0.953)0.101 $\pm$ 0.003 & (0.985)0.828 $\pm$0.015  \\
       40$-$50\%  & 318 $\pm$ 12   & (1.037)0.595 $\pm$ 0.004 & (0.964)0.108 $\pm$ 0.003 & (0.962)0.908 $\pm$0.019  \\
       50$-$60\%  & 183 $\pm$ 8    & (1.041)0.557 $\pm$ 0.005 & (0.975)0.115 $\pm$ 0.003 & (0.957)1.052 $\pm$ 0.024 \\
       60$-$70\%  & 96.3 $\pm$ 5.8 & (1.035)0.506 $\pm$ 0.008 & (1.000)0.129 $\pm$ 0.005 & (0.977)1.262 $\pm$ 0.043 \\
       70$-$80\%  & 44.9 $\pm$ 3.4 & (0.993)0.435 $\pm$ 0.011 & (1.058)0.147 $\pm$ 0.006 & (1.063)1.678 $\pm$ 0.088 \\
       80$-$90\%  & 17.5 $\pm$ 1.8 & (0.994)0.355 $\pm$ 0.016 & (1.066)0.161 $\pm$ 0.006 & (1.071)2.423 $\pm$ 0.208 \\
       \midrule
       \bottomrule
     \end{tabular}
     \label{table3}
   \end{table}

   \begin{figure}[htb]
     \centerline{
       \includegraphics[width=0.65\columnwidth]{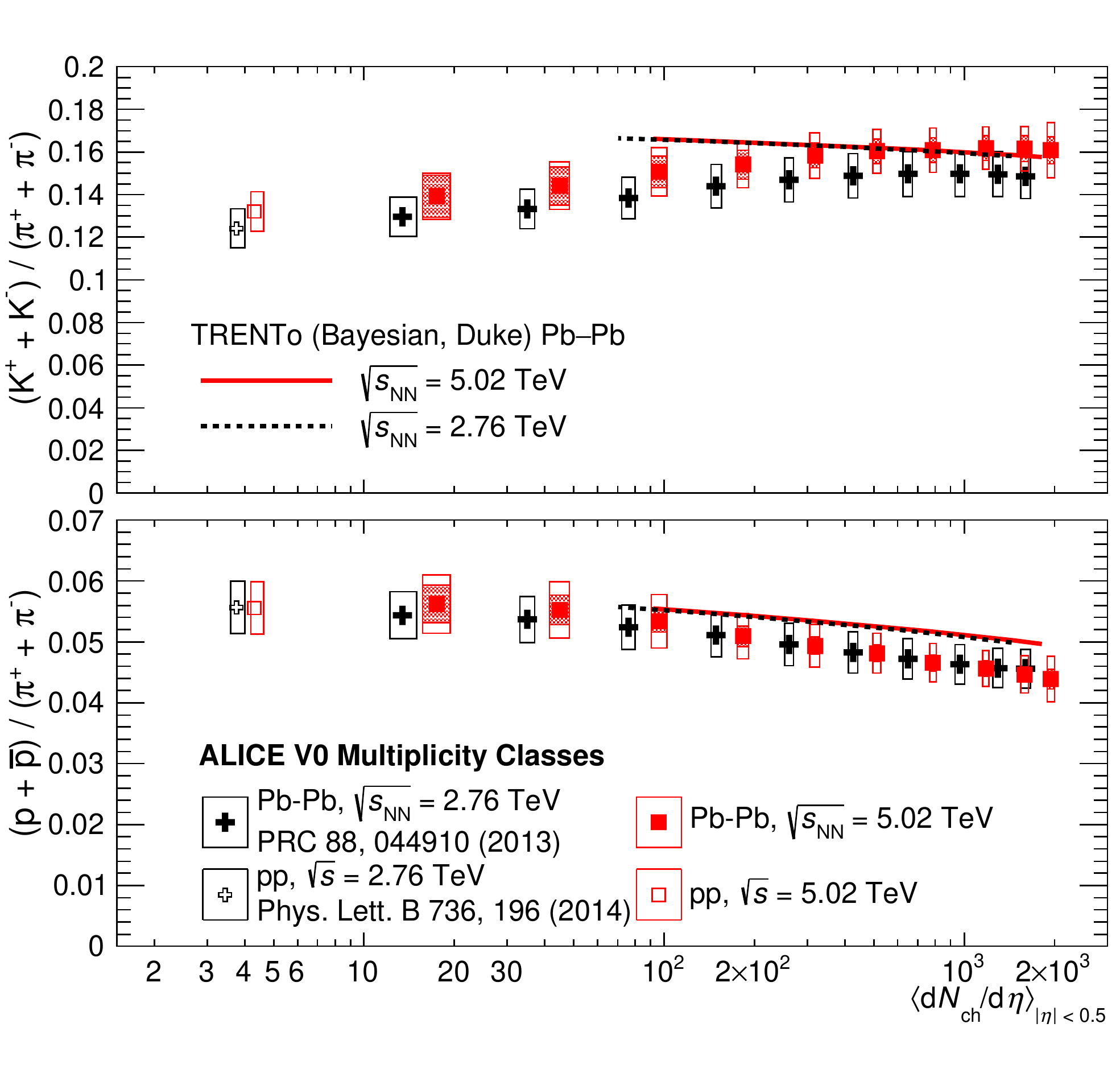}}
     \caption{Transverse momentum integrated K/\ppi (top) and p/\ppi (bottom) ratios as a function of \avdNchdeta in \pbpb collisions at \snnF~TeV, compared to \pbpb at 2.76 TeV \cite{Abelev:2013vea}.
       The values in pp collisions at \sF and 2.76 TeV are also shown.
       The empty boxes show the total systematic uncertainty; the shaded boxes indicate the contribution uncorrelated across centrality bins (not estimated in \pbpb collisions at \snnT~TeV).
     Continuous lines represent the Bayesian analysis predictions.}
     \label{ratiosvsmultilicity_bayesian}
   \end{figure}

   Figure~\ref{ratiosvsmultilicity_bayesian} shows the \pt-integrated particle ratios, K/\ppi and p/\ppi, as a function of \avdNchdeta in \pbpb at \snnT\,TeV and 5.02 TeV, and in inelastic pp collisions at \snnT\,TeV and 5.02 TeV.
%   The systematic uncertainties on the integrated ratios have been calculated assuming the uncertainties on the spectra of the different particle species as fully uncorrelated among each other.
   The systematic uncertainties on the integrated ratios have been evaluated using the uncertainties on the \pt-dependent ratios, taking into account the part of the uncertainties correlated among the different particle species.
   No significant energy dependence is observed, indicating that there is small or no dependence of the hadrochemistry on the collision energy.
   The K/\ppi ratio hints at a small increase with centrality.
   The effect is consistent with the observed increase of strange to non-strange hadron production in heavy-ion collisions compared to inelastic pp collisions~\cite{ALICE:2017jyt}.
   The p/\ppi ratio suggests a small decrease with centrality.
   Using the centrality uncorrelated uncertainties, the difference between the ratio in the most central (0$-$5\%) and peripheral (80$-$90\%) collisions is $\approx$~4.7 standard deviations, thus the difference is significant.
   The decreasing ratio is therefore consistent with the hypothesis of antibaryon-baryon annihilation in the hadronic phase~\cite{Becattini:2016xct, Steinheimer:2012rd,Karpenko:2012yf,Becattini:2012xb, Andronic:2018qqt,Stock:2018xaj}.
   The effect is expected to be less important for the more dilute system created in peripheral collisions.

   Recently, a new procedure has been implemented to quantitatively estimate properties of the quark--gluon plasma created in ultrarelativistic heavy-ion collisions utilizing Bayesian statistics and a multi-parameter model-to-data comparison  \cite{Bernhard:2016tnd}. 
   %,Bernhard:2018hnz}.
   The study is performed using a recently developed parametric initial condition model, T$_{\rm R}$ENTo (Reduced Thickness Event-by-event Nuclear Topology)~\cite{Moreland:2014oya}, which interpolates among a general class of energy-momentum distributions in the initial condition, and a modern hybrid model which couples viscous hydrodynamics to a hadronic cascade model.
   The model uses multiplicity, transverse momentum, and flow data from \pbpb collisions at \snnT~TeV to constrain the parametrized initial conditions and the temperature-dependent transport coefficients of the QGP.
   Based on this set of parameters, predictions for \pbpb collisions at \snnF~TeV are provided.
   The average transverse momentum and integrated yields in \pbpb collisions at \snnT~TeV are used as input to extract predictions at \snnF~TeV.
   The predictions from the multi-parameter Bayesian analysis are compared with data in \Figs~\ref{ptvsmultiplicity_bayesian} and \ref{ratiosvsmultilicity_bayesian}.
   The average transverse momentum as a function of \avdNchdeta is quite well reproduced by the model.
   The model predicts that the kaon-to-pion ratio should decrease with increasing charged particle multiplicity density while data show an increase with \avdNchdeta.
   Within uncertainties, the model agrees with the data for the most central \pbpb collisions.
   The trend of the proton-to-pion ratio is qualitatively well captured by the model but the values of the centrality-dependent ratios are overestimated.

   %\newpage
 \subsection{Intermediate transverse momentum}

\begin{figure}[htb]
     \centerline{
       \includegraphics[width=1.0\columnwidth]{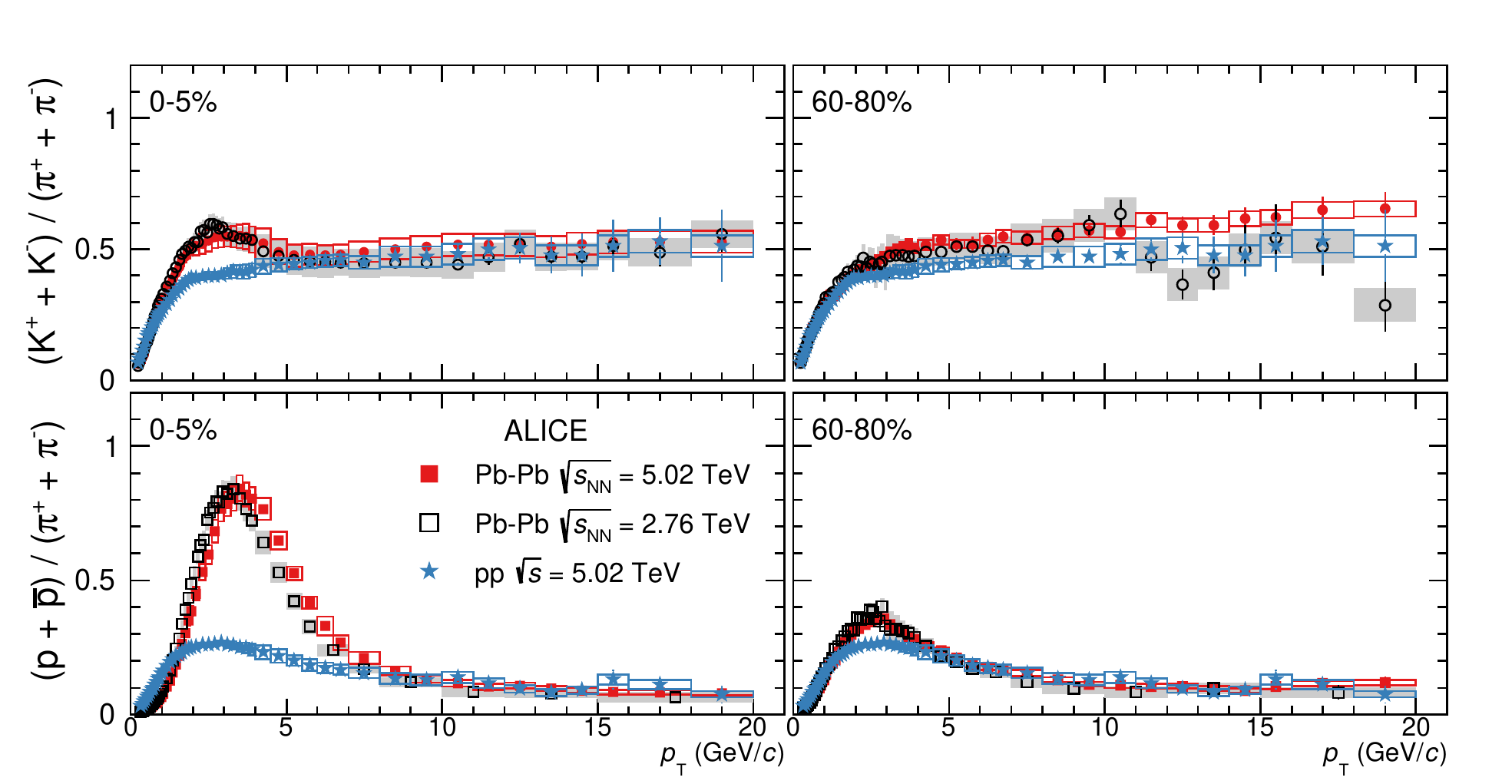}}
     \caption{Centrality dependence of the K/\ppi (top) and p/\ppi (bottom) ratios as a function of transverse momentum, measured in \pbpb collisions at \snnF and 2.76 TeV \cite{Adam:2015kca}.
       The ratios in pp collisions at \sF~TeV is also shown.
     The statistical and systematic uncertainties are shown as error bars and boxes around the data points, respectively.}
     \label{PtRatios}
   \end{figure}

   Figure~\ref{PtRatios} shows the K/\ppi and p/\ppi ratios as a function of \pt for \pbpb collisions at \snnT and 5.02~TeV.
   The results are also compared with inelastic pp collisions at \sF~TeV.
   Within uncertainties, in the K/\ppi ratio, no significant energy dependence is observed in heavy-ion data over the full \pt interval.
   The ratios measured in 60$-$80\% \pbpb collisions at both \snn agree within systematic uncertainties with that for inelastic pp collisions over the full \pt range.
   Given that in pp collisions at LHC energies the ratio as a function of \pt does not change with \s~\cite{Adam:2016dau}, and given the similarity between pp and peripheral \pbpb collisions, the large difference observed is likely a systematic effect of the measurement and not a physics effect. \\
   In general, the particle ratios exhibit a steep increase with \pt going from 0 to 3\,\gevc while for \pt larger than 10\,\gevc little or no \pt dependence is observed.
   Going from peripheral to the most central \pbpb collisions, the ratios in the region around $\pt\approx3$\,\gevc are continuously growing.
   A hint of an enhancement with respect to inelastic pp collisions is observed at $\pt\approx3$\,\gevc.
   As pointed out in a previous publication~\cite{Abelev:2013vea,Adam:2015kca}, the effect could be a consequence of radial flow which affects kaons more than pions.
     
  % Figure~\ref{PtRatios} shows the p/\ppi ratio as a function of \pt.
   The p/\ppi ratios measured in heavy-ion collisions exhibit a pronounced enhancement with respect to inelastic pp collisions, reaching a value of about 0.8 at \pt~=~3~\gevc.
   This is reminiscent of the increase in the baryon-to-meson ratio observed at RHIC in the intermediate \pt region \cite{Adare:2013esx,Abelev:2007ra}.
   Such an increase with \pt is due to the mass ordering induced by the radial flow (heavier particles are boosted to higher \pt by the collective motion) and it is an intrinsic feature of hydrodynamical models.
   It should be noted that this is also suggestive of the interplay of the hydrodynamic expansion of the system with the recombination picture as discussed in the introduction.
   However, since recombination mainly affects baryon-to-meson ratios, it would not explain the bump which is also observed in the kaon-to-pion ratio.
   The shift of the peak towards higher \pt in the proton-to-pion ratio is consistent with the larger radial flow
   measured in \pbpb at \snnF\,TeV compared to the one measured at \snnT\,TeV.
   The mass dependence of the radial flow explains also the observation that the maximum of the p/\ppi ratio is located at a larger \pt as compared to the K/\ppi ratio.
   The radial flow is expected to be stronger in the most central collisions, this explains the slight shift in the location of the maximum when central and peripheral data are compared.
   Finally, particle ratios at high \pt in \pbpb collisions at both energies become similar to those in pp collisions, suggesting that vacuum-like fragmentation processes dominate there \cite{Abelev:2014laa}.

 \subsection{Particle production at high transverse momentum}

\begin{figure}[htb]
     \centerline{
       \includegraphics[width=1.0\columnwidth]{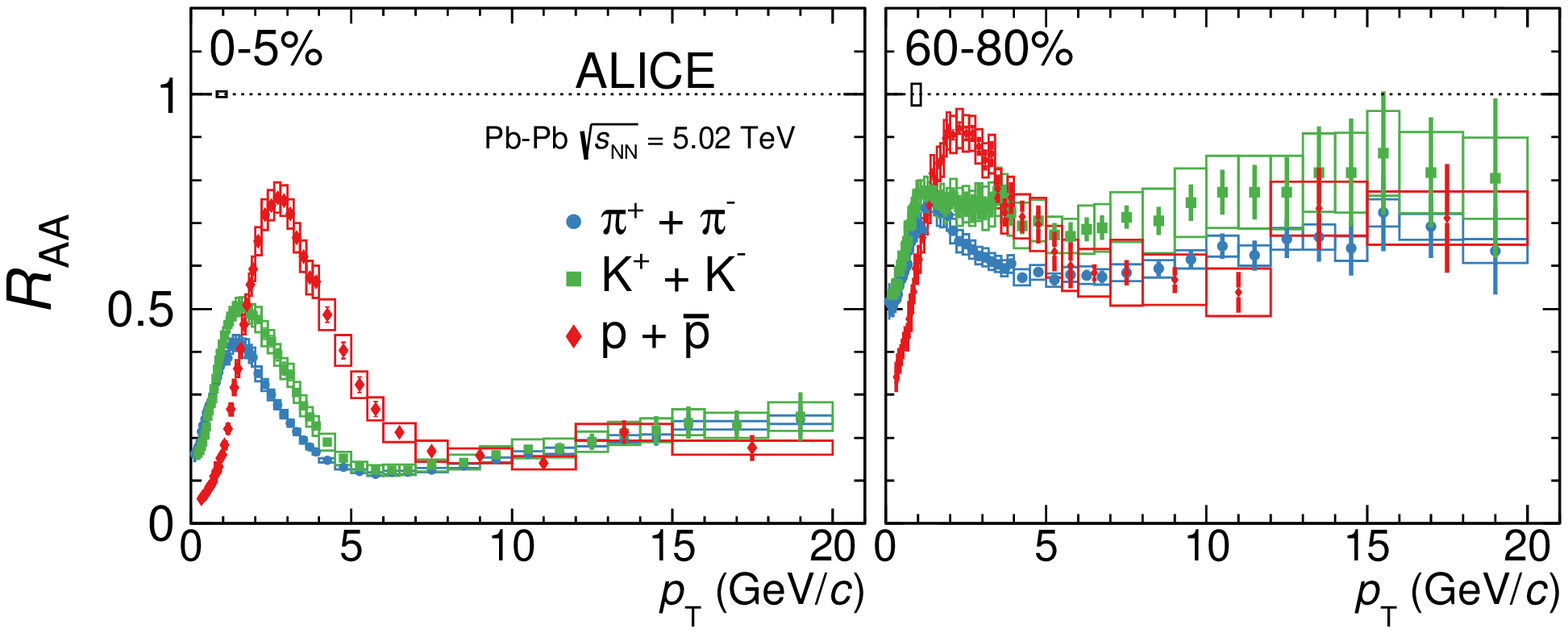}}
     \caption{Centrality dependence of the nuclear modification factor of charged \ppipm, \pkapm and \pprpm as a function of transverse momentum, measured in \pbpb collisions at \snnF~TeV.
       The statistical and systematic uncertainties are shown as error bars and boxes around the data points.
       The total normalization uncertainty (pp and \pbpb) is indicated in each panel by the vertical scale of the box centered at \pt = 1 GeV$/c$ and \raa = 1.}
     \label{RAA5TeV}
   \end{figure}

   Figure \ref{RAA5TeV} shows the centrality dependence of \raa as a function of \pt for charged pions, kaons and (anti-)protons.
   For \pt $<$ 10 \gevc, protons appear to be less suppressed than kaons and pions, consistent with the particle ratios shown in \Fig \ref{PtRatios}.
   The large difference between the suppression of different species is consistent with a mass ordering related to the radial flow. It is worth noting that 2.76~TeV measurements \cite{Adam:2017zbf} showed that the mesons, including $\phi$(1020), have smaller \raa than protons, indicating a baryon-meson ordering, so while there is a strong radial flow component, there are other effects affecting \raa in this \pt region. 
   At larger \pt, all particle species are equally suppressed.
   Despite the strong energy loss observed in the most central heavy-ion collisions, particle composition and ratios at high \pt are similar to those in vacuum.
   This suggests that jet quenching does not affect particle composition significantly.

   In the identified particle \raa for peripheral \pbpb collisions an apparent presence of jet quenching is observed ($\raa<1$), although for similar particle densities in smaller systems (like \ppb collisions) no jet quenching signatures have been reported~\cite{Adam:2014qja}.
   It has been argued that peripheral A$-$A collisions can be significantly affected by event selection and geometry biases~\cite{Morsch:2017brb}, leading to an apparent suppression for \raa even if jet quenching and shadowing are absent.
   The presence of biases on the \raa measurement in peripheral \pbpb collisions has been confirmed in \Ref~\cite{Acharya:2018njl}: the geometry bias sets in at mid-central collisions, reaching about 15\% for the 70$-$80\% \pbpb collisions.
   The additional effect of the selection bias becomes noticeable above the 60\% percentile and is significant above the 80\% percentile, where it is larger than 20\%.
   All hard probes should be similarly affected~\cite{Morsch:2017brb}, in particular, the leading pions, kaons and (anti-)protons reported in the present paper.

\begin{figure}[htb]
     \centerline{
       \includegraphics[width=1.0\columnwidth]{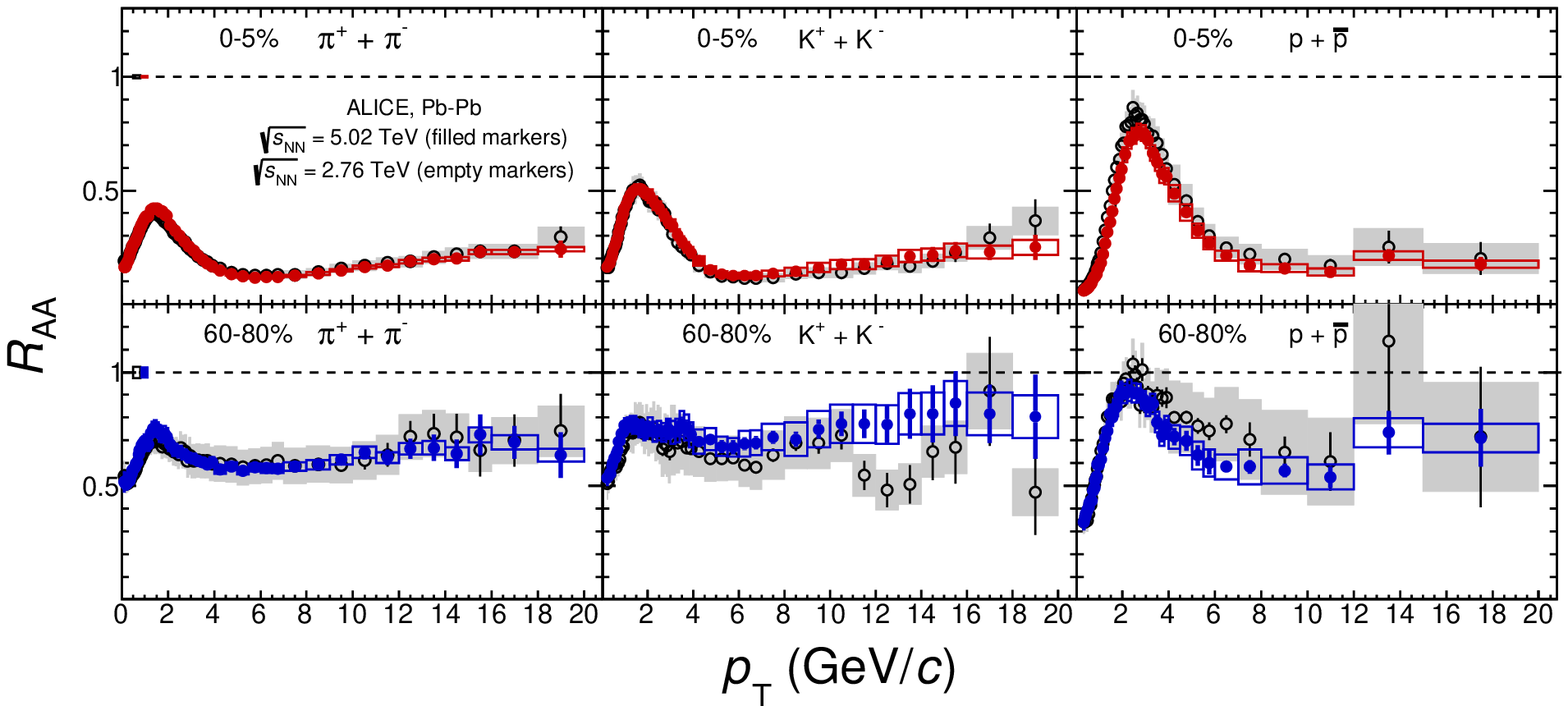}}
     \caption{Centrality dependence of the nuclear modification factor of charged \ppipm, \pkapm and \pprpm as a function of transverse momentum, measured in \pbpb collisions at \snnT \cite{Adam:2015kca} and 5.02 TeV, for 0$-$5\% and 60$-$80\% centrality classes.
       The statistical and systematic uncertainties are shown as error bars and boxes around the data points.
       The total normalization uncertainty (pp and \pbpb) is indicated in each panel by the vertical scale of the box centered at \pt = 1 \gevc and \raa = 1.}
     \label{RAATowEnergies}
   \end{figure}

   Figure~\ref{RAATowEnergies} shows the \raa for charged pions, kaons and (anti-)protons for central (0$-$5\%) and peripheral (60$-$80\%) \pbpb collisions at \snnT\,TeV \cite{Adam:2015kca} and \snnF\,TeV.
   No significant dependence on the collision energy is observed, as also been observed for unidentified charged particles~\cite{Acharya:2018qsh}.
   %, albeit the spectra in both pp and \pbpb collisions are substantially harder at \snnF\,TeV compared to  \snnT\,TeV, which may indicate a stronger parton energy loss and a larger energy density of the medium at the higher energy.

\section{Comparison to models}  \label{sec4}

 The results for identified particle production have been compared with the latest hydrodynamic model calculations based on the widely accepted ``standard" picture of heavy-ion collisions \cite{Busza:2018rrf}.
 These models all have similar ingredients: an initial state model provides the starting point for a viscous hydrodynamic calculation, chemical freeze-out occurs on a constant temperature hyper-surface, where local particle production is modeled with a statistical thermal model, and finally, the hadronic system is allowed to re-interact.
 The models used are: iEBE-VISHNU hybrid model~\cite{Shen:2014vra, Zhao:2017yhj}, McGill~\cite{McDonald:2016vlt} and EPOS \cite{Werner:2013tya}.
 In the following, specific features of each of them are described:

 \begin{itemize}
   \item The iEBE-VISHNU model is an event-by-event version of the VISHNU hybrid model~\cite{Song:2010aq}, which combines (2+1)-d viscous hydrodynamics VISH2+1~\cite{Song:2007fn,Song:2007ux} to describe the expansion of the sQGP fireball with a hadron cascade model (UrQMD)~\cite{Bleicher:1999xi,Bass:1998ca} to simulate the evolution of the system in the hadronic phase.
         The prediction of iEBE-VISHNU using either T$_{\rm R}$ENTo (Sec.\ \ref{sec3.1}) or AMPT (A Multi-Phase Transport Model)~\cite{Lin:2004en} as initial conditions gives a good description of flow measurements in \snnT TeV \pbpb collisions.
         T$_{\rm R}$ENTo parametrizes the initial entropy density via the reduced thickness function; AMPT constructs the initial energy density profiles using the energy decomposition of individual partons.
         Predictions by the iEBE-VISHNU hybrid model is available for \pt up to 3 \gevc.
   \item The McGill model initial conditions rely on a new formulation of the IP-Glasma model~\cite{Schenke:2012fw}, which provides realistic event-by-event fluctuations and non-zero pre-equilibrium flow at the early stage of heavy-ion collisions.
         Individual collision systems are evolved using relativistic hydrodynamics with non-zero shear and bulk viscosities~\cite{Ryu:2015vwa}.
         As the density of the system drops, fluid cells are converted into hadrons and further propagated microscopically using a hadronic cascade model~\cite{Bass:1998ca,Bleicher:1999xi}.
         The McGill predictions are available for \pt up to 4 \gevc and centralities 0-60\%.
   \item The EPOS model in the version EPOS3 is a phenomenological parton-based model that aims at modeling the full \pt range.
         EPOS is based on the theory of the Gribov-Regge multiple scattering, perturbative QCD and string fragmentation~\cite{Ryu:2015vwa}.
         However, dense regions in the system created in the collisions, the so-called core, is treated as a QGP and modeled with a hydrodynamic evolution followed by statistical hadronization.
         EPOS3 implements saturation in the initial state as predicted by the Color Glass Condensate model \cite{Gelis:2012ri}, a full viscous hydrodynamic simulation of the core, and a
         hadronic cascade, not present in the previous version of the model.
         EPOS3 implements also a new physics process that accounts for hydrodynamically expanding bulk matter, jets, and the interaction between the two, important for particle production at intermediate \pt \cite{Werner:2012xh} and reminiscent of the recombination mechanism \cite{Greco:2003xt,Fries:2003vb}.
         \\
 \end{itemize}

\begin{figure}[htb]
   \centerline{
     \includegraphics[width=1.0\columnwidth]{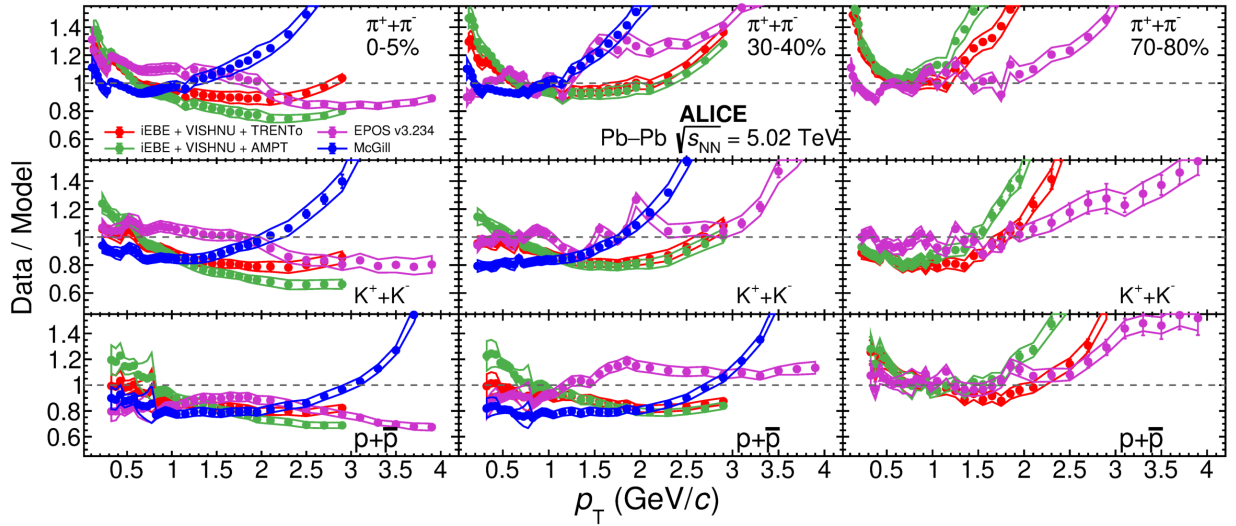}}
   \caption{Ratios of data to iEBE-VISHNU and McGill models (see text for details), for \ppipm, \pkapm and \pprpm \pt spectra in \pbpb collisions at \snnF\,TeV for centrality classes 0-5\%, 30-40\% and 70-80\%.
   The statistical and systematic uncertainties are shown as error bars and bands around the data points, respectively.}
   \label{spectramodels_zoomed}
 \end{figure}

 Figure \ref{spectramodels_zoomed} shows the ratios of the \pt spectra in \pbpb collisions at \snnF TeV to the models described above for \pt $<$ 4 \gevc.
 In the low \pt regime, one expects bulk particle production to dominate, so the absence of hard physics processes in the iEBE-VISHNU-T$_{\rm R}$ENTo, iEBE-VISHNU-AMPT, and McGill calculations is a minor issue.
 One observes that all models, in general, describe the spectra and the centrality dependence around \pt~$\approx$~1~\gevc within 20\%.
 For \pt $<$ 3 GeV/c the agreement with data is within 30\%.
 The models agree with the proton (kaon) data over a broader \pt range than for kaons (pions).
 This mass hierarchy is expected from the hydrodynamic expansion, which introduces a mass dependence via the flow velocity $-$ the larger the mass the larger the \pt boost.
 Similarly, it can be noticed that for the most central collisions the models describe the data over a broader \pt range than in peripheral ones.
 This is as expected from simple considerations.
 In central collisions, the system is larger and so the hydrodynamic expansion lasts longer, resulting in a stronger flow.
 At the same time, the fraction of the system involved in this expansion, the so-called core (e.g., the fraction of participant partons experiencing two or more binary collisions), is larger for the most central collisions.
 One can conclude that all four model calculations qualitatively describe the centrality dependence of radial flow and how it is imprinted on the different particle species.
 Like the simplified blast-wave fits in \Fig \ref{FitOverData}, the two iEBE-VISHNU calculations also have difficulties to describe the very low \pt (\pt~$<$~0.5~\gevc) pion spectra.

\begin{figure}[htb]
   \centerline{
     \includegraphics[width=1.0\columnwidth]{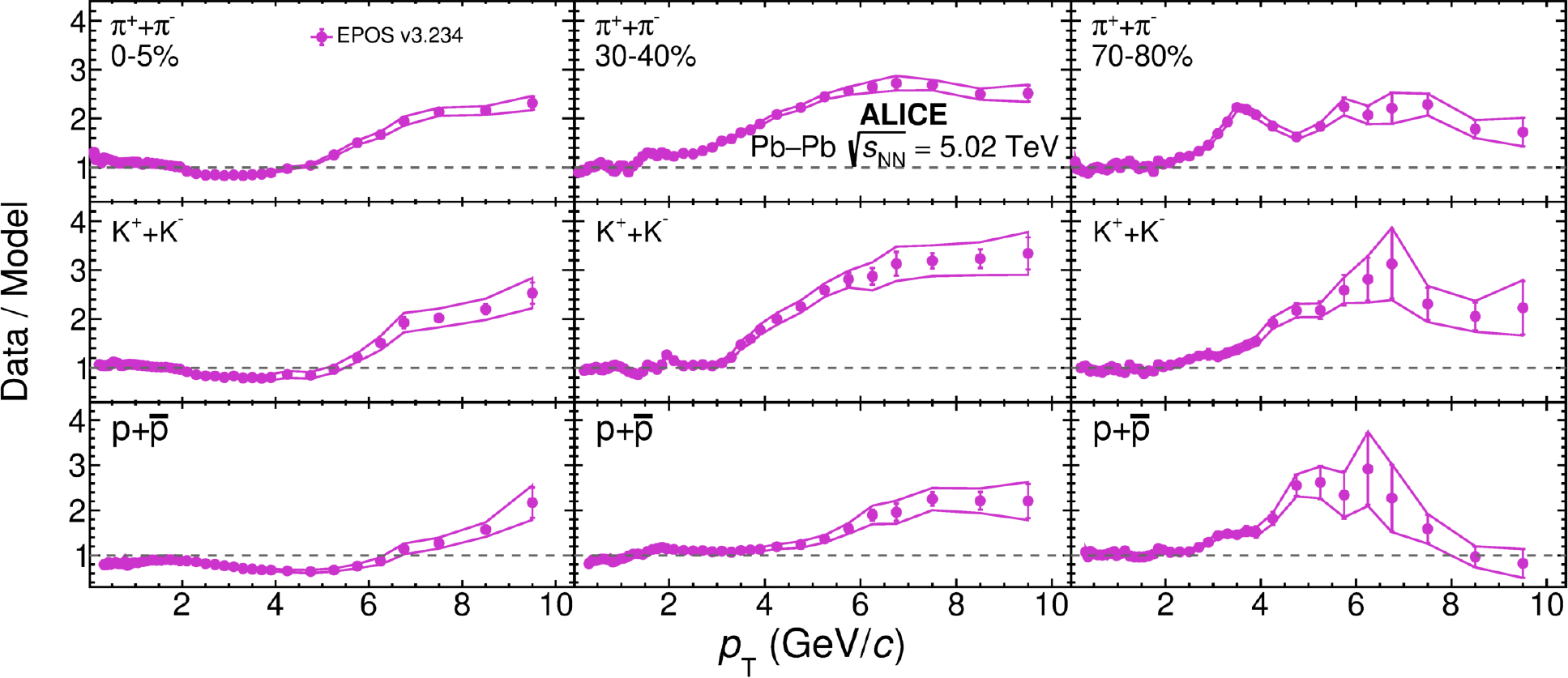}}
   \caption{Ratios of data to EPOS3 model (see text for details), for \ppipm, \pkapm and \pprpm \pt spectra in \pbpb collisions at \snnF\,TeV for centrality classes 0-5\%, 30-40\% and 70-80\%.
   The statistical and systematic uncertainties are shown as error bars and bands around the data points, respectively.}
   \label{spectramodels_eposonly}
 \end{figure}

 Figure \ref{spectramodels_eposonly} shows the ratios of the \pt spectra in \pbpb collisions at \snnF TeV to the EPOS3 model up to 10 \gevc in \pt.
 EPOS3 includes both soft and hard physics processes, which should give a better description of data at high \pt and in peripheral collisions.
 However, its agreement with data is not significantly better than for the other models in the same \pt interval (\pt $<$ 3\gevc) and at high \pt, it is about a factor 2 off with respect to data.

\begin{figure}[htb]
   \centerline{
     \includegraphics[width=1.0\columnwidth]{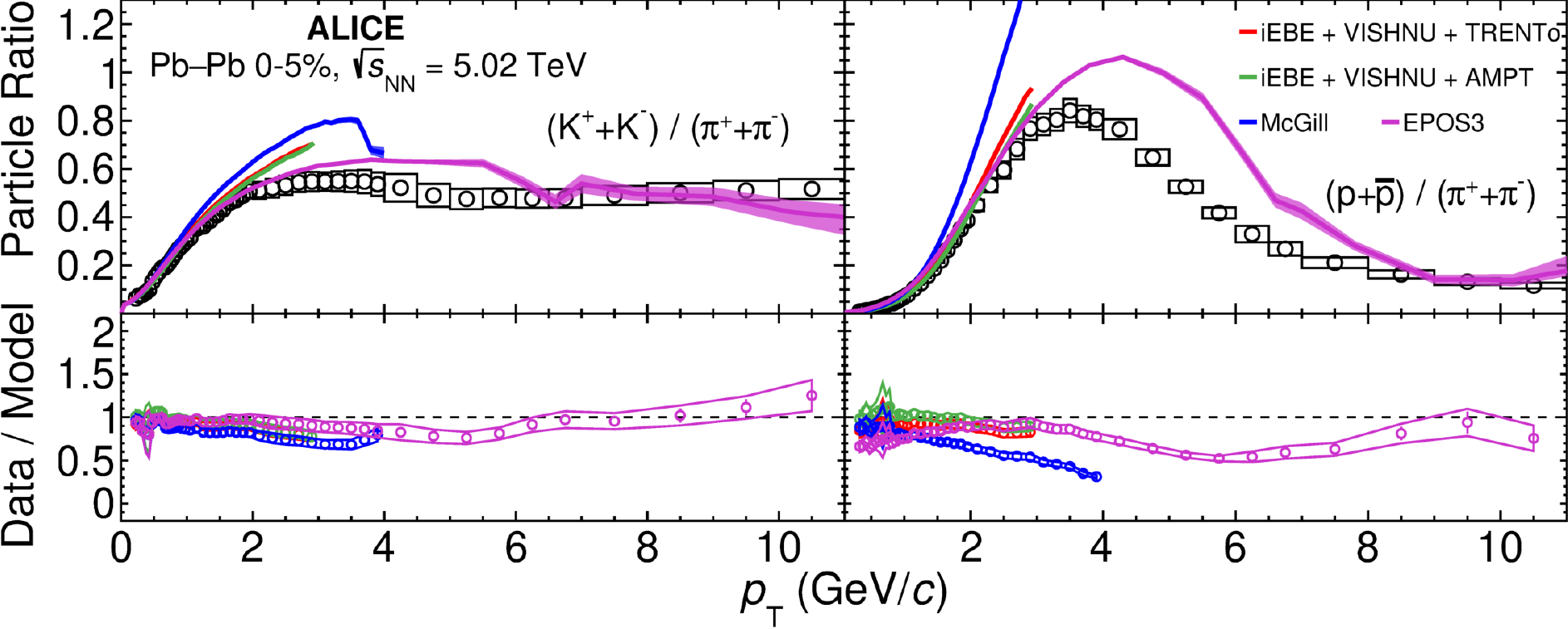}}
   \caption{(Top) K/\ppi and p/\ppi ratios in \pbpb collisions at \snnF~TeV in 0-5\% centrality class, compared to iEBE-VISHNU, McGill and EPOS3 model predictions, see text for details.
     The statistical and systematic uncertainties are shown as error bars and boxes around the data points, respectively.
     For model predictions the statistical uncertainties are represented by the band width.
   (Bottom) Data-to-model ratio, the statistical and systematic uncertainties are shown as error bars and bands around the data points, respectively.}
   \label{ratiomodels0_5}
 \end{figure}

\begin{figure}[htb]
   \centerline{
     \includegraphics[width=1.0\columnwidth]{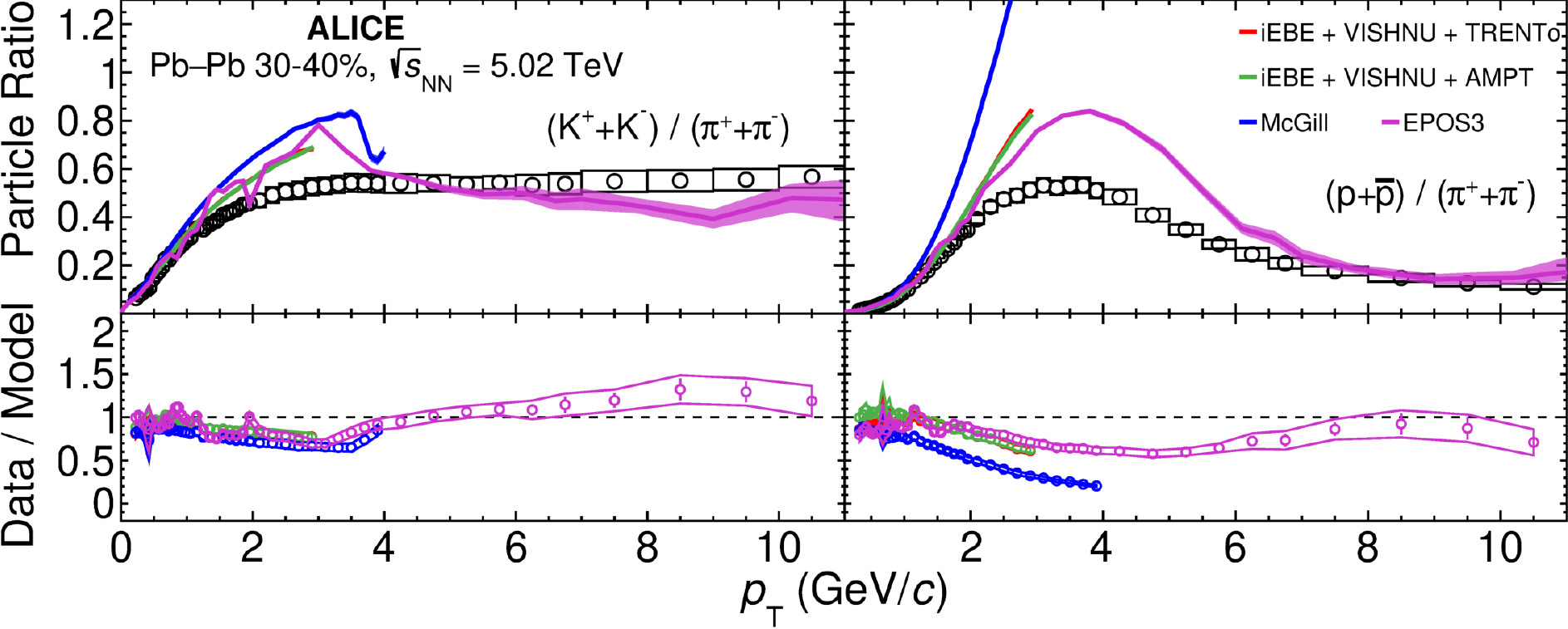}}
   \caption{(Top) K/\ppi and p/\ppi ratios in \pbpb collisions at \snnF~TeV in 30-40\% centrality class, compared to iEBE-VISHNU, McGill and EPOS3 model predictions, see text for details.
     The statistical and systematic uncertainties are shown as error bars and boxes around the data points, respectively.
     For model predictions the statistical uncertainties are represented by the band width.
   (Bottom) Data-to-model ratio, the statistical and systematic uncertainties are shown as error bars and bands around the data points, respectively.}
   \label{ratiomodels20_40}
 \end{figure}

\begin{figure}[htb]
   \centerline{
     \includegraphics[width=1.0\columnwidth]{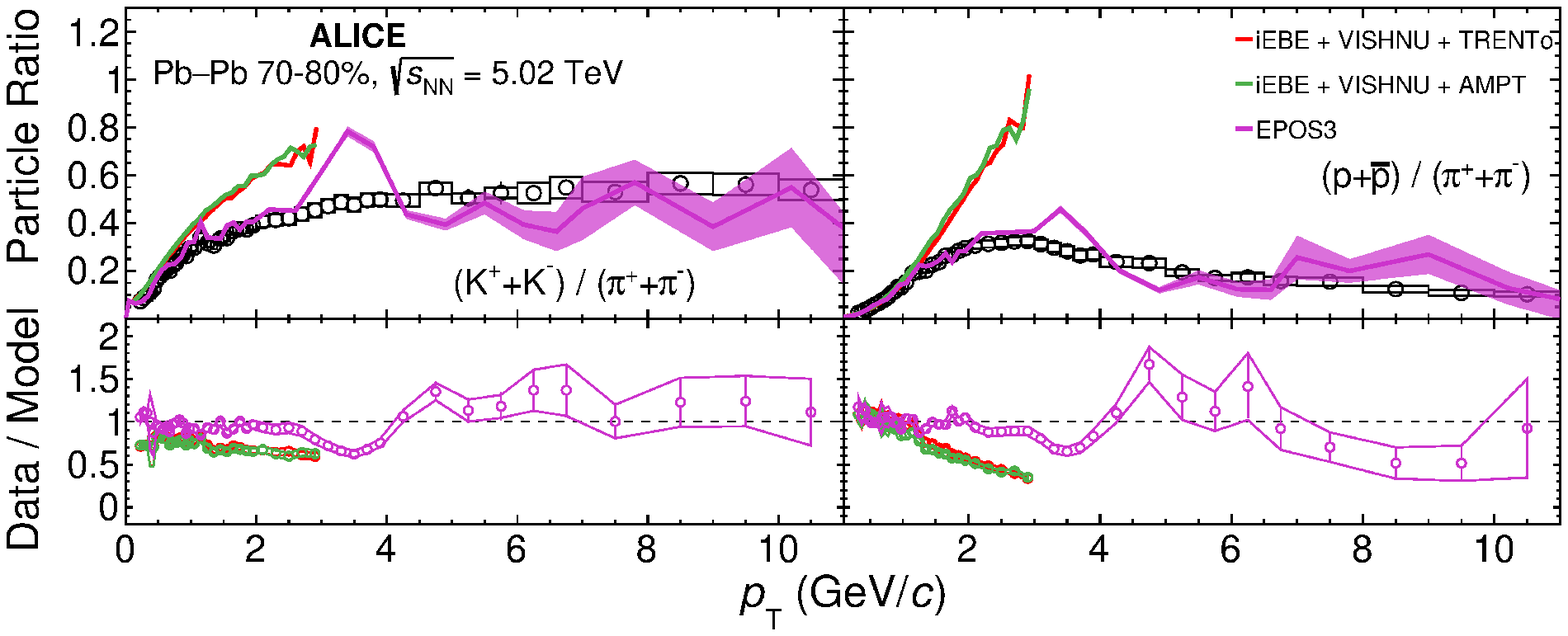}}
   \caption{(Top) K/\ppi and p/\ppi ratios in \pbpb collisions at \snnF~TeV in 70-80\% centrality class, compared to iEBE-VISHNU, McGill and EPOS3 model predictions, see text for details.
     The statistical and systematic uncertainties are shown as error bars and boxes around the data points, respectively.
     For model predictions the statistical uncertainties are represented by the band width.
   (Bottom) Data-to-model ratio, the statistical and systematic uncertainties are shown as error bars and bands around the data points, respectively.}
   \label{ratiomodels60_80}
 \end{figure}

 For completeness, \Figs \ref{ratiomodels0_5}, \ref{ratiomodels20_40} and \ref{ratiomodels60_80} show the comparison of the models with the \pt dependent particle ratios.
 The larger proton-to-pion ratio in EPOS3 than observed in the data can be understood as due to the underestimated pion yield in the model (see \Fig \ref{spectramodels_eposonly}).

\begin{figure}[htb]
   \centerline{
     \includegraphics[width=1.05\columnwidth]{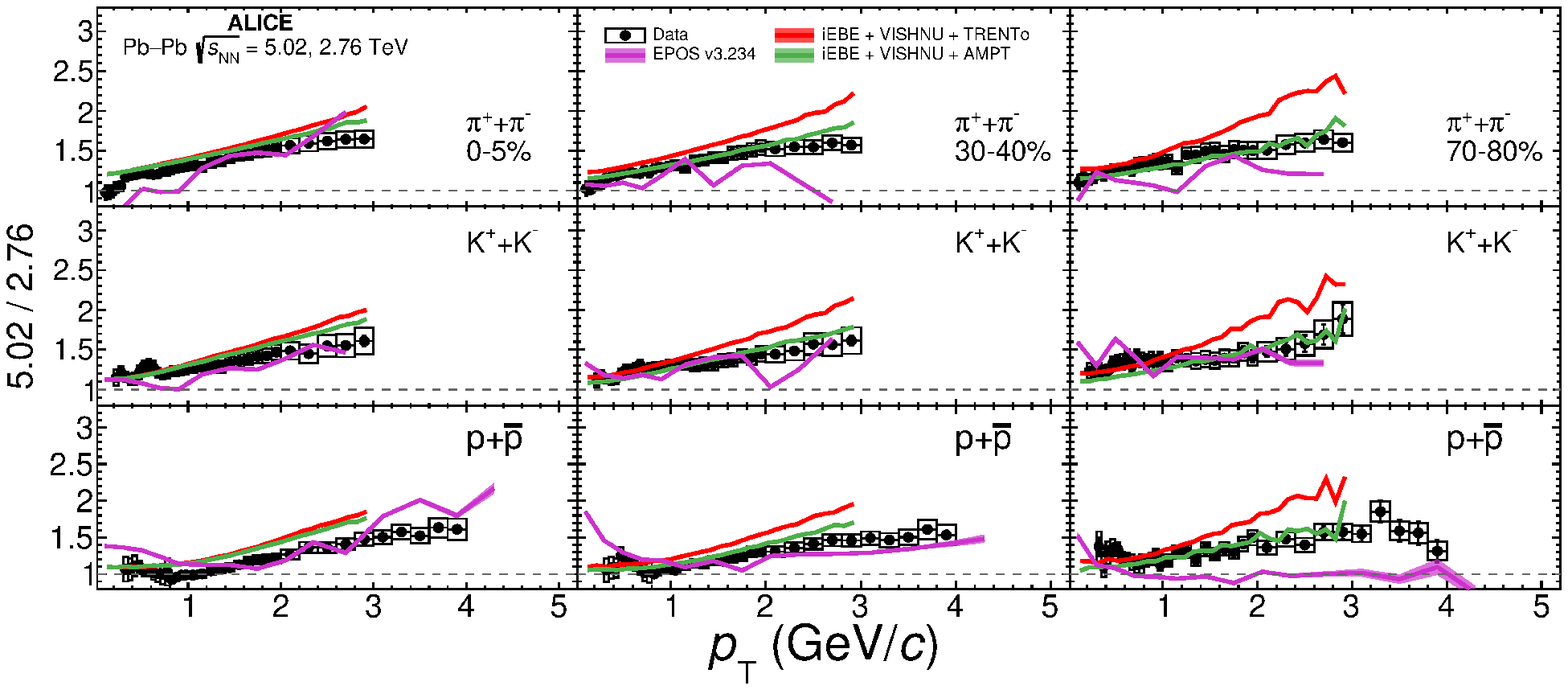}}
   \caption{\ppipm, \pkapm and \pprpm \pt spectra in \pbpb collisions at \snnF\,TeV divided for the same spectra measured in \pbpb collisions at \snnT\,TeV \cite{Adam:2015kca} for centrality classes 0-5\%, 30-40\% and 70-80\%, compared to iEBE-VISHNU, McGill and EPOS3 model predictions, see text for details.
     The statistical and systematic uncertainties are shown as error bars and boxes around the data points, respectively.
   For model predictions the statistical uncertainties are represented by the band width.}
   \label{doubleratio}
 \end{figure}

 In order to compare the energy evolution of the spectra between data and model,
 in \Fig \ref{doubleratio} is shown the ratio of the \ppipm, \pkapm and \pprpm \pt spectra measured at \snnF~TeV to those measured at \snnT~TeV, compared to the same ratios obtained from model predictions.
 %in \Fig \ref{doubleratio} is shown the double ratio: data to model at \snnF~TeV divided by the same ratio at \snn~=~2.76~TeV.
 For the McGill model, predictions at \snnT TeV are currently not available.
 For central collisions, the agreement of the energy evolution in data and predictions is very good for both VISHNU initial-state models, while for peripheral collisions the AMPT initial conditions are better.
 For EPOS3 instead, a good agreement with data can be observed for both central and peripheral collisions.
 The comparison of model predictions to the ALICE measurements of anisotropic flow \cite{Acharya:2018zuq, Abelev:2014pua, Abelev:2012di} can be useful in order to obtain tighter constraints on them.

 %\clearpage

\section{Conclusions}\label{sec5}

 In this paper, a comprehensive measurement of \ppipm, \pkapm and \pprpm production in inelastic pp and 0$-$90\% central \pbpb collisions at \snnF~TeV at the LHC is presented.
 A clear evolution of the spectra with centrality is observed, with a power-law-like behavior at high \pt and a flattening of the spectra at low \pt, confirming previous results obtained in \pbpb collisions at \snnT~TeV.
 These features are compatible with the development of a strong collective flow with centrality, which dominates the spectral shapes up to relatively high \pt in central collisions.
 The \pt-integrated particle ratios as a function of $\langle\rm{d}\it{N}_{\rm{ch}}$/d$\eta\rangle$ in \pbpb at \snnT\,TeV and 5.02 TeV as well as in inelastic pp collisions at \snnF\,TeV, have been compared.
 No significant energy dependence is observed, indicating that there is little or no dependence of the hadrochemistry on the collision energy.
 A blast-wave analysis of the \pt spectra gives an average transverse expansion velocity of $\langle \beta_{T} \rangle$~=~0.663~$\pm$~0.004 in the most central (0$-$5\%) \pbpb collisions that is ${\approx}~2\%$ larger than at \snnT\,TeV, with a difference of $\approx$ 2.4 standard deviations between the two energies. 
 The \pt-dependent particle ratios (p/\ppi, K/\ppi) show distinctive peaks at \pt~$\approx$~3~\gevc in central \pbpb collisions, more pronounced for the proton-to-pion ratio.
 Such an increase with \pt is due to the mass ordering induced by the radial flow that would affect heavier particles more than lighter ones.
 The \pt of the peak position increases slightly with energy, in particular for the proton-to-pion ratio, indicating that the initially hotter system is longer lived so that radial flow is stronger.
 At high \pt, both particle ratios at \snnF TeV are similar to those measured at \snnT TeV and in pp collisions, suggesting that vacuum-like fragmentation processes dominate there.
 No significant evolution of nuclear modification at high-\pt with the center-of-mass energy is observed.
 %although the spectra in both pp and \pbpb collisions are substantially harder at \snnF\,TeV compared to  \snnT\,TeV, indicating a stronger parton energy loss and a larger energy density of the medium at the higher energy.
 At high \pt, pions, kaons and (anti-)protons are equally suppressed as observed at \snnT~TeV.
 This suggests that the large energy loss leading to the suppression is not associated with strong mass ordering or large fragmentation differences between baryons and mesons.
 %Transverse momentum spectra and particle ratios in \pbpb collisions are compared to different model calculations based on the standard QGP picture.
 % For \pt~$<$~3~\gevc, all models agree with the data within 30\%, at \pt~$\approx$~1~\gevc they describe the spectra and the centrality dependence within 20\%.
 % They capture the essential features of the particle species and centrality dependence.
 Transverse momentum spectra and particle ratios in \pbpb collisions are compared to different model calculations based on the standard QGP picture, which are found to describe the observed trends satisfactorily. For \pt~$<$~3~\gevc, all models agree with the data within 30\%, at \pt~$\approx$~1~\gevc they describe the spectra and the centrality dependence within 20\%.

 %however, one cannot make a firm conclusion on which one is better.
 %To take full advantage of the precision dataset provided in this paper, the models have to implement the full suite of physics: soft production, hard production and energy loss.

 %
 %
 %\input{}               %%%%%%%%%%% put the body of the article here
 %
 %

 %%%%% acknowledgements
 \newenvironment{acknowledgement}{\relax}{\relax}
 \begin{acknowledgement}
   \section*{Acknowledgements}
   % Version: 2019-08-04

The ALICE Collaboration would like to thank all its engineers and technicians for their invaluable contributions to the construction of the experiment and the CERN accelerator teams for the outstanding performance of the LHC complex.
The ALICE Collaboration gratefully acknowledges the resources and support provided by all Grid centres and the Worldwide LHC Computing Grid (WLCG) collaboration.
The ALICE Collaboration acknowledges the following funding agencies for their support in building and running the ALICE detector:
A. I. Alikhanyan National Science Laboratory (Yerevan Physics Institute) Foundation (ANSL), State Committee of Science and World Federation of Scientists (WFS), Armenia;
Austrian Academy of Sciences, Austrian Science Fund (FWF): [M 2467-N36] and Nationalstiftung f\"{u}r Forschung, Technologie und Entwicklung, Austria;
Ministry of Communications and High Technologies, National Nuclear Research Center, Azerbaijan;
Conselho Nacional de Desenvolvimento Cient\'{\i}fico e Tecnol\'{o}gico (CNPq), Financiadora de Estudos e Projetos (Finep), Funda\c{c}\~{a}o de Amparo \`{a} Pesquisa do Estado de S\~{a}o Paulo (FAPESP) and Universidade Federal do Rio Grande do Sul (UFRGS), Brazil;
Ministry of Education of China (MOEC) , Ministry of Science \& Technology of China (MSTC) and National Natural Science Foundation of China (NSFC), China;
Ministry of Science and Education and Croatian Science Foundation, Croatia;
Centro de Aplicaciones Tecnol\'{o}gicas y Desarrollo Nuclear (CEADEN), Cubaenerg\'{\i}a, Cuba;
Ministry of Education, Youth and Sports of the Czech Republic, Czech Republic;
The Danish Council for Independent Research | Natural Sciences, the VILLUM FONDEN and Danish National Research Foundation (DNRF), Denmark;
Helsinki Institute of Physics (HIP), Finland;
Commissariat \`{a} l'Energie Atomique (CEA), Institut National de Physique Nucl\'{e}aire et de Physique des Particules (IN2P3) and Centre National de la Recherche Scientifique (CNRS) and R\'{e}gion des  Pays de la Loire, France;
Bundesministerium f\"{u}r Bildung und Forschung (BMBF) and GSI Helmholtzzentrum f\"{u}r Schwerionenforschung GmbH, Germany;
General Secretariat for Research and Technology, Ministry of Education, Research and Religions, Greece;
National Research, Development and Innovation Office, Hungary;
Department of Atomic Energy Government of India (DAE), Department of Science and Technology, Government of India (DST), University Grants Commission, Government of India (UGC) and Council of Scientific and Industrial Research (CSIR), India;
Indonesian Institute of Science, Indonesia;
Centro Fermi - Museo Storico della Fisica e Centro Studi e Ricerche Enrico Fermi and Istituto Nazionale di Fisica Nucleare (INFN), Italy;
Institute for Innovative Science and Technology , Nagasaki Institute of Applied Science (IIST), Japanese Ministry of Education, Culture, Sports, Science and Technology (MEXT) and Japan Society for the Promotion of Science (JSPS) KAKENHI, Japan;
Consejo Nacional de Ciencia (CONACYT) y Tecnolog\'{i}a, through Fondo de Cooperaci\'{o}n Internacional en Ciencia y Tecnolog\'{i}a (FONCICYT) and Direcci\'{o}n General de Asuntos del Personal Academico (DGAPA), Mexico;
Nederlandse Organisatie voor Wetenschappelijk Onderzoek (NWO), Netherlands;
The Research Council of Norway, Norway;
Commission on Science and Technology for Sustainable Development in the South (COMSATS), Pakistan;
Pontificia Universidad Cat\'{o}lica del Per\'{u}, Peru;
Ministry of Science and Higher Education and National Science Centre, Poland;
Korea Institute of Science and Technology Information and National Research Foundation of Korea (NRF), Republic of Korea;
Ministry of Education and Scientific Research, Institute of Atomic Physics and Ministry of Research and Innovation and Institute of Atomic Physics, Romania;
Joint Institute for Nuclear Research (JINR), Ministry of Education and Science of the Russian Federation, National Research Centre Kurchatov Institute, Russian Science Foundation and Russian Foundation for Basic Research, Russia;
Ministry of Education, Science, Research and Sport of the Slovak Republic, Slovakia;
National Research Foundation of South Africa, South Africa;
Swedish Research Council (VR) and Knut \& Alice Wallenberg Foundation (KAW), Sweden;
European Organization for Nuclear Research, Switzerland;
Suranaree University of Technology (SUT), National Science and Technology Development Agency (NSDTA) and Office of the Higher Education Commission under NRU project of Thailand, Thailand;
Turkish Atomic Energy Agency (TAEK), Turkey;
National Academy of  Sciences of Ukraine, Ukraine;
Science and Technology Facilities Council (STFC), United Kingdom;
National Science Foundation of the United States of America (NSF) and United States Department of Energy, Office of Nuclear Physics (DOE NP), United States of America.    %%%%%%% done by webmaster team
 \end{acknowledgement}

 %\section*{References}
 %%%%%%%% Bibliography (In case of using bibtex generate the bbl requested by arXiv)
 \bibliographystyle{utphys}   % Remember we use title in the biblio
 %\begin{thebibliography}{0}
 %\input {bibliography.tex}
 \bibliography{biblio}

\providecommand{\href}[2]{#2}\begingroup\raggedright\begin{thebibliography}{100}

\bibitem{Arsene:2004fa}
{\bfseries BRAHMS} Collaboration, I.~Arsene {\em et~al.}, ``{Quark gluon plasma
  and color glass condensate at RHIC? The Perspective from the BRAHMS
  experiment}'', \href{http://dx.doi.org/10.1016/j.nuclphysa.2005.02.130}{{\em
  Nucl. Phys.} {\bfseries A757} (2005) 1--27},
\href{http://arxiv.org/abs/nucl-ex/0410020}{{\ttfamily arXiv:nucl-ex/0410020
  [nucl-ex]}}.
%%CITATION = NUCL-EX/0410020;%%.

\bibitem{Adcox:2004mh}
{\bfseries PHENIX} Collaboration, K.~Adcox {\em et~al.}, ``{Formation of dense
  partonic matter in relativistic nucleus-nucleus collisions at RHIC:
  Experimental evaluation by the PHENIX collaboration}'',
  \href{http://dx.doi.org/10.1016/j.nuclphysa.2005.03.086}{{\em Nucl. Phys.}
  {\bfseries A757} (2005) 184--283},
\href{http://arxiv.org/abs/nucl-ex/0410003}{{\ttfamily arXiv:nucl-ex/0410003
  [nucl-ex]}}.
%%CITATION = NUCL-EX/0410003;%%.

\bibitem{Back:2004je}
B.~B. Back {\em et~al.}, ``{The PHOBOS perspective on discoveries at RHIC}'',
  \href{http://dx.doi.org/10.1016/j.nuclphysa.2005.03.084}{{\em Nucl. Phys.}
  {\bfseries A757} (2005) 28--101},
\href{http://arxiv.org/abs/nucl-ex/0410022}{{\ttfamily arXiv:nucl-ex/0410022
  [nucl-ex]}}.
%%CITATION = NUCL-EX/0410022;%%.

\bibitem{Adams:2005dq}
{\bfseries STAR} Collaboration, J.~Adams {\em et~al.}, ``{Experimental and
  theoretical challenges in the search for the quark gluon plasma: The STAR
  Collaboration's critical assessment of the evidence from RHIC collisions}'',
  \href{http://dx.doi.org/10.1016/j.nuclphysa.2005.03.085}{{\em Nucl. Phys.}
  {\bfseries A757} (2005) 102--183},
\href{http://arxiv.org/abs/nucl-ex/0501009}{{\ttfamily arXiv:nucl-ex/0501009
  [nucl-ex]}}.
%%CITATION = NUCL-EX/0501009;%%.

\bibitem{Schukraft:2011na}
{\bfseries ALICE} Collaboration, J.~Schukraft, ``{Heavy Ion physics with the
  ALICE experiment at the CERN LHC}'',
  \href{http://dx.doi.org/10.1098/rsta.2011.0469}{{\em Phil. Trans. Roy. Soc.
  Lond.} {\bfseries A370} (2012) 917--932},
\href{http://arxiv.org/abs/1109.4291}{{\ttfamily arXiv:1109.4291 [hep-ex]}}.
%%CITATION = ARXIV:1109.4291;%%.

\bibitem{Heinz:2013th}
U.~Heinz and R.~Snellings, ``{Collective flow and viscosity in relativistic
  heavy-ion collisions}'',
  \href{http://dx.doi.org/10.1146/annurev-nucl-102212-170540}{{\em Ann. Rev.
  Nucl. Part. Sci.} {\bfseries 63} (2013) 123--151},
\href{http://arxiv.org/abs/1301.2826}{{\ttfamily arXiv:1301.2826 [nucl-th]}}.
%%CITATION = ARXIV:1301.2826;%%.

\bibitem{Braun-Munzinger:2015hba}
P.~Braun-Munzinger, V.~Koch, T.~Schäfer, and J.~Stachel, ``{Properties of hot
  and dense matter from relativistic heavy ion collisions}'',
  \href{http://dx.doi.org/10.1016/j.physrep.2015.12.003}{{\em Phys. Rept.}
  {\bfseries 621} (2016) 76--126},
\href{http://arxiv.org/abs/1510.00442}{{\ttfamily arXiv:1510.00442 [nucl-th]}}.
%%CITATION = ARXIV:1510.00442;%%.

\bibitem{Adamczyk:2017iwn}
{\bfseries STAR} Collaboration, L.~Adamczyk {\em et~al.}, ``{Bulk Properties of
  the Medium Produced in Relativistic Heavy-Ion Collisions from the Beam Energy
  Scan Program}'', \href{http://dx.doi.org/10.1103/PhysRevC.96.044904}{{\em
  Phys. Rev.} {\bfseries C96} no.~4, (2017) 044904},
\href{http://arxiv.org/abs/1701.07065}{{\ttfamily arXiv:1701.07065 [nucl-ex]}}.
%%CITATION = ARXIV:1701.07065;%%.

\bibitem{Andronic:2017pug}
A.~Andronic, P.~Braun-Munzinger, K.~Redlich, and J.~Stachel, ``{Decoding the
  phase structure of QCD via particle production at high energy}'',
  \href{http://dx.doi.org/10.1038/s41586-018-0491-6}{{\em Nature} {\bfseries
  561} no.~7723, (2018) 321--330},
\href{http://arxiv.org/abs/1710.09425}{{\ttfamily arXiv:1710.09425 [nucl-th]}}.
%%CITATION = ARXIV:1710.09425;%%.

\bibitem{Becattini:2014hla}
F.~Becattini, E.~Grossi, M.~Bleicher, J.~Steinheimer, and R.~Stock,
  ``{Centrality dependence of hadronization and chemical freeze-out conditions
  in heavy ion collisions at $\sqrt s_{\rm{NN}}$ = 2.76 TeV}'',
  \href{http://dx.doi.org/10.1103/PhysRevC.90.054907}{{\em Phys. Rev.}
  {\bfseries C90} no.~5, (2014) 054907},
\href{http://arxiv.org/abs/1405.0710}{{\ttfamily arXiv:1405.0710 [nucl-th]}}.
%%CITATION = ARXIV:1405.0710;%%.

\bibitem{Becattini:2004rq}
F.~Becattini and L.~Ferroni, ``{Statistical hadronization and hadronic
  microcanonical ensemble 2.}'',
  \href{http://dx.doi.org/10.1140/epjc/s10052-010-1243-4,
  10.1140/epjc/s2004-02027-8}{{\em Eur. Phys. J.} {\bfseries C38} (2004)
  225--246}, \href{http://arxiv.org/abs/hep-ph/0407117}{{\ttfamily
  arXiv:hep-ph/0407117 [hep-ph]}}.
[Erratum: Eur. Phys. J.66,341(2010)].
%%CITATION = HEP-PH/0407117;%%.

\bibitem{Acharya:2017bso}
{\bfseries ALICE} Collaboration, S.~Acharya {\em et~al.}, ``{Production of
  $^{4}$He and $^{4}\overline{\textrm{He}}$ in Pb-Pb collisions at
  $\sqrt{s_{\mathrm{NN}}}$ = 2.76 TeV at the LHC}'',
  \href{http://dx.doi.org/10.1016/j.nuclphysa.2017.12.004}{{\em Nucl. Phys.}
  {\bfseries A971} (2018) 1--20},
\href{http://arxiv.org/abs/1710.07531}{{\ttfamily arXiv:1710.07531 [nucl-ex]}}.
%%CITATION = ARXIV:1710.07531;%%.

\bibitem{Bazavov:2014pvz}
{\bfseries HotQCD} Collaboration, A.~Bazavov {\em et~al.}, ``{Equation of state
  in ( 2+1 )-flavor QCD}'',
  \href{http://dx.doi.org/10.1103/PhysRevD.90.094503}{{\em Phys. Rev.}
  {\bfseries D90} (2014) 094503},
\href{http://arxiv.org/abs/1407.6387}{{\ttfamily arXiv:1407.6387 [hep-lat]}}.
%%CITATION = ARXIV:1407.6387;%%.

\bibitem{Abelev:2013vea}
{\bfseries ALICE} Collaboration, B.~Abelev {\em et~al.}, ``{Centrality
  dependence of $\pi$, K, p production in Pb-Pb collisions at
  $\sqrt{s_{\rm{NN}}}$ = 2.76 TeV}'',
  \href{http://dx.doi.org/10.1103/PhysRevC.88.044910}{{\em Phys. Rev.}
  {\bfseries C88} (2013) 044910},
\href{http://arxiv.org/abs/1303.0737}{{\ttfamily arXiv:1303.0737 [hep-ex]}}.
%%CITATION = ARXIV:1303.0737;%%.

\bibitem{Stock:2013iua}
R.~Stock, F.~Becattini, M.~Bleicher, T.~Kollegger, T.~Schuster, and
  J.~Steinheimer, ``{Hadronic Freeze-Out in A+A Collisions meets the Lattice
  QCD Parton-Hadron Transition Line}'', {\em PoS} {\bfseries CPOD2013} (2013)
  011,
\href{http://arxiv.org/abs/1306.4201}{{\ttfamily arXiv:1306.4201 [nucl-th]}}.
%%CITATION = ARXIV:1306.4201;%%.

\bibitem{Karpenko:2012yf}
I.~A. Karpenko, {\relax Yu}.~M. Sinyukov, and K.~Werner, ``{Uniform description
  of bulk observables in the hydrokinetic model of $A+A$ collisions at the BNL
  Relativistic Heavy Ion Collider and the CERN Large Hadron Collider}'',
  \href{http://dx.doi.org/10.1103/PhysRevC.87.024914}{{\em Phys. Rev.}
  {\bfseries C87} no.~2, (2013) 024914},
\href{http://arxiv.org/abs/1204.5351}{{\ttfamily arXiv:1204.5351 [nucl-th]}}.
%%CITATION = ARXIV:1204.5351;%%.

\bibitem{Becattini:2012xb}
F.~Becattini, M.~Bleicher, T.~Kollegger, T.~Schuster, J.~Steinheimer, and
  R.~Stock, ``{Hadron Formation in Relativistic Nuclear Collisions and the QCD
  Phase Diagram}'',
  \href{http://dx.doi.org/10.1103/PhysRevLett.111.082302}{{\em Phys. Rev.
  Lett.} {\bfseries 111} (2013) 082302},
\href{http://arxiv.org/abs/1212.2431}{{\ttfamily arXiv:1212.2431 [nucl-th]}}.
%%CITATION = ARXIV:1212.2431;%%.

\bibitem{Steinheimer:2012rd}
J.~Steinheimer, J.~Aichelin, and M.~Bleicher, ``{Nonthermal p/$\pi$ Ratio at
  LHC as a Consequence of Hadronic Final State Interactions}'',
  \href{http://dx.doi.org/10.1103/PhysRevLett.110.042501}{{\em Phys. Rev.
  Lett.} {\bfseries 110} no.~4, (2013) 042501},
\href{http://arxiv.org/abs/1203.5302}{{\ttfamily arXiv:1203.5302 [nucl-th]}}.
%%CITATION = ARXIV:1203.5302;%%.

\bibitem{Becattini:2016xct}
F.~Becattini, J.~Steinheimer, R.~Stock, and M.~Bleicher, ``{Hadronization
  conditions in relativistic nuclear collisions and the QCD pseudo-critical
  line}'', \href{http://dx.doi.org/10.1016/j.physletb.2016.11.033}{{\em Phys.
  Lett.} {\bfseries B764} (2017) 241--246},
\href{http://arxiv.org/abs/1605.09694}{{\ttfamily arXiv:1605.09694 [nucl-th]}}.
%%CITATION = ARXIV:1605.09694;%%.

\bibitem{Chatterjee:2013yga}
S.~Chatterjee, R.~M. Godbole, and S.~Gupta, ``{Strange freezeout}'',
  \href{http://dx.doi.org/10.1016/j.physletb.2013.11.008}{{\em Phys. Lett.}
  {\bfseries B727} (2013) 554--557},
\href{http://arxiv.org/abs/1306.2006}{{\ttfamily arXiv:1306.2006 [nucl-th]}}.
%%CITATION = ARXIV:1306.2006;%%.

\bibitem{Alba:2016hwx}
P.~Alba, V.~Vovchenko, M.~I. Gorenstein, and H.~Stoecker, ``{Flavor-dependent
  eigenvolume interactions in a hadron resonance gas}'',
  \href{http://dx.doi.org/10.1016/j.nuclphysa.2018.03.007}{{\em Nucl. Phys.}
  {\bfseries A974} (2018) 22--34},
\href{http://arxiv.org/abs/1606.06542}{{\ttfamily arXiv:1606.06542 [hep-ph]}}.
%%CITATION = ARXIV:1606.06542;%%.

\bibitem{Petran:2013lja}
M.~Petr\'a\ifmmode~\check{n}\else \v{n}\fi{}, J.~Letessier, V.~c.~v.
  Petr\'a\ifmmode~\check{c}\else \v{c}\fi{}ek, and J.~Rafelski, ``{Hadron
  production and quark-gluon plasma hadronization in Pb-Pb collisions at
  $\sqrt{s_{\rm{NN}}} = 2.76$ TeV}'',
  \href{http://dx.doi.org/10.1103/PhysRevC.88.034907}{{\em Phys. Rev.}
  {\bfseries C88} no.~3, (2013) 034907},
\href{http://arxiv.org/abs/1303.2098}{{\ttfamily arXiv:1303.2098 [hep-ph]}}.
%%CITATION = ARXIV:1303.2098;%%.

\bibitem{Bozek:2012qs}
P.~Bozek and I.~Wyskiel-Piekarska, ``{Particle spectra in Pb-Pb collisions at
  $\sqrt{s_{\rm{NN}}} = 2.76$ TeV}'',
  \href{http://dx.doi.org/10.1103/PhysRevC.85.064915}{{\em Phys. Rev.}
  {\bfseries C85} (2012) 064915},
\href{http://arxiv.org/abs/1203.6513}{{\ttfamily arXiv:1203.6513 [nucl-th]}}.
%%CITATION = ARXIV:1203.6513;%%.

\bibitem{Fries:2003kq}
R.~J. Fries, B.~Muller, C.~Nonaka, and S.~A. Bass, ``{Hadron production in
  heavy ion collisions: Fragmentation and recombination from a dense parton
  phase}'', \href{http://dx.doi.org/10.1103/PhysRevC.68.044902}{{\em Phys.
  Rev.} {\bfseries C68} (2003) 044902},
\href{http://arxiv.org/abs/nucl-th/0306027}{{\ttfamily arXiv:nucl-th/0306027
  [nucl-th]}}.
%%CITATION = NUCL-TH/0306027;%%.

\bibitem{Pop:2004dq}
V.~Topor~Pop, M.~Gyulassy, J.~Barrette, C.~Gale, X.~N. Wang, and N.~Xu,
  ``{Baryon junction loops in HIJING / B anti-B v2.0 and the baryon /meson
  anomaly at RHIC}'', \href{http://dx.doi.org/10.1103/PhysRevC.70.064906}{{\em
  Phys. Rev.} {\bfseries C70} (2004) 064906},
\href{http://arxiv.org/abs/nucl-th/0407095}{{\ttfamily arXiv:nucl-th/0407095
  [nucl-th]}}.
%%CITATION = NUCL-TH/0407095;%%.

\bibitem{Brodsky:2008qp}
S.~J. Brodsky and A.~Sickles, ``{The Baryon Anomaly: Evidence for Color
  Transparency and Direct Hadron Production at RHIC}'',
  \href{http://dx.doi.org/10.1016/j.physletb.2008.07.108}{{\em Phys. Lett.}
  {\bfseries B668} (2008) 111--115},
\href{http://arxiv.org/abs/0804.4608}{{\ttfamily arXiv:0804.4608 [hep-ph]}}.
%%CITATION = ARXIV:0804.4608;%%.

\bibitem{Abelev:2013xaa}
{\bfseries ALICE} Collaboration, B.~B. Abelev {\em et~al.}, ``{$K^0_S$ and
  $\Lambda$ production in Pb-Pb collisions at $\sqrt{s_{\rm{NN}}}$ = 2.76
  TeV}'', \href{http://dx.doi.org/10.1103/PhysRevLett.111.222301}{{\em Phys.
  Rev. Lett.} {\bfseries 111} (2013) 222301},
\href{http://arxiv.org/abs/1307.5530}{{\ttfamily arXiv:1307.5530 [nucl-ex]}}.
%%CITATION = ARXIV:1307.5530;%%.

\bibitem{Adam:2015kca}
{\bfseries ALICE} Collaboration, J.~Adam {\em et~al.}, ``{Centrality dependence
  of the nuclear modification factor of charged pions, kaons, and protons in
  Pb-Pb collisions at $\sqrt{s_{\rm{NN}}} =2.76$ TeV}'',
  \href{http://dx.doi.org/10.1103/PhysRevC.93.034913}{{\em Phys. Rev.}
  {\bfseries C93} no.~3, (2016) 034913},
\href{http://arxiv.org/abs/1506.07287}{{\ttfamily arXiv:1506.07287 [nucl-ex]}}.
%%CITATION = ARXIV:1506.07287;%%.

\bibitem{Shen:2014vra}
C.~Shen, Z.~Qiu, H.~Song, J.~Bernhard, S.~Bass, and U.~Heinz, ``{The
  iEBE-VISHNU code package for relativistic heavy-ion collisions}'',
  \href{http://dx.doi.org/10.1016/j.cpc.2015.08.039}{{\em Comput. Phys.
  Commun.} {\bfseries 199} (2016) 61--85},
\href{http://arxiv.org/abs/1409.8164}{{\ttfamily arXiv:1409.8164 [nucl-th]}}.
%%CITATION = ARXIV:1409.8164;%%.

\bibitem{Zhao:2017yhj}
W.~Zhao, H.-j. Xu, and H.~Song, ``{Collective flow in 2.76 A TeV and 5.02 A TeV
  Pb+Pb collisions}'',
  \href{http://dx.doi.org/10.1140/epjc/s10052-017-5186-x}{{\em Eur. Phys. J.}
  {\bfseries C77} no.~9, (2017) 645},
\href{http://arxiv.org/abs/1703.10792}{{\ttfamily arXiv:1703.10792 [nucl-th]}}.
%%CITATION = ARXIV:1703.10792;%%.

\bibitem{McDonald:2016vlt}
S.~McDonald, C.~Shen, F.~Fillion-Gourdeau, S.~Jeon, and C.~Gale,
  ``{Hydrodynamic predictions for Pb+Pb collisions at 5.02 TeV}'',
  \href{http://dx.doi.org/10.1103/PhysRevC.95.064913}{{\em Phys. Rev.}
  {\bfseries C95} no.~6, (2017) 064913},
\href{http://arxiv.org/abs/1609.02958}{{\ttfamily arXiv:1609.02958 [hep-ph]}}.
%%CITATION = ARXIV:1609.02958;%%.

\bibitem{Greco:2003xt}
V.~Greco, C.~M. Ko, and P.~Levai, ``{Parton coalescence and anti-proton / pion
  anomaly at RHIC}'',
  \href{http://dx.doi.org/10.1103/PhysRevLett.90.202302}{{\em Phys. Rev. Lett.}
  {\bfseries 90} (2003) 202302},
\href{http://arxiv.org/abs/nucl-th/0301093}{{\ttfamily arXiv:nucl-th/0301093
  [nucl-th]}}.
%%CITATION = NUCL-TH/0301093;%%.

\bibitem{Fries:2003vb}
R.~J. Fries, B.~Muller, C.~Nonaka, and S.~A. Bass, ``{Hadronization in heavy
  ion collisions: Recombination and fragmentation of partons}'',
  \href{http://dx.doi.org/10.1103/PhysRevLett.90.202303}{{\em Phys. Rev. Lett.}
  {\bfseries 90} (2003) 202303},
\href{http://arxiv.org/abs/nucl-th/0301087}{{\ttfamily arXiv:nucl-th/0301087
  [nucl-th]}}.
%%CITATION = NUCL-TH/0301087;%%.

\bibitem{Minissale:2015zwa}
V.~Minissale, F.~Scardina, and V.~Greco, ``{Hadrons from coalescence plus
  fragmentation in AA collisions at energies available at the BNL Relativistic
  Heavy Ion Collider to the CERN Large Hadron Collider}'',
  \href{http://dx.doi.org/10.1103/PhysRevC.92.054904}{{\em Phys. Rev.}
  {\bfseries C92} no.~5, (2015) 054904},
\href{http://arxiv.org/abs/1502.06213}{{\ttfamily arXiv:1502.06213 [nucl-th]}}.
%%CITATION = ARXIV:1502.06213;%%.

\bibitem{Abelev:2014laa}
{\bfseries ALICE} Collaboration, B.~B. Abelev {\em et~al.}, ``{Production of
  charged pions, kaons and protons at large transverse momenta in pp and
  Pb$-$Pb collisions at $\sqrt{s_{\rm{NN}}}$ =2.76 TeV}'',
  \href{http://dx.doi.org/10.1016/j.physletb.2014.07.011}{{\em Phys. Lett.}
  {\bfseries B736} (2014) 196--207},
\href{http://arxiv.org/abs/1401.1250}{{\ttfamily arXiv:1401.1250 [nucl-ex]}}.
%%CITATION = ARXIV:1401.1250;%%.

\bibitem{Burke:2013yra}
{\bfseries JET} Collaboration, K.~M. Burke {\em et~al.}, ``{Extracting the jet
  transport coefficient from jet quenching in high-energy heavy-ion
  collisions}'', \href{http://dx.doi.org/10.1103/PhysRevC.90.014909}{{\em Phys.
  Rev.} {\bfseries C90} no.~1, (2014) 014909},
\href{http://arxiv.org/abs/1312.5003}{{\ttfamily arXiv:1312.5003 [nucl-th]}}.
%%CITATION = ARXIV:1312.5003;%%.

\bibitem{Wang:2014xda}
X.-N. Wang, ``{What hard probes tell us about the quark-gluon plasma:
  Theory}'', \href{http://dx.doi.org/10.1016/j.nuclphysa.2014.09.065}{{\em
  Nucl. Phys.} {\bfseries A932} (2014) 1--8},
\href{http://arxiv.org/abs/1404.2327}{{\ttfamily arXiv:1404.2327 [nucl-th]}}.
%%CITATION = ARXIV:1404.2327;%%.

\bibitem{Abelev:2013qoq}
{\bfseries ALICE} Collaboration, B.~Abelev {\em et~al.}, ``{Centrality
  determination of Pb-Pb collisions at $\sqrt{s_{\rm{NN}}}$ = 2.76 TeV with
  ALICE}'', \href{http://dx.doi.org/10.1103/PhysRevC.88.044909}{{\em Phys.
  Rev.} {\bfseries C88} no.~4, (2013) 044909},
\href{http://arxiv.org/abs/1301.4361}{{\ttfamily arXiv:1301.4361 [nucl-ex]}}.
%%CITATION = ARXIV:1301.4361;%%.

\bibitem{Loizides:2017ack}
C.~Loizides, J.~Kamin, and D.~d'Enterria, ``{Improved Monte Carlo Glauber
  predictions at present and future nuclear colliders}'',
  \href{http://dx.doi.org/10.1103/PhysRevC.97.054910,
  10.1103/PhysRevC.99.019901}{{\em Phys. Rev.} {\bfseries C97} no.~5, (2018)
  054910}, \href{http://arxiv.org/abs/1710.07098}{{\ttfamily arXiv:1710.07098
  [nucl-ex]}}.
[erratum: Phys. Rev.C99,no.1,019901(2019)].
%%CITATION = ARXIV:1710.07098;%%.

\bibitem{Adler:2003au}
{\bfseries PHENIX} Collaboration, S.~S. Adler {\em et~al.}, ``{High
  $p_{\rm{T}}$ charged hadron suppression in Au + Au collisions at
  $\sqrt{s_{\rm{NN}}} = 200$ GeV}'',
  \href{http://dx.doi.org/10.1103/PhysRevC.69.034910}{{\em Phys. Rev.}
  {\bfseries {\bf C69}} (2004) 034910},
\href{http://arxiv.org/abs/nucl-ex/0308006}{{\ttfamily arXiv:nucl-ex/0308006
  [nucl-ex]}}.
%%CITATION = NUCL-EX/0308006;%%.

\bibitem{Adams:2003kv}
{\bfseries STAR} Collaboration, J.~Adams {\em et~al.}, ``{Transverse momentum
  and collision energy dependence of high $p_{\rm{T}}$ hadron suppression in
  Au+Au collisions at ultrarelativistic energies}'',
  \href{http://dx.doi.org/10.1103/PhysRevLett.91.172302}{{\em Phys. Rev. Lett.}
  {\bfseries {\bf 91}} (2003) 172302},
\href{http://arxiv.org/abs/nucl-ex/0305015}{{\ttfamily arXiv:nucl-ex/0305015
  [nucl-ex]}}.
%%CITATION = NUCL-EX/0305015;%%.

\bibitem{Aamodt:2010jd}
{\bfseries ALICE} Collaboration, K.~Aamodt {\em et~al.}, ``{Suppression of
  Charged Particle Production at Large Transverse Momentum in Central Pb-Pb
  Collisions at $\sqrt{s_{\rm{NN}}} =$ 2.76 TeV}'',
  \href{http://dx.doi.org/10.1016/j.physletb.2010.12.020}{{\em Phys. Lett.}
  {\bfseries {\bf B696}} (2011) 30--39},
\href{http://arxiv.org/abs/1012.1004}{{\ttfamily arXiv:1012.1004 [nucl-ex]}}.
%%CITATION = ARXIV:1012.1004;%%.

\bibitem{CMS:2012aa}
{\bfseries CMS} Collaboration, S.~Chatrchyan {\em et~al.}, ``{Study of
  high-$p_{\rm{T}}$ charged particle suppression in Pb$-$Pb compared to $pp$
  collisions at $\sqrt{s_{\rm{NN}}}=2.76$ TeV}'',
  \href{http://dx.doi.org/10.1140/epjc/s10052-012-1945-x}{{\em Eur. Phys. J.}
  {\bfseries {\bf C72}} (2012) 1945},
\href{http://arxiv.org/abs/1202.2554}{{\ttfamily arXiv:1202.2554 [nucl-ex]}}.
%%CITATION = ARXIV:1202.2554;%%.

\bibitem{Aad:2015wga}
{\bfseries ATLAS} Collaboration, G.~Aad {\em et~al.}, ``{Measurement of
  charged-particle spectra in Pb$-$Pb collisions at $\sqrt{{s}_\mathsf{{NN}}} =
  2.76$ TeV with the ATLAS detector at the LHC}'',
  \href{http://dx.doi.org/10.1007/JHEP09(2015)050}{{\em JHEP} {\bfseries {\bf
  09}} (2015) 050},
\href{http://arxiv.org/abs/1504.04337}{{\ttfamily arXiv:1504.04337 [hep-ex]}}.
%%CITATION = ARXIV:1504.04337;%%.

\bibitem{Abelev:2007ra}
{\bfseries STAR} Collaboration, B.~I. Abelev {\em et~al.}, ``{Energy dependence
  of $\pi^{\pm}$, p and anti-p transverse momentum spectra for Au+Au collisions
  at $\sqrt{s_{\rm{NN}}}$ = 62.4 and 200-GeV}'',
  \href{http://dx.doi.org/10.1016/j.physletb.2007.06.035}{{\em Phys. Lett.}
  {\bfseries B655} (2007) 104--113},
\href{http://arxiv.org/abs/nucl-ex/0703040}{{\ttfamily arXiv:nucl-ex/0703040
  [nucl-ex]}}.
%%CITATION = NUCL-EX/0703040;%%.

\bibitem{Agakishiev:2011dc}
{\bfseries STAR} Collaboration, G.~Agakishiev {\em et~al.}, ``{Identified
  hadron compositions in p+p and Au+Au collisions at high transverse momenta at
  $\sqrt{s_{_{NN}}} = 200$ GeV}'',
  \href{http://dx.doi.org/10.1103/PhysRevLett.108.072302}{{\em Phys. Rev.
  Lett.} {\bfseries 108} (2012) 072302},
\href{http://arxiv.org/abs/1110.0579}{{\ttfamily arXiv:1110.0579 [nucl-ex]}}.
%%CITATION = ARXIV:1110.0579;%%.

\bibitem{Qin:2015srf}
G.-Y. Qin and X.-N. Wang, ``{Jet quenching in high-energy heavy-ion
  collisions}'', \href{http://dx.doi.org/10.1142/S0218301315300143,
  10.1142/9789814663717_0007}{{\em Int. J. Mod. Phys.} {\bfseries {\bf E24}}
  no.~11, (2015) 1530014}, \href{http://arxiv.org/abs/1511.00790}{{\ttfamily
  arXiv:1511.00790 [hep-ph]}}.
[,309(2016)].
%%CITATION = ARXIV:1511.00790;%%.

\bibitem{Khachatryan:2016odn}
{\bfseries CMS} Collaboration, V.~Khachatryan {\em et~al.}, ``{Charged-particle
  nuclear modification factors in Pb$-$Pb and p$-$Pb collisions at $
  \sqrt{s_{\mathrm{N}\;\mathrm{N}}}=5.02 $ TeV}'',
  \href{http://dx.doi.org/10.1007/JHEP04(2017)039}{{\em JHEP} {\bfseries 04}
  (2017) 039},
\href{http://arxiv.org/abs/1611.01664}{{\ttfamily arXiv:1611.01664 [nucl-ex]}}.
%%CITATION = ARXIV:1611.01664;%%.

\bibitem{Khachatryan:2016jfl}
{\bfseries CMS} Collaboration, V.~Khachatryan {\em et~al.}, ``{Measurement of
  inclusive jet cross sections in $pp$ and PbPb collisions at
  $\sqrt{s_{\rm{NN}}}$ = 2.76 TeV}'',
  \href{http://dx.doi.org/10.1103/PhysRevC.96.015202}{{\em Phys. Rev.}
  {\bfseries C96} no.~1, (2017) 015202},
\href{http://arxiv.org/abs/1609.05383}{{\ttfamily arXiv:1609.05383 [nucl-ex]}}.
%%CITATION = ARXIV:1609.05383;%%.

\bibitem{Adare:2015cua}
{\bfseries PHENIX} Collaboration, A.~Adare {\em et~al.}, ``{Scaling properties
  of fractional momentum loss of high-$p_{\rm{T}}$ hadrons in nucleus-nucleus
  collisions at $\sqrt{s_{\rm{NN}}}$ from 62.4 GeV to 2.76 TeV}'',
  \href{http://dx.doi.org/10.1103/PhysRevC.93.024911}{{\em Phys. Rev.}
  {\bfseries C93} no.~2, (2016) 024911},
\href{http://arxiv.org/abs/1509.06735}{{\ttfamily arXiv:1509.06735 [nucl-ex]}}.
%%CITATION = ARXIV:1509.06735;%%.

\bibitem{Christiansen:2013hya}
P.~Christiansen, K.~Tywoniuk, and V.~Vislavicius, ``{Universal scaling
  dependence of QCD energy loss from data driven studies}'',
  \href{http://dx.doi.org/10.1103/PhysRevC.89.034912}{{\em Phys. Rev.}
  {\bfseries C89} no.~3, (2014) 034912},
\href{http://arxiv.org/abs/1311.1173}{{\ttfamily arXiv:1311.1173 [hep-ph]}}.
%%CITATION = ARXIV:1311.1173;%%.

\bibitem{Ortiz:2017cul}
A.~Ortiz and O.~Vazquez, ``{Energy density and path-length dependence of the
  fractional momentum loss in heavy-ion collisions at $\sqrt{s_{\rm{NN}}}$ from
  62.4 to 5020 GeV}'', \href{http://dx.doi.org/10.1103/PhysRevC.97.014910}{{\em
  Phys. Rev.} {\bfseries C97} no.~1, (2018) 014910},
\href{http://arxiv.org/abs/1708.07571}{{\ttfamily arXiv:1708.07571 [hep-ph]}}.
%%CITATION = ARXIV:1708.07571;%%.

\bibitem{Acharya:2018eaq}
{\bfseries ALICE} Collaboration, S.~Acharya {\em et~al.}, ``{Transverse
  momentum spectra and nuclear modification factors of charged particles in
  Xe-Xe collisions at $\sqrt{s_{\rm NN}}$ = 5.44 TeV}'',
  \href{http://dx.doi.org/10.1016/j.physletb.2018.10.052}{{\em Phys. Lett.}
  {\bfseries B788} (2019) 166--179},
\href{http://arxiv.org/abs/1805.04399}{{\ttfamily arXiv:1805.04399 [nucl-ex]}}.
%%CITATION = ARXIV:1805.04399;%%.

\bibitem{Aamodt:2008zz}
{\bfseries ALICE} Collaboration, K.~Aamodt {\em et~al.}, ``{The ALICE
  experiment at the CERN LHC}'',
\href{http://dx.doi.org/10.1088/1748-0221/3/08/S08002}{{\em JINST} {\bfseries
  3} (2008) S08002}.
%%CITATION = JINST,3,S08002;%%.

\bibitem{Abelev:2014ffa}
{\bfseries ALICE} Collaboration, B.~B. Abelev {\em et~al.}, ``{Performance of
  the ALICE Experiment at the CERN LHC}'',
  \href{http://dx.doi.org/10.1142/S0217751X14300440}{{\em Int. J. Mod. Phys.}
  {\bfseries A29} (2014) 1430044},
\href{http://arxiv.org/abs/1402.4476}{{\ttfamily arXiv:1402.4476 [nucl-ex]}}.
%%CITATION = ARXIV:1402.4476;%%.

\bibitem{Abbas:2013taa}
{\bfseries ALICE} Collaboration, E.~Abbas {\em et~al.}, ``{Performance of the
  ALICE VZERO system}'',
  \href{http://dx.doi.org/10.1088/1748-0221/8/10/P10016}{{\em JINST} {\bfseries
  8} (2013) P10016},
\href{http://arxiv.org/abs/1306.3130}{{\ttfamily arXiv:1306.3130 [nucl-ex]}}.
%%CITATION = ARXIV:1306.3130;%%.

\bibitem{Aamodt:2010aa}
{\bfseries ALICE} Collaboration, K.~Aamodt {\em et~al.}, ``{Alignment of the
  ALICE Inner Tracking System with cosmic-ray tracks}'',
  \href{http://dx.doi.org/10.1088/1748-0221/5/03/P03003}{{\em JINST} {\bfseries
  5} (2010) P03003},
\href{http://arxiv.org/abs/1001.0502}{{\ttfamily arXiv:1001.0502
  [physics.ins-det]}}.
%%CITATION = ARXIV:1001.0502;%%.

\bibitem{ALICE-PUBLIC-2017-005}
{\bfseries ALICE Collaboration} Collaboration, ``{The ALICE definition of
  primary particles}'',. \url{https://cds.cern.ch/record/2270008}.

\bibitem{Adam:2015qaa}
{\bfseries ALICE} Collaboration, J.~Adam {\em et~al.}, ``{Measurement of pion,
  kaon and proton production in proton$-$proton collisions at $\sqrt{s} = 7$
  TeV}'', \href{http://dx.doi.org/10.1140/epjc/s10052-015-3422-9}{{\em Eur.
  Phys. J.} {\bfseries C75} no.~5, (2015) 226},
\href{http://arxiv.org/abs/1504.00024}{{\ttfamily arXiv:1504.00024 [nucl-ex]}}.
%%CITATION = ARXIV:1504.00024;%%.

\bibitem{ALICE:2012xs}
{\bfseries ALICE} Collaboration, B.~Abelev {\em et~al.}, ``{Pseudorapidity
  density of charged particles in $p$ + Pb collisions at $\sqrt{s_{\rm{NN}}} =
  5.02$ TeV}'', \href{http://dx.doi.org/10.1103/PhysRevLett.110.032301}{{\em
  Phys. Rev. Lett.} {\bfseries 110} no.~3, (2013) 032301},
\href{http://arxiv.org/abs/1210.3615}{{\ttfamily arXiv:1210.3615 [nucl-ex]}}.
%%CITATION = ARXIV:1210.3615;%%.

\bibitem{Alme:2010ke}
J.~Alme {\em et~al.}, ``{The ALICE TPC, a large 3-dimensional tracking device
  with fast readout for ultra-high multiplicity events}'',
  \href{http://dx.doi.org/10.1016/j.nima.2010.04.042}{{\em Nucl. Instrum.
  Meth.} {\bfseries A622} (2010) 316--367},
\href{http://arxiv.org/abs/1001.1950}{{\ttfamily arXiv:1001.1950
  [physics.ins-det]}}.
%%CITATION = ARXIV:1001.1950;%%.

\bibitem{Akindinov:2013tea}
A.~Akindinov {\em et~al.}, ``{Performance of the ALICE Time-Of-Flight detector
  at the LHC}'',
\href{http://dx.doi.org/10.1140/epjp/i2013-13044-x}{{\em Eur. Phys. J. Plus}
  {\bfseries 128} (2013) 44}.
%%CITATION = EPHJP,128,44;%%.

\bibitem{Martinengo:2011zza}
{\bfseries ALICE} Collaboration, P.~Martinengo, ``{The ALICE high momentum
  particle identification system: An overview after the first Large Hadron
  Collider run}'',
\href{http://dx.doi.org/10.1016/j.nima.2010.10.038}{{\em Nucl. Instrum. Meth.}
  {\bfseries A639} (2011) 7--10}.
%%CITATION = NUIMA,A639,7;%%.

\bibitem{Abelev:2012wca}
{\bfseries ALICE} Collaboration, B.~Abelev {\em et~al.}, ``{Pion, Kaon, and
  Proton Production in Central Pb--Pb Collisions at $\sqrt{s_{\rm{NN}}} = 2.76$
  TeV}'', \href{http://dx.doi.org/10.1103/PhysRevLett.109.252301}{{\em Phys.
  Rev. Lett.} {\bfseries 109} (2012) 252301},
\href{http://arxiv.org/abs/1208.1974}{{\ttfamily arXiv:1208.1974 [hep-ex]}}.
%%CITATION = ARXIV:1208.1974;%%.

\bibitem{Abelev:2013haa}
{\bfseries ALICE} Collaboration, B.~B. Abelev {\em et~al.}, ``{Multiplicity
  Dependence of Pion, Kaon, Proton and Lambda Production in p-Pb Collisions at
  $\sqrt{s_{\rm{NN}}}$ = 5.02 TeV}'',
  \href{http://dx.doi.org/10.1016/j.physletb.2013.11.020}{{\em Phys. Lett.}
  {\bfseries B728} (2014) 25--38},
\href{http://arxiv.org/abs/1307.6796}{{\ttfamily arXiv:1307.6796 [nucl-ex]}}.
%%CITATION = ARXIV:1307.6796;%%.

\bibitem{Adam:2016dau}
{\bfseries ALICE} Collaboration, J.~Adam {\em et~al.}, ``{Multiplicity
  dependence of charged pion, kaon, and (anti)proton production at large
  transverse momentum in p-Pb collisions at $\mathbf{\sqrt{{\textit s}_{\rm
  NN}}}$ = 5.02 TeV}'',
  \href{http://dx.doi.org/10.1016/j.physletb.2016.07.050}{{\em Phys. Lett.}
  {\bfseries B760} (2016) 720--735},
\href{http://arxiv.org/abs/1601.03658}{{\ttfamily arXiv:1601.03658 [nucl-ex]}}.
%%CITATION = ARXIV:1601.03658;%%.

\bibitem{Abelevetal:2014dna}
{\bfseries ALICE} Collaboration, B.~Abelev {\em et~al.}, ``{Technical Design
  Report for the Upgrade of the ALICE Inner Tracking System}'',
\href{http://dx.doi.org/10.1088/0954-3899/41/8/087002}{{\em J. Phys.}
  {\bfseries G41} (2014) 087002}.
%%CITATION = JPAGA,G41,087002;%%.

\bibitem{Adam:2016ilk}
{\bfseries ALICE} Collaboration, J.~Adam {\em et~al.}, ``{Determination of the
  event collision time with the ALICE detector at the LHC}'',
  \href{http://dx.doi.org/10.1140/epjp/i2017-11279-1}{{\em Eur. Phys. J. Plus}
  {\bfseries 132} no.~2, (2017) 99},
\href{http://arxiv.org/abs/1610.03055}{{\ttfamily arXiv:1610.03055
  [physics.ins-det]}}.
%%CITATION = ARXIV:1610.03055;%%.

\bibitem{DiBari:2003wy}
{\bfseries ALICE} Collaboration, D.~Di~Bari, ``{The pattern recognition method
  for the CsI-RICH detector in ALICE}'',
\href{http://dx.doi.org/10.1016/S0168-9002(03)00292-4}{{\em Nucl. Instrum.
  Meth.} {\bfseries A502} (2003) 300--304}.
%%CITATION = NUIMA,A502,300;%%.

\bibitem{Skands:2014pea}
P.~Skands, S.~Carrazza, and J.~Rojo, ``{Tuning PYTHIA 8.1: the Monash 2013
  Tune}'', \href{http://dx.doi.org/10.1140/epjc/s10052-014-3024-y}{{\em Eur.
  Phys. J.} {\bfseries C74} no.~8, (2014) 3024},
\href{http://arxiv.org/abs/1404.5630}{{\ttfamily arXiv:1404.5630 [hep-ph]}}.
%%CITATION = ARXIV:1404.5630;%%.

\bibitem{Wang:1991hta}
X.-N. Wang and M.~Gyulassy, ``{HIJING: A Monte Carlo model for multiple jet
  production in p p, p A and A A collisions}'',
\href{http://dx.doi.org/10.1103/PhysRevD.44.3501}{{\em Phys. Rev.} {\bfseries
  D44} (1991) 3501--3516}.
%%CITATION = PHRVA,D44,3501;%%.

\bibitem{Brun:1119728}
R.~Brun, F.~Bruyant, M.~Maire, A.~C. McPherson, and P.~Zanarini, {\em {GEANT 3:
  user's guide Geant 3.10, Geant 3.11; rev. version}}.
\newblock CERN, Geneva, 1987.
\newblock \url{https://cds.cern.ch/record/1119728}.

\bibitem{Agostinelli:2002hh}
{\bfseries GEANT4} Collaboration, S.~Agostinelli {\em et~al.}, ``{GEANT4: A
  Simulation toolkit}'',
\href{http://dx.doi.org/10.1016/S0168-9002(03)01368-8}{{\em Nucl. Instrum.
  Meth.} {\bfseries A506} (2003) 250--303}.
%%CITATION = NUIMA,A506,250;%%.

\bibitem{Aamodt:2010dx}
{\bfseries ALICE} Collaboration, K.~Aamodt {\em et~al.}, ``{Midrapidity
  antiproton-to-proton ratio in pp collisions at $\sqrt{s} = 0.9$ and $7$~TeV
  measured by the ALICE experiment}'',
  \href{http://dx.doi.org/10.1103/PhysRevLett.105.072002}{{\em Phys. Rev.
  Lett.} {\bfseries 105} (2010) 072002},
\href{http://arxiv.org/abs/1006.5432}{{\ttfamily arXiv:1006.5432 [hep-ex]}}.
%%CITATION = ARXIV:1006.5432;%%.

\bibitem{Battistoni:2007zzb}
G.~Battistoni, S.~Muraro, P.~R. Sala, F.~Cerutti, A.~Ferrari, S.~Roesler,
  A.~Fasso, and J.~Ranft, ``{The FLUKA code: Description and benchmarking}'',
  \href{http://dx.doi.org/10.1063/1.2720455}{{\em AIP Conf. Proc.} {\bfseries
  896} (2007) 31--49}.
[,31(2007)].
%%CITATION = APCPC,896,31;%%.

\bibitem{Adam:2016acv}
{\bfseries ALICE} Collaboration, J.~Adam {\em et~al.}, ``{Particle
  identification in ALICE: a Bayesian approach}'',
  \href{http://dx.doi.org/10.1140/epjp/i2016-16168-5}{{\em Eur. Phys. J. Plus}
  {\bfseries 131} no.~5, (2016) 168},
\href{http://arxiv.org/abs/1602.01392}{{\ttfamily arXiv:1602.01392
  [physics.data-an]}}.
%%CITATION = ARXIV:1602.01392;%%.

\bibitem{Beole:1998yq}
{\bfseries ALICE} Collaboration, S.~Beole {\em et~al.},
``{ALICE technical design report: Detector for high momentum PID}'',.
%%CITATION = CERN-LHCC-98-19;%%.

\bibitem{Kretzer:2000yf}
S.~Kretzer, ``{Fragmentation functions from flavor inclusive and flavor tagged
  e$^{+}$e$^{-}$ annihilations}'',
  \href{http://dx.doi.org/10.1103/PhysRevD.62.054001}{{\em Phys. Rev.}
  {\bfseries D62} (2000) 054001},
\href{http://arxiv.org/abs/hep-ph/0003177}{{\ttfamily arXiv:hep-ph/0003177
  [hep-ph]}}.
%%CITATION = HEP-PH/0003177;%%.

\bibitem{Schnedermann:1993ws}
E.~Schnedermann, J.~Sollfrank, and U.~W. Heinz, ``{Thermal phenomenology of
  hadrons from 200 AGeV S+S collisions}'',
  \href{http://dx.doi.org/10.1103/PhysRevC.48.2462}{{\em Phys. Rev.} {\bfseries
  C48} (1993) 2462--2475},
\href{http://arxiv.org/abs/nucl-th/9307020}{{\ttfamily arXiv:nucl-th/9307020
  [nucl-th]}}.
%%CITATION = NUCL-TH/9307020;%%.

\bibitem{Tsallis:1987eu}
C.~Tsallis, ``{Possible Generalization of Boltzmann-Gibbs Statistics}'',
\href{http://dx.doi.org/10.1007/BF01016429}{{\em J. Statist. Phys.} {\bfseries
  52} (1988) 479--487}.
%%CITATION = JSTPB,52,479;%%.

\bibitem{Abelev:2006cs}
{\bfseries STAR} Collaboration, B.~I. Abelev {\em et~al.}, ``{Strange particle
  production in p+p collisions at $\sqrt{s}$ = 200 GeV}'',
  \href{http://dx.doi.org/10.1103/PhysRevC.75.064901}{{\em Phys. Rev.}
  {\bfseries C75} (2007) 064901},
\href{http://arxiv.org/abs/nucl-ex/0607033}{{\ttfamily arXiv:nucl-ex/0607033
  [nucl-ex]}}.
%%CITATION = NUCL-EX/0607033;%%.

\bibitem{Heinz:2004qz}
U.~W. Heinz, ``{Concepts of heavy ion physics}'', in {\em {2002 European School
  of high-energy physics, Pylos, Greece, 25 Aug-7 Sep 2002: Proceedings}},
  pp.~165--238.
\newblock 2004.
\newblock \href{http://arxiv.org/abs/hep-ph/0407360}{{\ttfamily
  arXiv:hep-ph/0407360 [hep-ph]}}.
\newblock
\url{http://doc.cern.ch/yellowrep/CERN-2004-001}.
\newblock
%%CITATION = HEP-PH/0407360;%%.

\bibitem{Morsch:2017brb}
C.~Loizides and A.~Morsch, ``{Absence of jet quenching in peripheral
  nucleus-nucleus collisions}'',
  \href{http://dx.doi.org/10.1016/j.physletb.2017.09.002}{{\em Phys. Lett.}
  {\bfseries B773} (2017) 408--411},
\href{http://arxiv.org/abs/1705.08856}{{\ttfamily arXiv:1705.08856 [nucl-ex]}}.
%%CITATION = ARXIV:1705.08856;%%.

\bibitem{Adam:2016ddh}
{\bfseries ALICE} Collaboration, J.~Adam {\em et~al.}, ``{Centrality dependence
  of the pseudorapidity density distribution for charged particles in Pb-Pb
  collisions at $\sqrt{s_{\rm NN}}=5.02$ TeV}'',
  \href{http://dx.doi.org/10.1016/j.physletb.2017.07.017}{{\em Phys. Lett.}
  {\bfseries B772} (2017) 567--577},
\href{http://arxiv.org/abs/1612.08966}{{\ttfamily arXiv:1612.08966 [nucl-ex]}}.
%%CITATION = ARXIV:1612.08966;%%.

\bibitem{ALICE-PUBLIC-2015-008}
{\bfseries ALICE} Collaboration, ``{Centrality dependence of the
  charged-particle multiplicity density at midrapidity in Pb-Pb collisions at
  $\sqrt{s_{\rm NN}}$ = 5.02 TeV}'',. \url{https://cds.cern.ch/record/2118084}.

\bibitem{ALICE:2017jyt}
{\bfseries ALICE} Collaboration, J.~Adam {\em et~al.}, ``{Enhanced production
  of multi-strange hadrons in high-multiplicity proton-proton collisions}'',
  \href{http://dx.doi.org/10.1038/nphys4111}{{\em Nature Phys.} {\bfseries 13}
  (2017) 535--539},
\href{http://arxiv.org/abs/1606.07424}{{\ttfamily arXiv:1606.07424 [nucl-ex]}}.
%%CITATION = ARXIV:1606.07424;%%.

\bibitem{Andronic:2018qqt}
A.~Andronic, P.~Braun-Munzinger, B.~Friman, P.~M. Lo, K.~Redlich, and
  J.~Stachel, ``{The thermal proton yield anomaly in Pb-Pb collisions at the
  LHC and its resolution}'',
\href{http://arxiv.org/abs/1808.03102}{{\ttfamily arXiv:1808.03102 [hep-ph]}}.
%%CITATION = ARXIV:1808.03102;%%.

\bibitem{Stock:2018xaj}
R.~Stock, F.~Becattini, M.~Bleicher, and J.~Steinheimer, ``{The QCD Phase
  Diagram from Statistical Model Analysis}'',
  \href{http://dx.doi.org/10.1016/j.nuclphysa.2018.11.019}{{\em Nucl. Phys.}
  {\bfseries A982} (2019) 827--830},
\href{http://arxiv.org/abs/1811.07766}{{\ttfamily arXiv:1811.07766 [nucl-th]}}.
%%CITATION = ARXIV:1811.07766;%%.

\bibitem{Bernhard:2016tnd}
J.~E. Bernhard, J.~S. Moreland, S.~A. Bass, J.~Liu, and U.~Heinz, ``{Applying
  Bayesian parameter estimation to relativistic heavy-ion collisions:
  simultaneous characterization of the initial state and quark-gluon plasma
  medium}'', \href{http://dx.doi.org/10.1103/PhysRevC.94.024907}{{\em Phys.
  Rev.} {\bfseries C94} no.~2, (2016) 024907},
\href{http://arxiv.org/abs/1605.03954}{{\ttfamily arXiv:1605.03954 [nucl-th]}}.
%%CITATION = ARXIV:1605.03954;%%.

\bibitem{Moreland:2014oya}
J.~S. Moreland, J.~E. Bernhard, and S.~A. Bass, ``{Alternative ansatz to
  wounded nucleon and binary collision scaling in high-energy nuclear
  collisions}'', \href{http://dx.doi.org/10.1103/PhysRevC.92.011901}{{\em Phys.
  Rev.} {\bfseries C92} no.~1, (2015) 011901},
\href{http://arxiv.org/abs/1412.4708}{{\ttfamily arXiv:1412.4708 [nucl-th]}}.
%%CITATION = ARXIV:1412.4708;%%.

\bibitem{Adare:2013esx}
{\bfseries PHENIX} Collaboration, A.~Adare {\em et~al.}, ``{Spectra and ratios
  of identified particles in Au+Au and $d$+Au collisions at
  $\sqrt{s_{\rm{NN}}}=200$ GeV}'',
  \href{http://dx.doi.org/10.1103/PhysRevC.88.024906}{{\em Phys. Rev.}
  {\bfseries C88} no.~2, (2013) 024906},
\href{http://arxiv.org/abs/1304.3410}{{\ttfamily arXiv:1304.3410 [nucl-ex]}}.
%%CITATION = ARXIV:1304.3410;%%.

\bibitem{Adam:2017zbf}
{\bfseries ALICE} Collaboration, J.~Adam {\em et~al.}, ``{K$^{*}(892)^{0}$ and
  $\phi(1020)$ meson production at high transverse momentum in pp and Pb-Pb
  collisions at $\sqrt{s_\mathrm{NN}}$ = 2.76 TeV}'',
  \href{http://dx.doi.org/10.1103/PhysRevC.95.064606}{{\em Phys. Rev.}
  {\bfseries C95} no.~6, (2017) 064606},
\href{http://arxiv.org/abs/1702.00555}{{\ttfamily arXiv:1702.00555 [nucl-ex]}}.
%%CITATION = ARXIV:1702.00555;%%.

\bibitem{Adam:2014qja}
{\bfseries ALICE} Collaboration, J.~Adam {\em et~al.}, ``{Centrality dependence
  of particle production in p-Pb collisions at $\sqrt{s_{\rm{NN}}}$= 5.02
  TeV}'', \href{http://dx.doi.org/10.1103/PhysRevC.91.064905}{{\em Phys. Rev.}
  {\bfseries C91} no.~6, (2015) 064905},
\href{http://arxiv.org/abs/1412.6828}{{\ttfamily arXiv:1412.6828 [nucl-ex]}}.
%%CITATION = ARXIV:1412.6828;%%.

\bibitem{Acharya:2018njl}
{\bfseries ALICE} Collaboration, S.~Acharya {\em et~al.}, ``{Analysis of the
  apparent nuclear modification in peripheral Pb-Pb collisions at 5.02 TeV}'',
\href{http://arxiv.org/abs/1805.05212}{{\ttfamily arXiv:1805.05212 [nucl-ex]}}.
%%CITATION = ARXIV:1805.05212;%%.

\bibitem{Acharya:2018qsh}
{\bfseries ALICE} Collaboration, S.~Acharya {\em et~al.}, ``{Transverse
  momentum spectra and nuclear modification factors of charged particles in pp,
  p-Pb and Pb-Pb collisions at the LHC}'',
  \href{http://dx.doi.org/10.1007/JHEP11(2018)013}{{\em JHEP} {\bfseries 11}
  (2018) 013},
\href{http://arxiv.org/abs/1802.09145}{{\ttfamily arXiv:1802.09145 [nucl-ex]}}.
%%CITATION = ARXIV:1802.09145;%%.

\bibitem{Busza:2018rrf}
W.~Busza, K.~Rajagopal, and W.~van~der Schee, ``{Heavy Ion Collisions: The Big
  Picture, and the Big Questions}'',
  \href{http://dx.doi.org/10.1146/annurev-nucl-101917-020852}{{\em Ann. Rev.
  Nucl. Part. Sci.} {\bfseries 68} (2018) 339--376},
\href{http://arxiv.org/abs/1802.04801}{{\ttfamily arXiv:1802.04801 [hep-ph]}}.
%%CITATION = ARXIV:1802.04801;%%.

\bibitem{Werner:2013tya}
K.~Werner, B.~Guiot, I.~Karpenko, and T.~Pierog, ``{Analysing radial flow
  features in p-Pb and p-p collisions at several TeV by studying identified
  particle production in EPOS3}'',
  \href{http://dx.doi.org/10.1103/PhysRevC.89.064903}{{\em Phys. Rev.}
  {\bfseries C89} no.~6, (2014) 064903},
\href{http://arxiv.org/abs/1312.1233}{{\ttfamily arXiv:1312.1233 [nucl-th]}}.
%%CITATION = ARXIV:1312.1233;%%.

\bibitem{Song:2010aq}
H.~Song, S.~A. Bass, and U.~Heinz, ``{Viscous QCD matter in a hybrid
  hydrodynamic+Boltzmann approach}'',
  \href{http://dx.doi.org/10.1103/PhysRevC.83.024912}{{\em Phys. Rev.}
  {\bfseries C83} (2011) 024912},
\href{http://arxiv.org/abs/1012.0555}{{\ttfamily arXiv:1012.0555 [nucl-th]}}.
%%CITATION = ARXIV:1012.0555;%%.

\bibitem{Song:2007fn}
H.~Song and U.~W. Heinz, ``{Suppression of elliptic flow in a minimally viscous
  quark-gluon plasma}'',
  \href{http://dx.doi.org/10.1016/j.physletb.2007.11.019}{{\em Phys. Lett.}
  {\bfseries B658} (2008) 279--283},
\href{http://arxiv.org/abs/0709.0742}{{\ttfamily arXiv:0709.0742 [nucl-th]}}.
%%CITATION = ARXIV:0709.0742;%%.

\bibitem{Song:2007ux}
H.~Song and U.~W. Heinz, ``{Causal viscous hydrodynamics in 2+1 dimensions for
  relativistic heavy-ion collisions}'',
  \href{http://dx.doi.org/10.1103/PhysRevC.77.064901}{{\em Phys. Rev.}
  {\bfseries C77} (2008) 064901},
\href{http://arxiv.org/abs/0712.3715}{{\ttfamily arXiv:0712.3715 [nucl-th]}}.
%%CITATION = ARXIV:0712.3715;%%.

\bibitem{Bleicher:1999xi}
M.~Bleicher {\em et~al.}, ``{Relativistic hadron hadron collisions in the
  ultrarelativistic quantum molecular dynamics model}'',
  \href{http://dx.doi.org/10.1088/0954-3899/25/9/308}{{\em J. Phys.} {\bfseries
  G25} (1999) 1859--1896},
\href{http://arxiv.org/abs/hep-ph/9909407}{{\ttfamily arXiv:hep-ph/9909407
  [hep-ph]}}.
%%CITATION = HEP-PH/9909407;%%.

\bibitem{Bass:1998ca}
S.~A. Bass {\em et~al.}, ``{Microscopic models for ultrarelativistic heavy ion
  collisions}'', \href{http://dx.doi.org/10.1016/S0146-6410(98)00058-1}{{\em
  Prog. Part. Nucl. Phys.} {\bfseries 41} (1998) 255--369},
  \href{http://arxiv.org/abs/nucl-th/9803035}{{\ttfamily arXiv:nucl-th/9803035
  [nucl-th]}}.
[Prog. Part. Nucl. Phys.41,225(1998)].
%%CITATION = NUCL-TH/9803035;%%.

\bibitem{Lin:2004en}
Z.-W. Lin, C.~M. Ko, B.-A. Li, B.~Zhang, and S.~Pal, ``{A Multi-phase transport
  model for relativistic heavy ion collisions}'',
  \href{http://dx.doi.org/10.1103/PhysRevC.72.064901}{{\em Phys. Rev.}
  {\bfseries C72} (2005) 064901},
\href{http://arxiv.org/abs/nucl-th/0411110}{{\ttfamily arXiv:nucl-th/0411110
  [nucl-th]}}.
%%CITATION = NUCL-TH/0411110;%%.

\bibitem{Schenke:2012fw}
B.~Schenke, P.~Tribedy, and R.~Venugopalan, ``{Event-by-event gluon
  multiplicity, energy density, and eccentricities in ultrarelativistic
  heavy-ion collisions}'',
  \href{http://dx.doi.org/10.1103/PhysRevC.86.034908}{{\em Phys. Rev.}
  {\bfseries C86} (2012) 034908},
\href{http://arxiv.org/abs/1206.6805}{{\ttfamily arXiv:1206.6805 [hep-ph]}}.
%%CITATION = ARXIV:1206.6805;%%.

\bibitem{Ryu:2015vwa}
S.~Ryu, J.~F. Paquet, C.~Shen, G.~S. Denicol, B.~Schenke, S.~Jeon, and C.~Gale,
  ``{Importance of the Bulk Viscosity of QCD in Ultrarelativistic Heavy-Ion
  Collisions}'', \href{http://dx.doi.org/10.1103/PhysRevLett.115.132301}{{\em
  Phys. Rev. Lett.} {\bfseries 115} no.~13, (2015) 132301},
\href{http://arxiv.org/abs/1502.01675}{{\ttfamily arXiv:1502.01675 [nucl-th]}}.
%%CITATION = ARXIV:1502.01675;%%.

\bibitem{Gelis:2012ri}
F.~Gelis, ``{Color Glass Condensate and Glasma}'',
  \href{http://dx.doi.org/10.1142/S0217751X13300019}{{\em Int. J. Mod. Phys.}
  {\bfseries A28} (2013) 1330001},
\href{http://arxiv.org/abs/1211.3327}{{\ttfamily arXiv:1211.3327 [hep-ph]}}.
%%CITATION = ARXIV:1211.3327;%%.

\bibitem{Werner:2012xh}
K.~Werner, I.~Karpenko, M.~Bleicher, T.~Pierog, and S.~Porteboeuf-Houssais,
  ``{Jets, Bulk Matter, and their Interaction in Heavy Ion Collisions at
  Several TeV}'', \href{http://dx.doi.org/10.1103/PhysRevC.85.064907}{{\em
  Phys. Rev.} {\bfseries C85} (2012) 064907},
\href{http://arxiv.org/abs/1203.5704}{{\ttfamily arXiv:1203.5704 [nucl-th]}}.
%%CITATION = ARXIV:1203.5704;%%.

\bibitem{Acharya:2018zuq}
{\bfseries ALICE} Collaboration, S.~Acharya {\em et~al.}, ``{Anisotropic flow
  of identified particles in Pb-Pb collisions at $
  {\sqrt{s}}_{\mathrm{NN}}=5.02 $ TeV}'',
  \href{http://dx.doi.org/10.1007/JHEP09(2018)006}{{\em JHEP} {\bfseries 09}
  (2018) 006},
\href{http://arxiv.org/abs/1805.04390}{{\ttfamily arXiv:1805.04390 [nucl-ex]}}.
%%CITATION = ARXIV:1805.04390;%%.

\bibitem{Abelev:2014pua}
{\bfseries ALICE} Collaboration, B.~B. Abelev {\em et~al.}, ``{Elliptic flow of
  identified hadrons in Pb-Pb collisions at $ \sqrt{s_{\mathrm{NN}}}=2.76 $
  TeV}'', \href{http://dx.doi.org/10.1007/JHEP06(2015)190}{{\em JHEP}
  {\bfseries 06} (2015) 190},
\href{http://arxiv.org/abs/1405.4632}{{\ttfamily arXiv:1405.4632 [nucl-ex]}}.
%%CITATION = ARXIV:1405.4632;%%.

\bibitem{Abelev:2012di}
{\bfseries ALICE} Collaboration, B.~Abelev {\em et~al.}, ``{Anisotropic flow of
  charged hadrons, pions and (anti-)protons measured at high transverse
  momentum in Pb-Pb collisions at $\sqrt{s_{NN}}$=2.76 TeV}'',
  \href{http://dx.doi.org/10.1016/j.physletb.2012.12.066}{{\em Phys. Lett.}
  {\bfseries B719} (2013) 18--28},
\href{http://arxiv.org/abs/1205.5761}{{\ttfamily arXiv:1205.5761 [nucl-ex]}}.
%%CITATION = ARXIV:1205.5761;%%.

\end{thebibliography}\endgroup
 %\end{thebibliography}

 %%%%%%%%% appendix with author list
 \newpage
 \appendix
\section{The ALICE Collaboration}
 \label{app:collab}
% Collaboration: CERN-LHC-ALICE
% Generation Date is 2019-Aug-04

% How to use:
%%%%%%%%% appendix with author list
%\appendix
%\section{The ALICE Collaboration}
%\label{app:collab}
%\input{Alice_Authorslist_XXXX-Axx-XX.tex}
\begingroup
\small
\begin{flushleft}
S.~Acharya\Irefn{org141}\And 
D.~Adamov\'{a}\Irefn{org93}\And 
S.P.~Adhya\Irefn{org141}\And 
A.~Adler\Irefn{org73}\And 
J.~Adolfsson\Irefn{org79}\And 
M.M.~Aggarwal\Irefn{org98}\And 
G.~Aglieri Rinella\Irefn{org34}\And 
M.~Agnello\Irefn{org31}\And 
N.~Agrawal\Irefn{org10}\textsuperscript{,}\Irefn{org48}\textsuperscript{,}\Irefn{org53}\And 
Z.~Ahammed\Irefn{org141}
% ...
\And
\renewcommand\labelenumi{\textsuperscript{\theenumi}~}

\section*{Affiliation notes}
\renewcommand\theenumi{\roman{enumi}}
\begin{Authlist}
\item \Adef{org*}Deceased
\item \Adef{orgI}Dipartimento DET del Politecnico di Torino, Turin, Italy
\item \Adef{orgII}M.V. Lomonosov Moscow State University, D.V. Skobeltsyn Institute of Nuclear, Physics, Moscow, Russia
\item \Adef{orgIII}Department of Applied Physics, Aligarh Muslim University, Aligarh, India
\item \Adef{orgIV}Institute of Theoretical Physics, University of Wroclaw, Poland
\end{Authlist}

\section*{Collaboration Institutes}
\renewcommand\theenumi{\arabic{enumi}~}
\begin{Authlist}
\item \Idef{org1}A.I. Alikhanyan National Science Laboratory (Yerevan Physics Institute) Foundation, Yerevan, Armenia
\item \Idef{org2}Bogolyubov Institute for Theoretical Physics, National Academy of Sciences of Ukraine, Kiev, Ukraine
\item \Idef{org3}Bose Institute, Department of Physics  and Centre for Astroparticle Physics and Space Science (CAPSS), Kolkata, India
\item \Idef{org4}Budker Institute for Nuclear Physics, Novosibirsk, Russia
\item \Idef{org5}California Polytechnic State University, San Luis Obispo, California, United States
\item \Idef{org6}Central China Normal University, Wuhan, China
\item \Idef{org7}Centre de Calcul de l'IN2P3, Villeurbanne, Lyon, France
\item \Idef{org8}Centro de Aplicaciones Tecnol\'{o}gicas y Desarrollo Nuclear (CEADEN), Havana, Cuba
\item \Idef{org9}Centro de Investigaci\'{o}n y de Estudios Avanzados (CINVESTAV), Mexico City and M\'{e}rida, Mexico
\item \Idef{org10}Centro Fermi - Museo Storico della Fisica e Centro Studi e Ricerche ``Enrico Fermi', Rome, Italy
\item \Idef{org11}Chicago State University, Chicago, Illinois, United States
\item \Idef{org12}China Institute of Atomic Energy, Beijing, China
\item \Idef{org13}Chonbuk National University, Jeonju, Republic of Korea
\item \Idef{org14}Comenius University Bratislava, Faculty of Mathematics, Physics and Informatics, Bratislava, Slovakia
\item \Idef{org15}COMSATS University Islamabad, Islamabad, Pakistan
\item \Idef{org16}Creighton University, Omaha, Nebraska, United States
\item \Idef{org17}Department of Physics, Aligarh Muslim University, Aligarh, India
\item \Idef{org18}Department of Physics, Pusan National University, Pusan, Republic of Korea
\item \Idef{org19}Department of Physics, Sejong University, Seoul, Republic of Korea
\item \Idef{org20}Department of Physics, University of California, Berkeley, California, United States
\item \Idef{org21}Department of Physics, University of Oslo, Oslo, Norway
\item \Idef{org22}Department of Physics and Technology, University of Bergen, Bergen, Norway
\item \Idef{org23}Dipartimento di Fisica dell'Universit\`{a} 'La Sapienza' and Sezione INFN, Rome, Italy
\item \Idef{org24}Dipartimento di Fisica dell'Universit\`{a} and Sezione INFN, Cagliari, Italy
\item \Idef{org25}Dipartimento di Fisica dell'Universit\`{a} and Sezione INFN, Trieste, Italy
\item \Idef{org26}Dipartimento di Fisica dell'Universit\`{a} and Sezione INFN, Turin, Italy
\item \Idef{org27}Dipartimento di Fisica e Astronomia dell'Universit\`{a} and Sezione INFN, Bologna, Italy
\item \Idef{org28}Dipartimento di Fisica e Astronomia dell'Universit\`{a} and Sezione INFN, Catania, Italy
\item \Idef{org29}Dipartimento di Fisica e Astronomia dell'Universit\`{a} and Sezione INFN, Padova, Italy
\item \Idef{org30}Dipartimento di Fisica `E.R.~Caianiello' dell'Universit\`{a} and Gruppo Collegato INFN, Salerno, Italy
\item \Idef{org31}Dipartimento DISAT del Politecnico and Sezione INFN, Turin, Italy
\item \Idef{org32}Dipartimento di Scienze e Innovazione Tecnologica dell'Universit\`{a} del Piemonte Orientale and INFN Sezione di Torino, Alessandria, Italy
\item \Idef{org33}Dipartimento Interateneo di Fisica `M.~Merlin' and Sezione INFN, Bari, Italy
\item \Idef{org34}European Organization for Nuclear Research (CERN), Geneva, Switzerland
\item \Idef{org35}Faculty of Electrical Engineering, Mechanical Engineering and Naval Architecture, University of Split, Split, Croatia
\item \Idef{org36}Faculty of Engineering and Science, Western Norway University of Applied Sciences, Bergen, Norway
\item \Idef{org37}Faculty of Nuclear Sciences and Physical Engineering, Czech Technical University in Prague, Prague, Czech Republic
\item \Idef{org38}Faculty of Science, P.J.~\v{S}af\'{a}rik University, Ko\v{s}ice, Slovakia
\item \Idef{org39}Frankfurt Institute for Advanced Studies, Johann Wolfgang Goethe-Universit\"{a}t Frankfurt, Frankfurt, Germany
\item \Idef{org40}Gangneung-Wonju National University, Gangneung, Republic of Korea
\item \Idef{org41}Gauhati University, Department of Physics, Guwahati, India
\item \Idef{org42}Helmholtz-Institut f\"{u}r Strahlen- und Kernphysik, Rheinische Friedrich-Wilhelms-Universit\"{a}t Bonn, Bonn, Germany
\item \Idef{org43}Helsinki Institute of Physics (HIP), Helsinki, Finland
\item \Idef{org44}High Energy Physics Group,  Universidad Aut\'{o}noma de Puebla, Puebla, Mexico
\item \Idef{org45}Hiroshima University, Hiroshima, Japan
\item \Idef{org46}Hochschule Worms, Zentrum  f\"{u}r Technologietransfer und Telekommunikation (ZTT), Worms, Germany
\item \Idef{org47}Horia Hulubei National Institute of Physics and Nuclear Engineering, Bucharest, Romania
\item \Idef{org48}Indian Institute of Technology Bombay (IIT), Mumbai, India
\item \Idef{org49}Indian Institute of Technology Indore, Indore, India
\item \Idef{org50}Indonesian Institute of Sciences, Jakarta, Indonesia
\item \Idef{org51}INFN, Laboratori Nazionali di Frascati, Frascati, Italy
\item \Idef{org52}INFN, Sezione di Bari, Bari, Italy
\item \Idef{org53}INFN, Sezione di Bologna, Bologna, Italy
\item \Idef{org54}INFN, Sezione di Cagliari, Cagliari, Italy
\item \Idef{org55}INFN, Sezione di Catania, Catania, Italy
\item \Idef{org56}INFN, Sezione di Padova, Padova, Italy
\item \Idef{org57}INFN, Sezione di Roma, Rome, Italy
\item \Idef{org58}INFN, Sezione di Torino, Turin, Italy
\item \Idef{org59}INFN, Sezione di Trieste, Trieste, Italy
\item \Idef{org60}Inha University, Incheon, Republic of Korea
\item \Idef{org61}Institut de Physique Nucl\'{e}aire d'Orsay (IPNO), Institut National de Physique Nucl\'{e}aire et de Physique des Particules (IN2P3/CNRS), Universit\'{e} de Paris-Sud, Universit\'{e} Paris-Saclay, Orsay, France
\item \Idef{org62}Institute for Nuclear Research, Academy of Sciences, Moscow, Russia
\item \Idef{org63}Institute for Subatomic Physics, Utrecht University/Nikhef, Utrecht, Netherlands
\item \Idef{org64}Institute of Experimental Physics, Slovak Academy of Sciences, Ko\v{s}ice, Slovakia
\item \Idef{org65}Institute of Physics, Homi Bhabha National Institute, Bhubaneswar, India
\item \Idef{org66}Institute of Physics of the Czech Academy of Sciences, Prague, Czech Republic
\item \Idef{org67}Institute of Space Science (ISS), Bucharest, Romania
\item \Idef{org68}Institut f\"{u}r Kernphysik, Johann Wolfgang Goethe-Universit\"{a}t Frankfurt, Frankfurt, Germany
\item \Idef{org69}Instituto de Ciencias Nucleares, Universidad Nacional Aut\'{o}noma de M\'{e}xico, Mexico City, Mexico
\item \Idef{org70}Instituto de F\'{i}sica, Universidade Federal do Rio Grande do Sul (UFRGS), Porto Alegre, Brazil
\item \Idef{org71}Instituto de F\'{\i}sica, Universidad Nacional Aut\'{o}noma de M\'{e}xico, Mexico City, Mexico
\item \Idef{org72}iThemba LABS, National Research Foundation, Somerset West, South Africa
\item \Idef{org73}Johann-Wolfgang-Goethe Universit\"{a}t Frankfurt Institut f\"{u}r Informatik, Fachbereich Informatik und Mathematik, Frankfurt, Germany
\item \Idef{org74}Joint Institute for Nuclear Research (JINR), Dubna, Russia
\item \Idef{org75}Korea Institute of Science and Technology Information, Daejeon, Republic of Korea
\item \Idef{org76}KTO Karatay University, Konya, Turkey
\item \Idef{org77}Laboratoire de Physique Subatomique et de Cosmologie, Universit\'{e} Grenoble-Alpes, CNRS-IN2P3, Grenoble, France
\item \Idef{org78}Lawrence Berkeley National Laboratory, Berkeley, California, United States
\item \Idef{org79}Lund University Department of Physics, Division of Particle Physics, Lund, Sweden
\item \Idef{org80}Nagasaki Institute of Applied Science, Nagasaki, Japan
\item \Idef{org81}Nara Women{'}s University (NWU), Nara, Japan
\item \Idef{org82}National and Kapodistrian University of Athens, School of Science, Department of Physics , Athens, Greece
\item \Idef{org83}National Centre for Nuclear Research, Warsaw, Poland
\item \Idef{org84}National Institute of Science Education and Research, Homi Bhabha National Institute, Jatni, India
\item \Idef{org85}National Nuclear Research Center, Baku, Azerbaijan
\item \Idef{org86}National Research Centre Kurchatov Institute, Moscow, Russia
\item \Idef{org87}Niels Bohr Institute, University of Copenhagen, Copenhagen, Denmark
\item \Idef{org88}Nikhef, National institute for subatomic physics, Amsterdam, Netherlands
\item \Idef{org89}NRC Kurchatov Institute IHEP, Protvino, Russia
\item \Idef{org90}NRC «Kurchatov Institute»  - ITEP, Moscow, Russia
\item \Idef{org91}NRNU Moscow Engineering Physics Institute, Moscow, Russia
\item \Idef{org92}Nuclear Physics Group, STFC Daresbury Laboratory, Daresbury, United Kingdom
\item \Idef{org93}Nuclear Physics Institute of the Czech Academy of Sciences, \v{R}e\v{z} u Prahy, Czech Republic
\item \Idef{org94}Oak Ridge National Laboratory, Oak Ridge, Tennessee, United States
\item \Idef{org95}Ohio State University, Columbus, Ohio, United States
\item \Idef{org96}Petersburg Nuclear Physics Institute, Gatchina, Russia
\item \Idef{org97}Physics department, Faculty of science, University of Zagreb, Zagreb, Croatia
\item \Idef{org98}Physics Department, Panjab University, Chandigarh, India
\item \Idef{org99}Physics Department, University of Jammu, Jammu, India
\item \Idef{org100}Physics Department, University of Rajasthan, Jaipur, India
\item \Idef{org101}Physikalisches Institut, Eberhard-Karls-Universit\"{a}t T\"{u}bingen, T\"{u}bingen, Germany
\item \Idef{org102}Physikalisches Institut, Ruprecht-Karls-Universit\"{a}t Heidelberg, Heidelberg, Germany
\item \Idef{org103}Physik Department, Technische Universit\"{a}t M\"{u}nchen, Munich, Germany
\item \Idef{org104}Politecnico di Bari, Bari, Italy
\item \Idef{org105}Research Division and ExtreMe Matter Institute EMMI, GSI Helmholtzzentrum f\"ur Schwerionenforschung GmbH, Darmstadt, Germany
\item \Idef{org106}Rudjer Bo\v{s}kovi\'{c} Institute, Zagreb, Croatia
\item \Idef{org107}Russian Federal Nuclear Center (VNIIEF), Sarov, Russia
\item \Idef{org108}Saha Institute of Nuclear Physics, Homi Bhabha National Institute, Kolkata, India
\item \Idef{org109}School of Physics and Astronomy, University of Birmingham, Birmingham, United Kingdom
\item \Idef{org110}Secci\'{o}n F\'{\i}sica, Departamento de Ciencias, Pontificia Universidad Cat\'{o}lica del Per\'{u}, Lima, Peru
\item \Idef{org111}Shanghai Institute of Applied Physics, Shanghai, China
\item \Idef{org112}St. Petersburg State University, St. Petersburg, Russia
\item \Idef{org113}Stefan Meyer Institut f\"{u}r Subatomare Physik (SMI), Vienna, Austria
\item \Idef{org114}SUBATECH, IMT Atlantique, Universit\'{e} de Nantes, CNRS-IN2P3, Nantes, France
\item \Idef{org115}Suranaree University of Technology, Nakhon Ratchasima, Thailand
\item \Idef{org116}Technical University of Ko\v{s}ice, Ko\v{s}ice, Slovakia
\item \Idef{org117}Technische Universit\"{a}t M\"{u}nchen, Excellence Cluster 'Universe', Munich, Germany
\item \Idef{org118}The Henryk Niewodniczanski Institute of Nuclear Physics, Polish Academy of Sciences, Cracow, Poland
\item \Idef{org119}The University of Texas at Austin, Austin, Texas, United States
\item \Idef{org120}Universidad Aut\'{o}noma de Sinaloa, Culiac\'{a}n, Mexico
\item \Idef{org121}Universidade de S\~{a}o Paulo (USP), S\~{a}o Paulo, Brazil
\item \Idef{org122}Universidade Estadual de Campinas (UNICAMP), Campinas, Brazil
\item \Idef{org123}Universidade Federal do ABC, Santo Andre, Brazil
\item \Idef{org124}University of Cape Town, Cape Town, South Africa
\item \Idef{org125}University of Houston, Houston, Texas, United States
\item \Idef{org126}University of Jyv\"{a}skyl\"{a}, Jyv\"{a}skyl\"{a}, Finland
\item \Idef{org127}University of Liverpool, Liverpool, United Kingdom
\item \Idef{org128}University of Science and Techonology of China, Hefei, China
\item \Idef{org129}University of South-Eastern Norway, Tonsberg, Norway
\item \Idef{org130}University of Tennessee, Knoxville, Tennessee, United States
\item \Idef{org131}University of the Witwatersrand, Johannesburg, South Africa
\item \Idef{org132}University of Tokyo, Tokyo, Japan
\item \Idef{org133}University of Tsukuba, Tsukuba, Japan
\item \Idef{org134}Universit\'{e} Clermont Auvergne, CNRS/IN2P3, LPC, Clermont-Ferrand, France
\item \Idef{org135}Universit\'{e} de Lyon, Universit\'{e} Lyon 1, CNRS/IN2P3, IPN-Lyon, Villeurbanne, Lyon, France
\item \Idef{org136}Universit\'{e} de Strasbourg, CNRS, IPHC UMR 7178, F-67000 Strasbourg, France, Strasbourg, France
\item \Idef{org137}Universit\'{e} Paris-Saclay Centre d'Etudes de Saclay (CEA), IRFU, D\'{e}partment de Physique Nucl\'{e}aire (DPhN), Saclay, France
\item \Idef{org138}Universit\`{a} degli Studi di Foggia, Foggia, Italy
\item \Idef{org139}Universit\`{a} degli Studi di Pavia, Pavia, Italy
\item \Idef{org140}Universit\`{a} di Brescia, Brescia, Italy
\item \Idef{org141}Variable Energy Cyclotron Centre, Homi Bhabha National Institute, Kolkata, India
\item \Idef{org142}Warsaw University of Technology, Warsaw, Poland
\item \Idef{org143}Wayne State University, Detroit, Michigan, United States
\item \Idef{org144}Westf\"{a}lische Wilhelms-Universit\"{a}t M\"{u}nster, Institut f\"{u}r Kernphysik, M\"{u}nster, Germany
\item \Idef{org145}Wigner Research Centre for Physics, Hungarian Academy of Sciences, Budapest, Hungary
\item \Idef{org146}Yale University, New Haven, Connecticut, United States
\item \Idef{org147}Yonsei University, Seoul, Republic of Korea
\end{Authlist}
\endgroup
  %%%%%%% done by webmaster team
\end{document}